\documentclass[useAMS,usenatbib]{mn2e}

\usepackage{graphics} 
\usepackage{graphicx}
\usepackage{amsxtra}
\usepackage{array,longtable}
\usepackage{color}
\newcommand{\ser}{S\'ersic}

\title[2MASS photometry of edge-on spiral galaxies. I. Sample and general
results]{2MASS photometry of edge-on spiral galaxies. I. Sample and general
results}
\author[A. V. Mosenkov, N. Ya. Sotnikova and V. P. Reshetnikov]{A. V.
Mosenkov, N. Ya. Sotnikova\thanks{E-mail: nsot@astro.spbu.ru} and 
V. P. Reshetnikov
\\
St.Petersburg State University, Universitetskij pr. 28, 198504
St.Petersburg, Stary Peterhof, Russia \\ and Isaac Newton Institute
of Chile, St.Petersburg Branch}

\begin{document}

\date{Accepted 2009 September 6. Received September 6; in original form 2009 July 2}

\pagerange{\pageref{firstpage}--\pageref{lastpage}} \pubyear{2009}

\maketitle

\label{firstpage}
\begin{abstract}
A sample of edge-on spiral galaxies aimed at a thorough study of the main 
structural and photometric parameters of edge-on galaxies both of early
and late types is presented. The data were taken from the Two Micron All
Sky Survey (2MASS) in the $J$, $H$ and 
$K_s$ filters. The sources were selected according to their 
angular size mainly on the basis of the 2MASS-selected Flat Galaxy Catalog 
(2MFGC). The sample consists of 175 galaxies in the $K_s$-filter, 
169 galaxies in the $H$-filter and 165 galaxies in the $J$-filter. We present 
bulge and disc decompositions of each galaxy image. All 
galaxies have been modelled with a \ser\ bulge and exponential disc with 
the BUDDA v2.1 package. Bulge and disc sizes, profile shapes, surface 
brightnesses are provided. 

Our sample is the biggest up-to-date sample of edge-on galaxies with derived 
structural parameters for discs and bulges. In this paper we present the 
general results of the study of this sample. We determine several scaling 
relations for bulges and discs which 
indicate a tight link between their formation and evolution. We 
show that galaxies with bulges fitted by the \ser\ index $n \la 2$ have quite 
different distributions of their structural parameters 
than galaxies with $n \ga 2$ bulges. At a first 
approximation the \ser\ index threshold $n \simeq 2$ can be used to identify 
pseudobulges and classical bulges. Thus, the difference in parameter 
distributions and scaling relations for these subsamples
suggests that two or more processes are responsible for disk galaxy 
formation. The main conclusions 
of our general statistical analysis of the sample are:

(1) The distribution of the apparent bulge axis ratio $q_\mathrm{b}$ for 
the subsample with $n \la 2$ can be attributed to triaxial, nearly prolate 
bulges that are seen from different projections, while $n \ga 2$ bulges seem 
to be oblate spheroids with moderate flattening. 
Triaxiality of late-type bulges may be due to the presence of a bar 
that thickened in the vertical direction during its secular evolution. 

(2) For the sample galaxies, the effective radius of the bulge 
$r_\mathrm{e,b}$, the disc scalelength $h$ and the disc scaleheight 
$z_0$ are well correlated. However, there is a clear trend for the ratio 
$r_\mathrm{e,b}/ h$ to increase with $n$. As $n$ is an indicator of the Hubble 
type, such a trend unambiguously rules out the widely discussed hypothesis of 
a scale-free Hubble sequence. The found correlation between $z_0$ and 
$r_\mathrm{e,b}$ is new and was not described earlier. 

(3) There is a hint that the fundamental planes of discs, which links only 
disc parameters and the maximum rotational velocity of gas, are different 
for galaxies with different bulges. This may indicate a real difference of 
discs in galaxies with low and high concentration bulges.

(4) The most surprising result arises from the investigation of the 
Photometric Plane of sample bulges. It turns that the plane is not flat and 
has a prominent curvature towards small values of $n$. For bulges this 
fact was not noticed earlier. 

(5) The clear relation between the flattening of stellar discs 
$h/z_0$ and the relative mass of a spherical component, including a dark halo, 
is confirmed not for bulgeless galaxies but for galaxies with massive 
bulges.

Many of our results are in good agreement with the results of other authors, 
several ones are new. 
Thus, our sample is very useful for further detailed studying and modelling
of the edge-on spiral galaxies.
\end{abstract}

\begin{keywords}
galaxies: fundamental parameters --- galaxies: general --- galaxies: infrared
--- galaxies: spiral edge-on galaxies --- galaxies: photometry.
\end{keywords}

\section{Introduction}

The photometric and kinematic study of spiral galaxies provides a method to tackle
the problems concerning their formation and evolution. A lot of
observational data has been obtained in recent decades, especially with the
advent of CCD arrays. Over the last years, significant progress has been
made in both observational and theoretical studies aimed at understanding
the structure and evolution of galaxies. Modern sky surveys (2MASS, SDSS, etc.)
provide information on the characteristics and spatial distribution of millions 
of galaxies. The large available amount of information has altered the face 
of modern extragalactic astronomy. Many important tasks, including detailed 
photometric studies of the structure of nearby galaxies, can be now solved 
without additional observations by telescopes. Here we focus on the surface 
photometry of spiral galaxies on the basis of the 2MASS near-infrared survey.

The parametrization of galaxies is a way to describe the multitude of 
normal galaxies without any peculiarities within their structure.
Various fitting functions are used to derive structural parameters of bulges 
and discs. Traditional fitting functions for stellar discs  include 
an exponential function (de Vaucouleurs 1959, Freeman 1970) and 
an inner-truncated exponential function (Kormendy 1977). Bulges 
are well described by the so-called \ser\ profile that includes the de 
Vaucouleurs profile as a special case (S\'ersic 1968). The de Vaucouleurs 
law, or $r^{1/4}$-law, was obtained empirically for elliptical galaxies 
(de Vaucouleurs 1948) and later for bulges of early-type galaxies 
(de Vaucouleurs 1959). Later it was shown that the \ser\ law is preferable 
to fit spheroidal components of galaxies (e.g. Andredakis et al. 1995). 
The $z$-distribution of surface brightness of edge-on discs is exponential or 
the so called ``sech$^2$-distribution'' (van der Kruit \& Searle 1981). 
(All the mentioned functions are described in Section~3 of this paper.)

In order to derive structural parameters of bulges and discs, we must perform
photometric bulge/disc decomposition of galaxy images. There are numerous 
methods of such decomposition. The one-dimensional analysis of the 
surface-brightness distribution is the most commonly used  method
(e.g. Boroson 1981; Kent 1985; 
Baggett, Baggett \& Anderson 1998; Bizyaev \& Mitronova 2002, 
hereafter BM02). But as was shown by Byun \& Freeman (1995) the one-dimensional 
decomposition procedure leads to strong systematic errors. The two-dimensional 
approach provides much better estimates of the structural parameters for each 
component of a galaxy. This method is powerful when we need to separate the 
non-axisymmetric structures (bars, rings, spiral arms) from axisymmetric 
components (disc and bulge).

In recent decades a number of studies concerning structural properties of 
galaxies has been performed. The results of decompositions of near face-on 
disc dominated galaxies as well as of early-type galaxies were presented, 
for instance, by de Jong (1996). It was concluded that there is a correlation 
between structural parameters of discs and bulges. 
M\"ollenhoff \& Heidt (2001) found similar correlations and speculated
about the formation and evolution of spiral galaxies. Graham (2001) performed 
a detailed analysis of low-inclination spiral galaxies and described many 
significant properties of their bulges and discs. Other studies dealt with 
the correlation between the scalelengths of bulges and discs 
(e.g. Courteau et al. 1996; de Jong 1996), 
the presence of additional morphological components such as bars, lenses, 
rings, inner bars and inner discs 
(e.g. de Souza, Gadotti \& dos Anjos 2004, SGA04 hereafter; 
Castro-Rodr\'iguez \& Garz\'on 2003; Laurikainen, Salo \& Buta 2005; 
Gadotti 2009) 
and the revision of morphological classification of galaxies (SGA04). The 
fundamental plane (FP) of discs as well as the photometric plane of bulges 
were also widely discussed 
(e.g. Khosroshahi et al. 2000a,b; M\"ollenhoff \& Heidt 2001; 
Moriondo, Giovanelli \& Haynes 1999). 

The main aim of our project is to perform decompositions of a large number of 
edge-on galaxies both of early and late types into bulges and discs. We chose 
this type of galaxies because they provide a unique possibility to obtain 
information about the vertical structure of galactic discs. In the edge-on view 
non-axisymmetric features such as bars, rings or spiral arms are not seen. 
On the other hand, the vertical distribution of light in edge-on galaxies 
allows us to study a thin disc, a thick disc and a stellar halo (see e.g. 
Yoachim \& Dalcanton 2005; Seth, Dalcanton \& de Jong 2005). 

Unfortunately, there are some difficulties in studying edge-on galaxies. 
In the optical band we need to take into account the dust extinction 
in the plane of a disc. The distribution of dust is often unknown and it is 
hard to obtain true values of disc and bulge parameters. The advantage of NIR 
observations is that the general extinction is drastically reduced, 
although we can not exclude it completely. Nevertheless it is better to perform 
the decomposition of galaxies by using their images in red and infrared bands. 
Using three IR bands for galaxy decomposition and subsequent statistical 
analysis of galaxy properties in all three bands encourages us in our 
suggestion that extinction problems are not very severe (at least
for general statistical correlations which are the subject of the
present work). 
However, edge-on galaxies are rare in occurrence 
in contrast to face-on galaxies or galaxies with intermediate
inclination angles. Therefore, in order to create a large sample of edge-on 
galaxies it is necessary to use special catalogues which include a lot of 
edge-on spiral galaxies such as the Revised Flat Galaxy Catalog 
(Karachentsev et al. 1999).

A detailed study of the disks of edge-on galaxies is described in many
papers (e.g., van der Kruit \& Searle 1981; Reshetnikov \& Combes 1997;
de Grijs 1998). In a more recent study, 
Dalcanton \& Bernstein (2000, 2002) examined a sample of extremely late-type, 
edge-on spirals with no apparent bulges and concluded that thick disks 
are common around galaxy disks of all masses. BM02 also analyzed a sample of 
late-type edge-on galaxies in the $J$, $H$ and $K_s$-bands to compare the mean 
ratios $h/z_0$ in these bands. They have noted a strong correlation between 
the central surface brightness of the disk and the $h/z_0$ ratio: the thinner 
the galaxy, the lower the central surface brightness reduced to the face-on 
inclination. 
Kregel, van der Kruit \& de Grijs (2002), Kregel (2003),  
Kregel, van der Kruit \& Freeman (2005) and Zasov et al. (2002) presented 
a complex study (dynamical properties of the stellar discs, three-dimensional 
disc structure, stellar kinematics, rotation curves) of several dozens of 
edge-on spirals. Unfortunately, there are no extended studies of structural 
parameters of edge-on galaxies both of early and late types which join 
analysis of their bulges and discs. Our study is intended to fill this gap. 
We analyse the statistical properties of the sample edge-on galaxies and 
present correlations between the global structural parameters of {\it bulges} 
and {\it discs}, including those that describe their vertical structure, 
in the near-infrared bands. 

This paper is organized as follows. 
In Section~2 we present the detailed description of our sample of spiral 
edge-on galaxies which includes the selection criteria 
and a brief description of the data sources. We also describe 
the completeness of the sample and its main properties. 
In Section~3 we use the two-dimensional least-squares algorithm to decompose 
the galaxy images into a bulge and a disc. We recall some aspects of the code 
BUDDA (SGA04) which we used as the decomposition tool. 
In Section~4, we present the results of image decomposition in the 
$J$, $H$ and $K_s$-bands. We compare our results with previous results from 
the literature and analyse the reliability of the results of the BUDDA 
decompositions. 
In Section~5 we describe general properties of edge-on spiral galaxies and
discuss some correlations between galactic subsystems. Some of them are 
well-known but there are also some new results. 
In Section~6 we summarize our main conclusions. 
In the forthcoming papers we are going to discuss properties of edge-on 
galaxies in more detail.

\section[]{The Sample}

We intended to find edge-on galaxies with a wide range of bulges: from
bulge-dominated spirals to galaxies with small bulges. We used
the 2MASS-selected Flat Galaxy Catalog (2MFGC) (Mitronova et al. 2003) 
as a source of objects and the 2MASS survey (Skrutskie et al. 2006)
as a source of image data. 
We also used the Revised Flat Galaxies Catalog (RFGC) by 
Karachentsev et al. (1999) to add late-type galaxies to our sample.

The Two Micron All Sky Survey is one of the best known digital surveys in a
wavelength range close to the optical one. This survey covers the whole sky
in the filters $J$ (1.25 $\mu$m), $H$ (1.65 $\mu$m), and $K_s$ (2.16 $\mu$m).
Unfortunately, the 2MASS survey has a weak sensitivity to late-type galaxies, 
especially of low surface brightness (Jarrett 2000). This is due to the
high brightness of the night sky in the near-IR range and short exposures. 
For this reason the periphery of the discs of spiral galaxies is 
generally unseen beyond the isophotes fainter than 
$\mu_K=20$ mag arcsec$^{-2}$. We did not study truncation radii for the 
2MASS-galaxies and did not take them into account performing the 
decomposition.

The sample was selected mainly from the 2MFGC. This catalogue contains 
18020 disc-like galaxies covering all the celestial sphere. The objects were 
taken from the Extended Source Catalog of the 2MASS (XSC 2MASS) according 
to their 2MASS axial ratio $a/b \geq 3$. The 2MFGC gives value of axial 
ratio $b/a$, fiducial elliptical Kron magnitude, position angle, 
the $J$-band concentration index ($IC$) and other useful information. 
Concentration index is the ratio of the radii of circles that contain 
3/4 and 1/4 of the measured flux in the same band, respectively. 
$IC$ together with $b/a$ are known to serve as dividing lines between 
early- and late-type galaxies (see e.g. Strateva et al. 2001; 
Nakamura et al. 2003). We decided to select galaxies both of early and 
late types in almost equal quantities. Our criteria are
\begin{itemize}
\item[--] axial ratio for the combined $J+H+K$ image, $sba>0.2$;
\item[--] $K$-band fiducial elliptical Kron radius, $R_K\geq30''$;
\item[--] Hubble type ranging from S0 to late types; 
\item[--] non-interacting galaxies;
\item[--] concentration index $IC>2.0$.
\end{itemize} 

The latter item means that we exclude galaxies with
strongly peculiar surface brightness distributions since for pure
exponential disks $IC=2.8$ and for $r^{1/4}$ ellipticals $IC=7$
(Fraser 1972). 

From the set of 3167 objects which contains galaxies with pronounced
bulges (Sa, Sb types) and galaxies with small bulges (Sc, Scd types) we 
selected a sufficient number of regular spiral galaxies to make statistical 
analysis of edge-on galaxies properties. We did not add to the sample 
galaxies that are clearly lopsided or warped (with some exceptions below). 
In addition, on the basis of careful visual inspection we rejected 
non-edge-on and peculiar objects. Also, we added 13 bright galaxies that 
were absent in the 2MFGC and several edge-on galaxies which were not 
selected according with foregoing criteria but were included in the sample 
of RFGC galaxies by BM02. We chose them to compare the results of our 
decomposition with those obtained by BM02.

The final sample consists of 67 early-type galaxies (S0-Sab), 61 late-type 
galaxies (Sbc-Sd) and 47 galaxies of intermediate type Sb. The absolute 
numbers of galaxies of various morphological types (taken from the LEDA) and 
their percentages are listed in the Table~\ref{Types}. The full list of the 
sample galaxies studied here is presented in the Appendix, where we provide the 
basic information about them. According to LEDA 23 galaxies have bars, 
6 galaxies have rings, and 3 objects are interacting galaxies. 
We did not take into account additional components like bars performing 
the decomposition, but checking all correlations between structural 
parameters, we did not find any strong deviation of the above-mentioned objects from 
derived trends. 

We also did not reject 3 interacting 
galaxies and several other galaxies with some unusual structural features 
(double exponential discs, rings, etc., which will be described in our 
resulting tables) to have about 10 unusual spiral edge-on galaxies for 
further work. 

\begin{figure}
 \begin{center}
 \includegraphics[width=6cm]{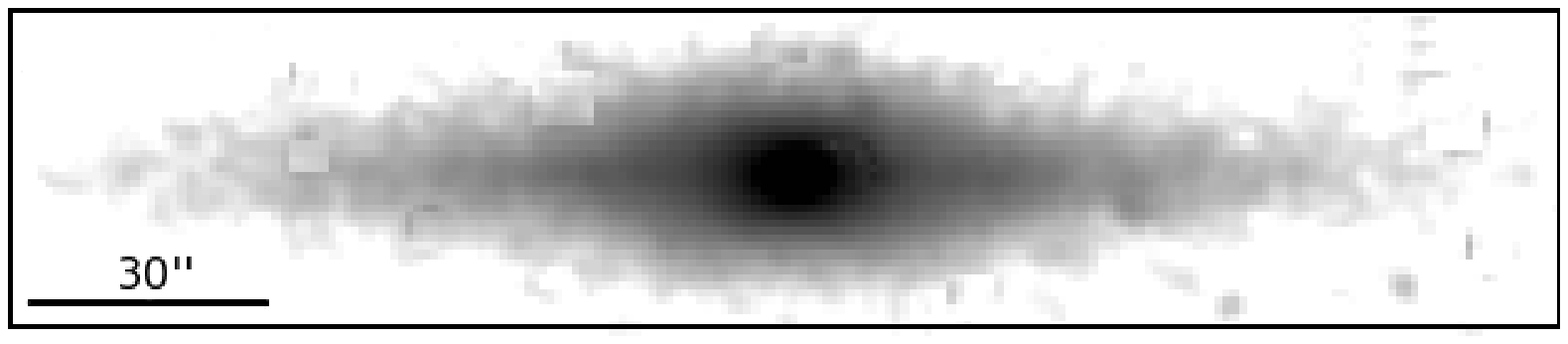}
 \includegraphics[width=6cm]{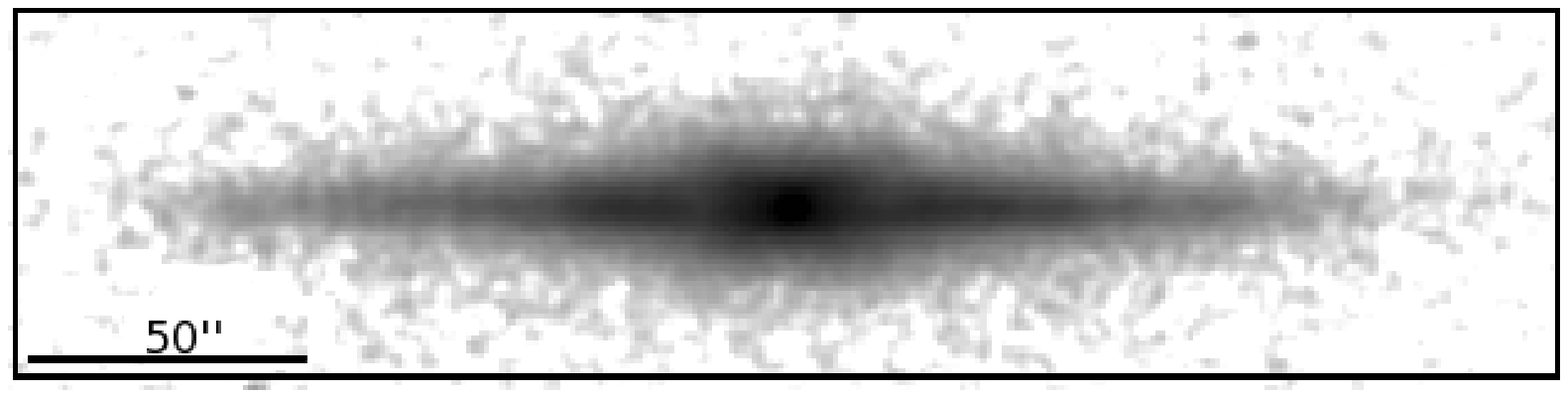}
 \includegraphics[width=6cm]{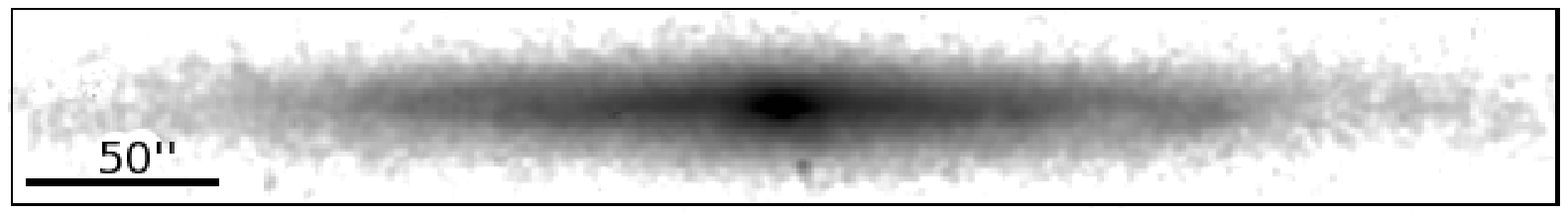}
 \end{center}
 \caption{Examples of the sample galaxies in the $K_s$-band. The upper image is
a typical galaxy of S0-a type: ES0 311-G012. The middle image is NGC 4013 of
Sb type. The bottom image is an example of Sc galaxy: 
NGC 5907.}
 \label{GALexamples}
 \end{figure}

As can be seen in the
Table~\ref{Types} the sample contains a substantial percentage of 
lenticular galaxies. It is necessary to notice that Hubble classification 
for edge-on galaxies is very subjective because the  spiral
arm structure is not seen in this case. The only way to classify edge-on 
galaxies according to morphological types is the estimation of the ratio of 
the total luminosities of a bulge and a disc (hereafter $B/D$). We assume that 
a large fraction of classified S0 galaxies are Sa galaxies in  fact. 
In Fig.~\ref{GALexamples} the images of typical sample galaxies are shown.
According to the LEDA data, our sample galaxies are, on
average, true highly-inclined galaxies with the mean inclination 
$i = 87.^{\circ}8 \pm 4.^{\circ}8$. 
Of course, this estimate can be too optimistic but a moderate deviation 
from edge-on orientation does not significantly change the slope of the 
vertical surface brightness distribution (Barteldrees \& Dettmar 1994). 

\begin{table}
 \centering
 \begin{minipage}{49mm}
 \parbox[t]{49mm}{\caption{Morphological types of the sample galaxies and
their fractions} \label{Types}} 
 \begin{tabular}{ccc}
 \hline 
 \hline
 Type & Number & Percentages\\  \hline
 S0, S0-a & 41 & 23.4  \tabularnewline
 Sa & 11 & 6.3 \tabularnewline
 Sab & 15 & 8.6 \tabularnewline
 Sb & 47 & 26.9 \tabularnewline
 Sbc & 21 & 12.0 \tabularnewline
 Sc & 36 & 20.5 \tabularnewline
 Scd & 4 & 2.3 \tabularnewline
 Total & 175 & 100 \tabularnewline
 \hline
\end{tabular}
\end{minipage}
\end{table}

The distances of the galaxies are taken from NASA/IPAC Extragalactic
Database (NED) with the Hubble constant $H_0=73$ km s$^{-1}$/Mpc, 
$\Omega_\mathrm{{matter}}=0.27$, and $\Omega_{\mathrm{vacuum}}=0.73$). 
The luminosity distances are in the range from 4.3 Mpc (NGC~891) to 171 Mpc 
(ESO~251-G028) with a sample median value of 40 Mpc. Three galaxies 
(ESO~013-G024, ESO~555-G023, ESO~555-G032) have no measured or estimated 
distances. We do not take them into account in statistical analysis where 
the distances are required.

To investigate some kinematical and dynamical properties of the sample
galaxies we added information about maximum rotation velocity 
for each galaxy if it was presented in the LEDA data base 
(see the section 5.3 for more details).

The sample consists of 175 galaxies in the $K_s$-filter, 169 galaxies in 
the $H$-filter and 165 galaxies in the $J$-filter. (Several galaxies have 
prominent dust lanes within their discs and even in the $J$ and $H$ 
passbands absorption of the dust is significant.)

In order to evaluate the completeness of the sample, we use the method 
proposed by Schmidt (1968) (the so-called $V/V_\mathrm{max}$ completness 
test). Following Thuan \& Seitzer (1979) we used this method for our 
angular-size-limited sample. The full sample is incomplete but the subsample 
of galaxies with angular radius $r\geq 60''$ is complete 
($V/V_\mathrm{max}=0.49 \pm 0.03$). This subsample consists of 92 galaxies 
(47 early-type galaxies and 45 late-type galaxies). 

The images used in this work were taken from the database of the 
All-Sky Release Survey Atlas. The Atlas' images are FITS standard images in 
512 $\times$ 1024 pixel format (with 1.$^{''}$00 pixel$^{-1}$). 
Description of the images 
such as the pixel gain, readout noise, seeing FWHM ($\sim2.''5$), sky value 
and also zero points are available in the 2MASS webpages. This information 
has been used in making decompositions of galaxy images. Each image was 
thoroughly examined in order to mask background stars and galaxies and other 
contaminating sources. All the images were rotated to align the galaxy major 
axis parallel to the horizontal image borders. These procedures were carried 
out in three filters so all images of the same galaxy in the $J$, $H$ and 
$K_s$ bands have the same extension and position angles. The smallest image 
of our sample is 48$\times$18 pixels for ESO 555-G032 and the largest image is 
869$\times$165 pixels for NGC 4565. The median value of the size of galaxy 
images is 119$\times$27 pixels, that is enough to determine the main 
photometric parameters of galaxy discs and bulges. The forthcoming analysis 
was performed with the MIDAS (developed by the European Southern Observatory) 
and the BUDDA packages.

\section[]{Two-dimensional bulge/disc decomposition}

BUDDA (Bulge/Disc Decomposition Analysis) is a code devised to perform 
two-dimensional decomposition of galactic images. This code was produced 
by R.E. de Souza, D.A. Gadotti and S. dos Anjos (for more information about 
BUDDA see SGA04). Since the year of 2004 BUDDA has been publicly 
available to the astronomical community\footnote{see 
http://www.mpa-garching.mpg.de/$\sim$dimitri/budda.html}. We use BUDDA v2.1
to produce the fits, allowing the inclusion of bulges, discs, bars and AGN in the
models.

The input data for the code is an image of a galaxy, consisting of two major 
components: a bulge and a disc. The disc is represented by an exponential 
distribution of the luminosity density $I(r,z)$ with the central luminosity 
density $I(0,0)$, the scalelength $h$ and the `isothermal' scaleheight 
$z_0$:
\begin{equation}
 I(r,z) = I(0,0) \, e^{-r/h} \, \hbox{sech}^2(z / z_0) \, ,
\label{form1}
\end{equation}
where $(r,z)$ are the cylindrical coordinates. The disc is assumed to be
axisymmetric and transparent. 

The surface brightness distribution for the face-on disk
($i = 0^\circ$) can be expressed as follows:
\begin{equation}
 I_\mathrm{d}(r) = I_\mathrm{0,d} \, e^{-r / h}, \,
\label{form2}
\end{equation}
where $I_\mathrm{0} = 2 \, z_0 \, I(0,0)$. 
The same expression in magnitudes per arcsec$^2$ is:
\begin{equation}
 \mu_\mathrm{d}(r) = S_\mathrm{0,d} + 1.086 \, r/h, \,
\label{form3}
\end{equation}
where $S_\mathrm{0,d}$ is the central surface brightness of the face-on disc. 
For edge-on galaxies ($i = 90^\circ$) the following expression is valid: 
\begin{equation}
 I(r,z) = I(0,0) \, 
 \frac{r}{h} \, K_1\left(\frac{r}{h}\right) \, \hbox{sech}^2(z / z_0)\, ,
\label{form4}
\end{equation}
where $K_1$ is the modified Bessel function of the  first order. 

The bulge surface brightness profile is described by the \ser\ law 
(S\'ersic 1968):
\begin{equation}
 I_\mathrm{b}(r) = I_\mathrm{0,b} \, 
 e^{-\nu_\mathrm{n}[(r/r_\mathrm{e,b})^{1/n}]}\, ,
\end{equation}
where $r_\mathrm{e,b}$ is the effective radius of the bulge, i.e., the radius 
of a circle that contains 50\% of the total galaxy luminosity, 
$I_\mathrm{0,b}$ is the bulge central surface brightness, $n$ is the 
\ser\ index, defining the shape of the profile, and the parameter 
$\nu_n$ ensures that $r_\mathrm{e,b}$ is the half-light radius. In magnitudes 
per arcsec$^2$ the expression looks as follows:
\begin{equation}
 \mu_\mathrm{b}(r) = \mu_\mathrm{0,b} + 
 \frac{2.5\nu_\mathrm{n}}{\ln 10} \, 
 \left(\frac{r}{r_\mathrm{e,b}}\right)^{1/n}\, ,
\label{form6}
\end{equation}
where $\mu_\mathrm{0,b}$ is the bulge effective surface brightness expressed
in mag per arcsec$^2$, i.e., the surface brightness at $r_\mathrm{e,b}$. 
The BUDDA code uses a numerical approximation of 
$\nu_n \simeq  (0.868n-0.142) \, \ln 10$.  The \ser\ index of $n=4$ 
represents the de Vaucouleurs profile which was very popular to describe the 
surface brightness distribution of bright elliptical galaxies and of bulges 
in early-type spirals. The \ser\ index of $n=1$ represents the exponential 
profile of bulges in late-type spirals, of galactic discs and of dwarf 
elliptical galaxies (see e.g. de Vaucouleurs 1948, 1959; Freeman 1970; 
Graham 2001 and references therein). 

It is also useful to determine such parameters as ellipticities 
($\epsilon = 1 - b/a$) of a bulge $\epsilon_\mathrm{b}$ and a disc 
$\epsilon_\mathrm{d}$, as well as their position angles 
$P.A._\mathrm{b}$ and $P.A._\mathrm{d}$. To account for the effect of seeing 
there is a Moffat smearing (Moffat 1969) of the brightness profiles in the 
code controlled by a parameter associated with the seeing radius 
$a_\mathrm{s}$. There is also included a correction term of the sky level 
$\Delta_\mathrm{sky}$. The isophotes of bulges and discs can be described 
by generalized ellipses with the ellipse index parameter $\gamma$ 
controlling their shape:
\begin{equation}
 \left(\frac{|x|}{a}\right)^{\gamma + 1} + 
 \left(\frac{|y|}{b}\right)^{\gamma + 1} = 1, 
\label{form36}
\end{equation}
where $x$ and $y$ are the pixel coordinates of the ellipse points, $b$ and
$a$ are the sizes of its semi-major and semi-minor axes, respectively. When 
$\gamma = 1$ the ellipse is simple, while for $\gamma > 1$ the ellipse is 
boxy and for $\gamma < 1$ the ellipse is discy.

For edge-on galaxies BUDDA derives the disc scaleheight $z_0$ instead of 
the disc ellipticity $\epsilon_\mathrm{d}$. Thus, a 
total number of the main parameters for the bulge/disc model is 11: the
center coordinates $x_0$ and $y_0$, the central surface brightnesses of
a bulge $I_\mathrm{0,b}$ (or the effective surface brightness of bulge 
$I_\mathrm{e,b}$ as used in BUDDA) and  a disc $I_\mathrm{0,d}$ (both in ADU),
$r_\mathrm{e,b}$, $h$, $P.A._\mathrm{b}$, $P.A._\mathrm{d}$, the
disc scaleheight $z_0$, the bulge ellipticity 
$\epsilon_\mathrm{b}$ and the \ser\ index of a bulge $n$.

Total luminosity of a disc can be expressed as:
\begin{equation}
  L_\mathrm{D} = 
  2\pi \, I_\mathrm{0,d} \, h^2\, , 
\label{form37}
\end{equation}
while the total luminosity of a bulge is
\begin{equation}
  L_\mathrm{B} = \frac{2\pi n}{\nu_\mathrm{n}^{2n}} \, 
  \Gamma (2n) \, I_\mathrm{0,b} \, r_\mathrm{e,b}^2 \, 
  (1 - \epsilon_\mathrm{b})\,,
\label{form38}
\end{equation}
where $\Gamma$ means the gamma function.

The code BUDDA allows the user to add some additional components in a galaxy model
(bars, AGNs and double exponential discs), but we made decompositions 
of all our galaxies into a bulge and a disc only. There are some difficulties 
to determine the presence of a bar or any other substructures in edge-on 
perspective. The observed surface brightness distribution is a result of 
luminosity integration along the line of sight and such integration can smooth 
and hide such subcomponents. It is also hard to make the first approximation 
of the values of parameters which characterize these additional features. 

\begin{center}
 \begin{figure*}
 \begin{center}
  \includegraphics[width=5.4cm]{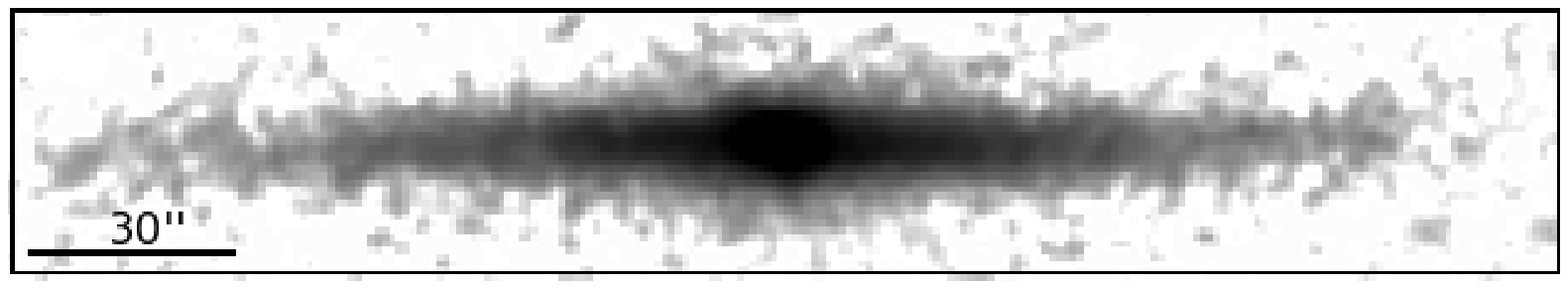}
  \includegraphics[width=5.4cm]{fig2b.ps}
  \includegraphics[width=5.4cm]{fig2c.ps}
  \includegraphics[width=6.0cm, angle=-90]{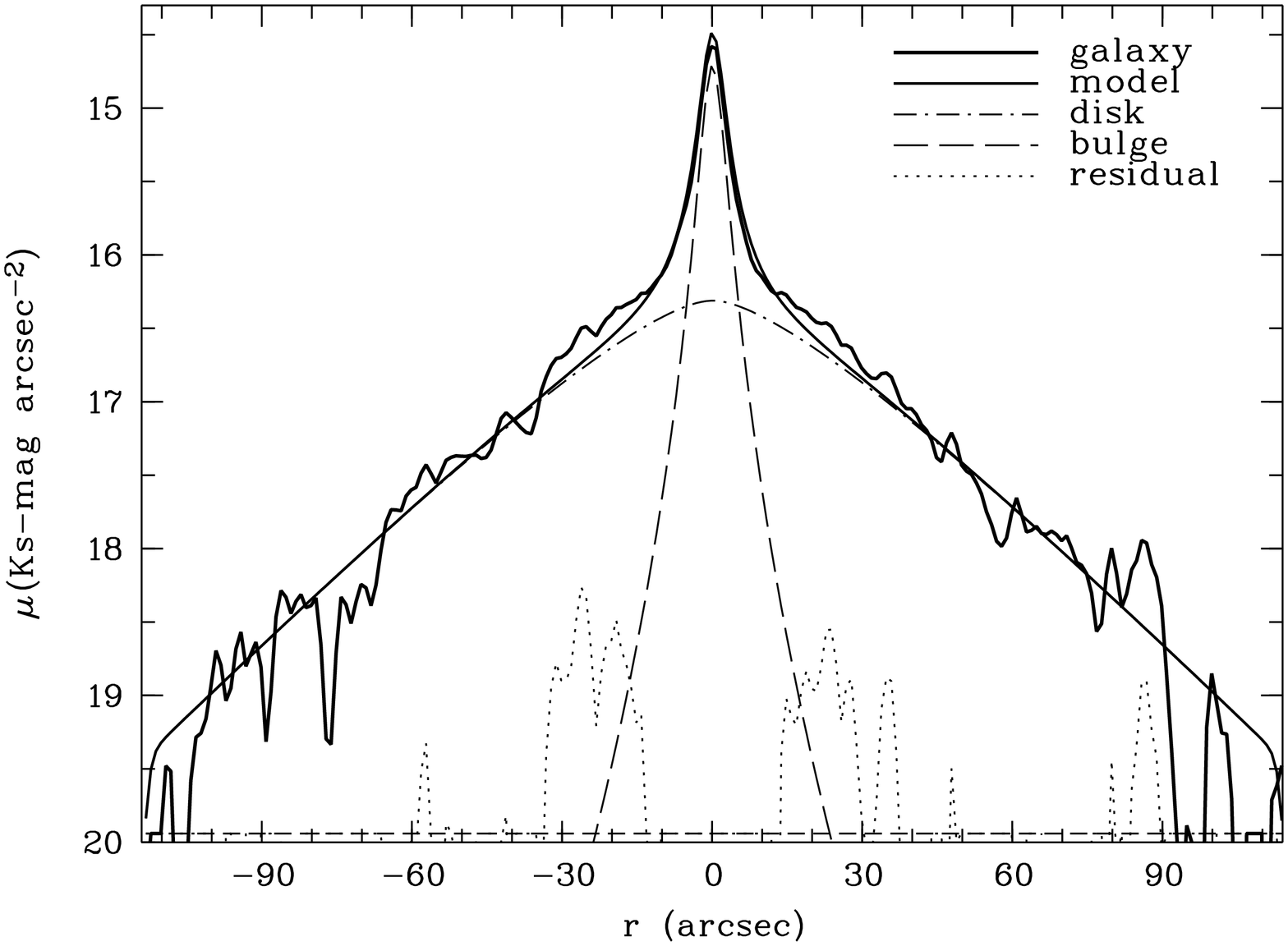}
  \includegraphics[width=6.0cm, angle=-90]{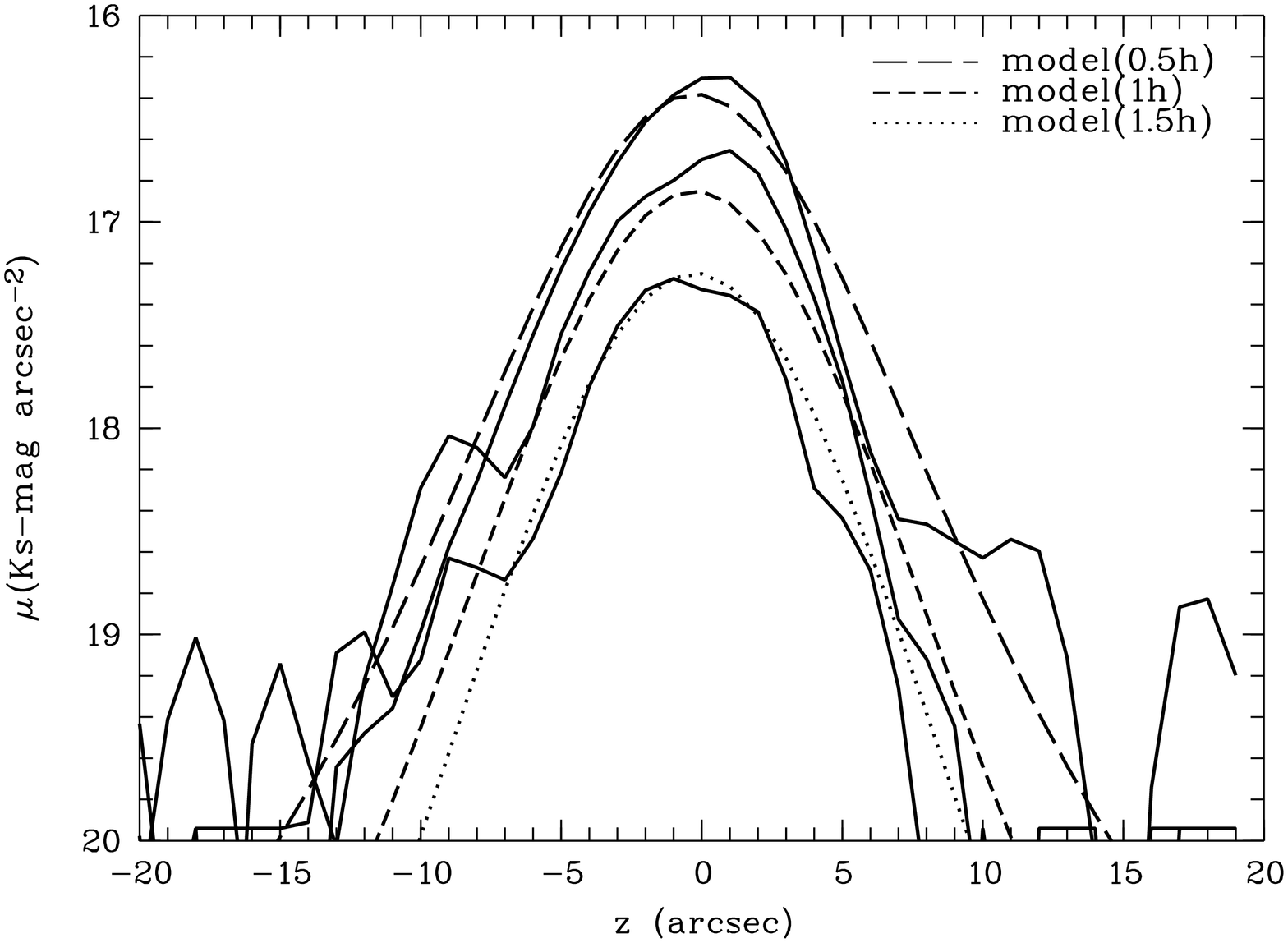}
 \end{center}
 \caption{{An example of the edge-on spiral galaxy ESO 240-G011 in
the $K_s$-band. The images on the top show the galaxy (top left), the model 
(top middle) and the residual image (top right). The residua were obtained 
after subtraction of the model from the image. The bottom panels
show the surface brightness profiles of the galaxy, the model and the residual 
images are indicated. The profiles are drawn along major- (left) and minor  
axes (right) at $r=0.5h$, $1.0h$ and $1.5h$.}}
 \label{example}
\end{figure*}
\end{center}

Fig.~\ref{example} illustrates typical results of our decomposition for the
Sb-type galaxy ESO 240-G011. The size of the image is 228$\times$40 pixels. 
The sky background, Galactic stars, bright HII regions and nearby companions 
were removed. The derived parameters are given in 
Table~\ref{example2}. As seen in Fig.~\ref{example} the fit is fairly good. 

\begin{table}
 \centering
 \begin{minipage}{77mm}
  \parbox[t]{77mm} {\caption{Parameters of the two-component model for 
ES0~240-G011 ($K_s$ band)} 
  \label{example2}}
  \begin{tabular}{llll}
  \hline 
  \hline 
 $D$ (Mpc) \quad & 35.7 & $\mu_\mathrm{e,b}$ \quad & $17.09\pm0.09$ \tabularnewline
  \quad&   & $r_\mathrm{e,b}$ (kpc) \quad & $1.34\pm0.09$ \tabularnewline
 $\mu_\mathrm{0,d}$ \quad & $16.22 \pm 0.16$ & $n$ \quad &$2.4 \pm0.3$ \tabularnewline
 $h$ (kpc) \quad & $5.1 \pm 0.9$ & $q_\mathrm{b}$ \quad &  $0.69\pm0.04$ \tabularnewline
 $z_0$ (kpc)  \quad &$0.80\pm0.13$ & $B/D$  \quad&0.38 \tabularnewline
 $h/z_0$ \quad &  $6.4$ & $v_\mathrm{rot}$ (km/s) \quad & $267.5\pm3.3$ \tabularnewline
 \hline
\end{tabular}
\end{minipage}
\label{Texample}
\end{table}

\section[]{Decomposition and Results}

To start the galaxy decomposition, we must determine the first-order 
approximations for the input parameters (see previous section). 
To obtain the first-order approximation for our 2D fits, we have used 
parameters derived from 1D fits of the major and minor axes profiles. 

In almost all cases the ellipse index $\gamma$ of a bulge was taken to be 
equal $1.0 \pm 0.01$ (for simplicity we consider bulges with elliptical 
isophotes). A galaxy with complex morphology of a bulge requires more 
detailed investigation with higher resolution images than those that are 
given by 2MASS. 
The result of decompositions into a bulge and a disc is the list of all 
structural parameters of each object of the sample in 3 photometric bands. 
The final values of $\mu_\mathrm{e,b}$ and $\mu_\mathrm{0,d}$ were corrected 
for the Galactic extinction according to Schlegel, Finkbeiner \& Davis (1998).

To check the reliability of our 2D decompositions, we subtracted the 
full apparent magnitudes of our models from the apparent magnitudes 
presented in the 2MASS catalogue. The mean values and standard deviations of 
these subtractions for the sample galaxies are:
$\Delta J = 0.^m011\pm0.^m057\,,$
$\Delta H = 0.^m021\pm0.^m049\,,$
$\Delta K_s = 0.^m014\pm0.^m053\,.$

\begin{center}
 \begin{figure}
  \includegraphics[width=5.0cm, angle=-90]{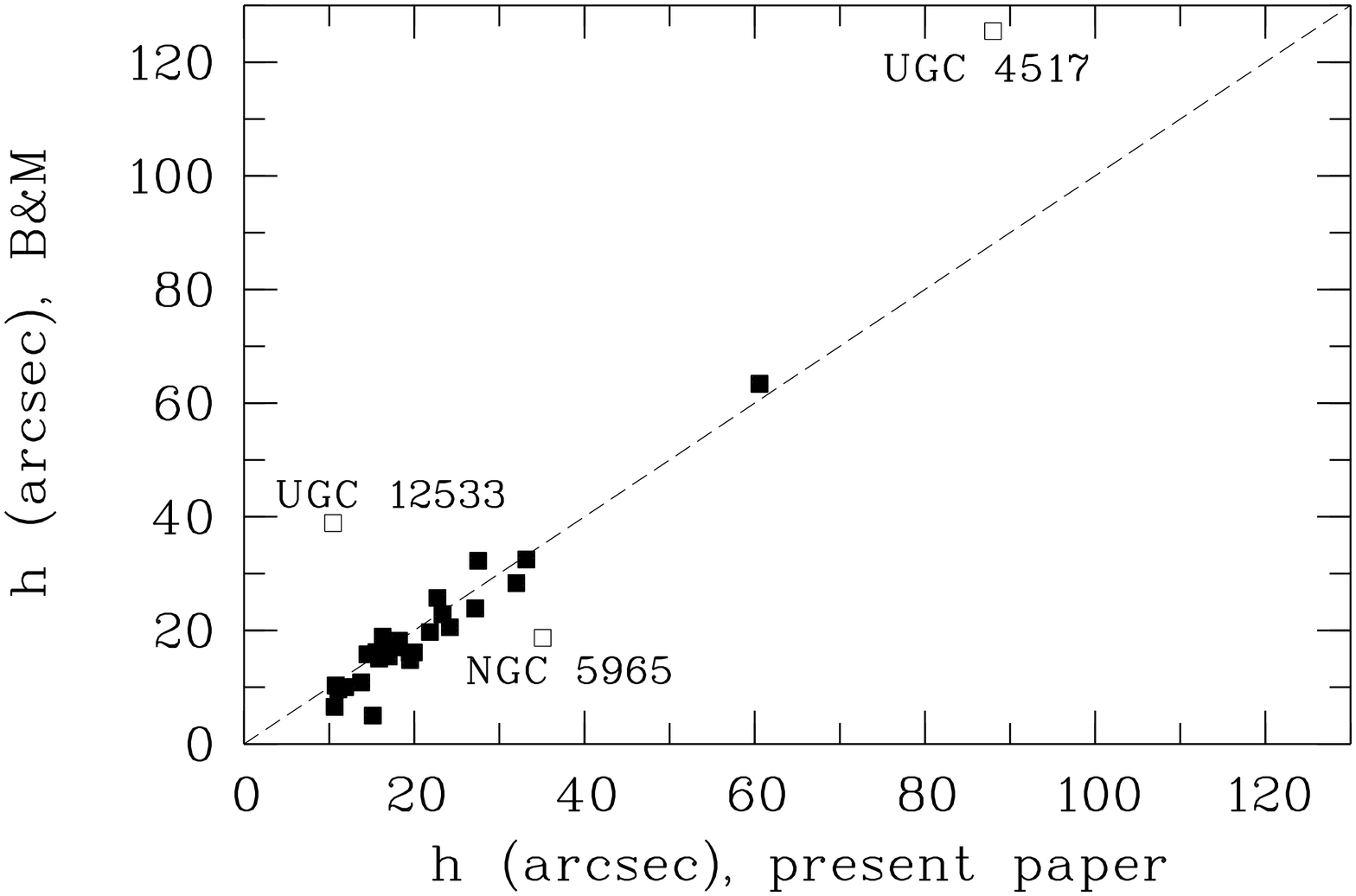}
  \includegraphics[width=5.0cm, angle=-90]{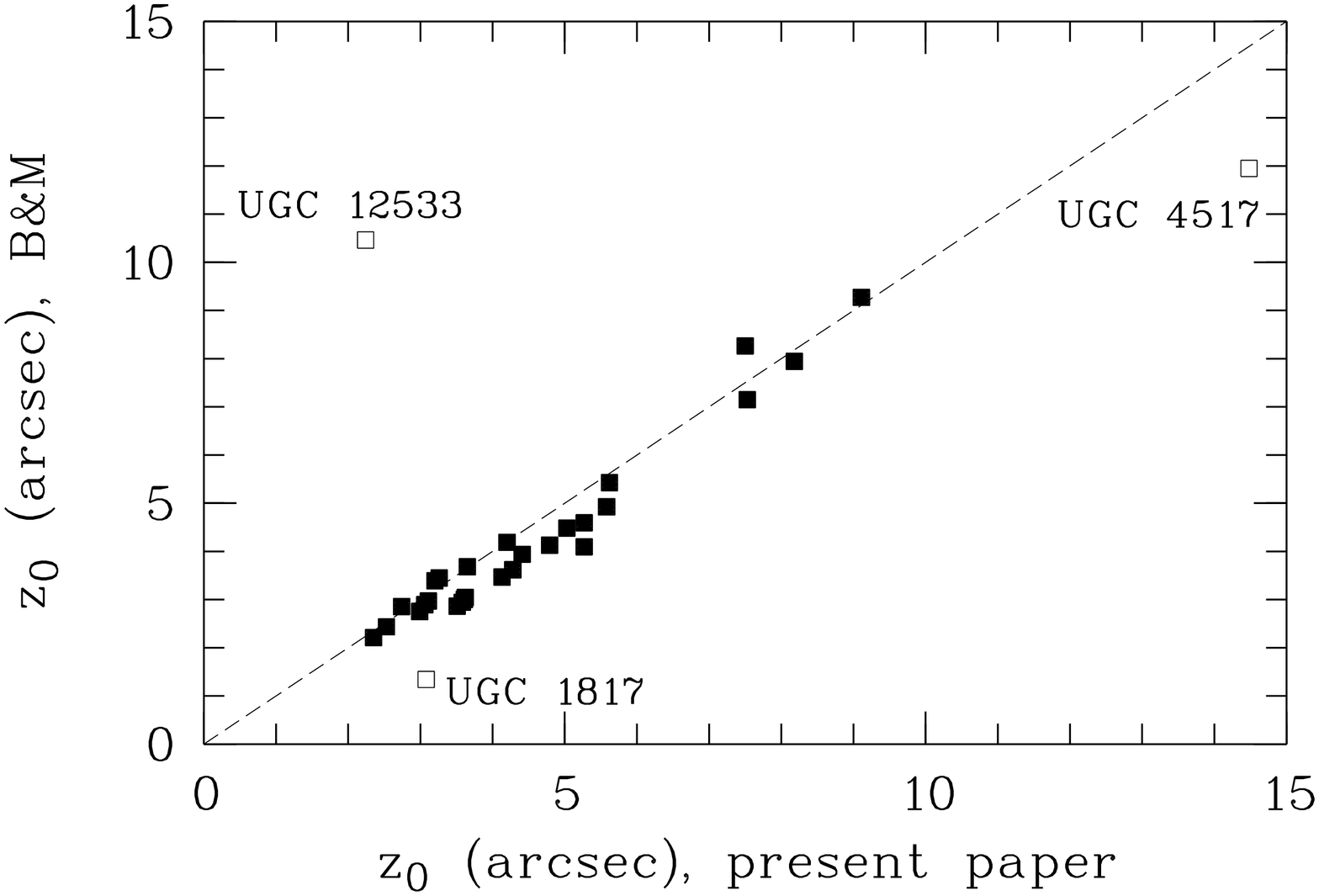}
  \includegraphics[width=5.0cm, angle=-90]{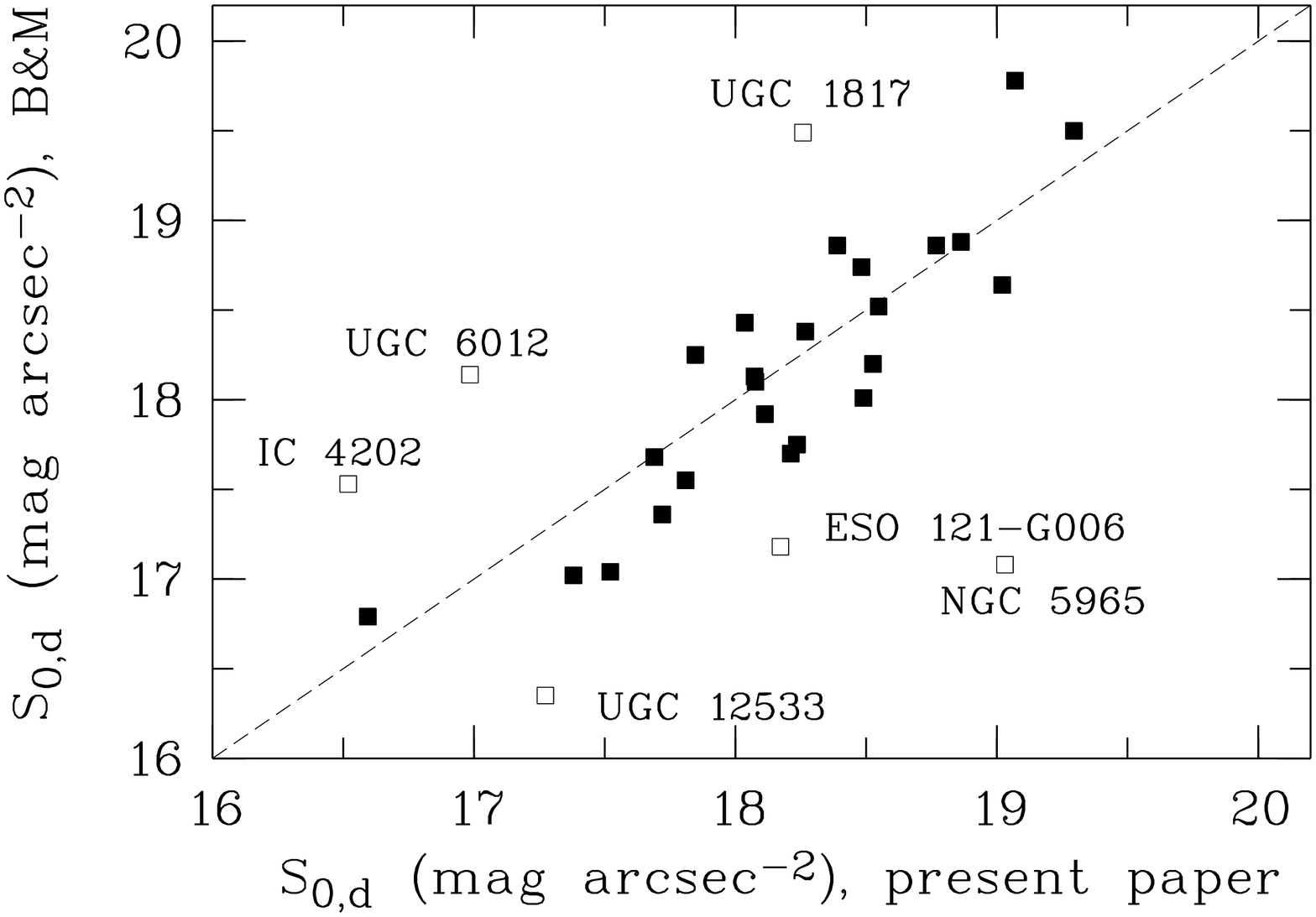}
 \caption{\small{Comparison of the radial scalelengths (the upper frame), 
the vertical scaleheights (the middle frame) and the deprojected central 
surface brightnesses (the bottom frame) between data of BM02 and the data 
presented in this paper. Open squares are related to the galaxies with some 
peculiarities and features within their structure as noted in Sect.4.}}
\label{Mitronova}
\end{figure}
\end{center}

Our sample has 30 galaxies in common with the sample of BM02. The method used 
by these authors is based on several photometric cuts parallel to the minor axis 
of each galaxy and parallel to its major axis. Their sample includes flat 
galaxies mainly of late types in the $K_s$ filter. Fig.~\ref{Mitronova} shows 
a comparison of radial scalelengths (the upper frame), vertical scale 
heights (the middle frame) and deprojected central surface brightness of  
discs (the bottom frame). Open squares are related to the objects which have 
some peculiarities or features in structure (NGC\,5965, UGC\,6012, UGC\,12533 are
with possible bars, IC\,4202, UGC\,4517, ESO\,121-G006 are with strong dust lanes,
and UGC\,1817 is a bulgeless galaxy in a group). 
These objects do not lie on the diagonal. The mean values and standard
deviations of difference between parameters obtained by us and by BM02 are:
$$\langle h-h_\mathrm{BM} \rangle = 2.\arcsec46\pm2.\arcsec11\, ,$$
$$\langle z_0-z_\mathrm{0,BM} \rangle = 0.\arcsec41\pm0.\arcsec29\, ,$$
$$\langle S_0-S_\mathrm{0,BM} \rangle = 0.^m28\pm0.^m19\,.$$
As one can see (Fig.~\ref{Mitronova}), our results and those of BM02 are in 
good agreement.

\begin{figure}
 \centering
 \includegraphics[width=5.0cm, angle=-90]{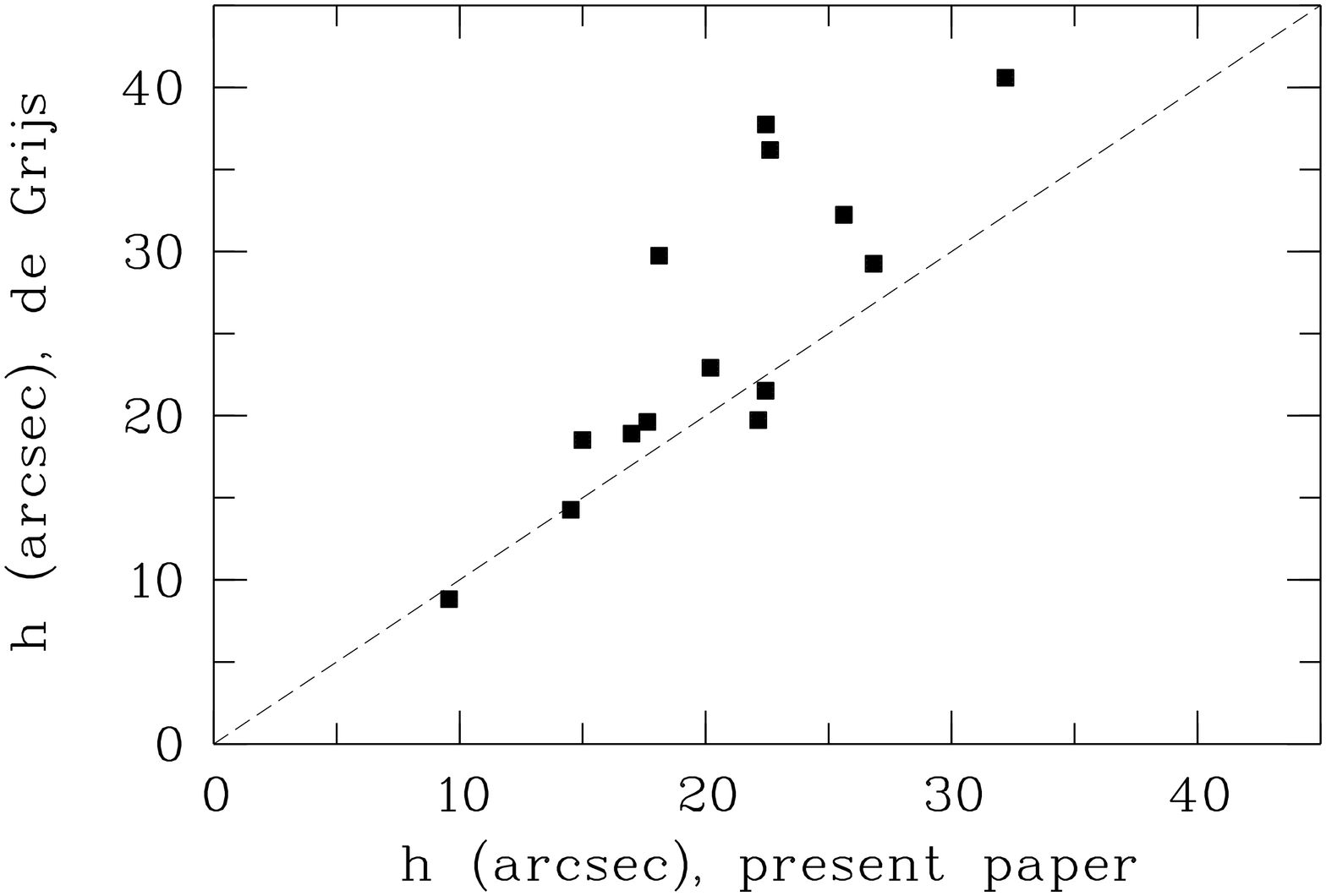}
 \caption{\small{Comparison between radial scalelengths obtained by  
de Grijs (1998) and in this work.}}
\label{de Grijs}
\end{figure}

De Grijs (1998) presented detailed surface photometry for a sample of
edge-on galaxies in the $B$, $I$ and $K$ bands. Fig.~\ref{de Grijs} 
shows a comparison of radial scalelengths for 14 joint galaxies ($K$-band).  
Some systematics are evident which can be attributed to different
decomposition procedures.

Table~\ref{Bedregal} compares decomposition results for 5 galaxies in the
present work and in the paper by 
Bedregal, Aragon-Salamanca \& Merrifield (2006). As one can see, the results 
are in agreement that is typical for such a 
comparison\footnote{For NGC~5308 and NGC~1184 Bedregal et al. (2006) give 
the formal errors for the bulge effective radii that are very large and 
asymmetric. Unfortunately, they did not provide any explanation of these 
facts. For both galaxies, the bulge effective radii from our study and those 
by Bedregal et al. (2006) are quite different.}. 
It proves the reliability of the obtained results and makes them suitable in 
our further statistical analysis.

\begin{table*}
 \centering
 \begin{minipage}{125mm}
 \parbox[t]{125mm}{\caption{Comparison between structural parameters obtained 
 in this work and by Bedregal et al. (2006) in the $K_s$-band
 (parameters given in arcsec were rewritten in kpc according to distances 
 adopted by Bedregal et al.).}\label{Bedregal}}
  \begin{tabular}{lcccccc}
  \hline 
  \hline
     & \multicolumn{3}{c}{This paper} & \multicolumn{3}{c}{Bedregal et al. 
     (2002)} \\  
     \hline  
 Name & $h$ & $r_\mathrm{e,b}$ & $n$ & $h$ & $r_\mathrm{e,b}$ & $n$ \\  
        & (kpc) & (kpc) &  & (kpc) & (kpc) \\  
     \hline 
 NGC 5308 & $2.80 \pm$ 0.89 & $1.53 \pm 0.10$ & $3.74 \pm 0.47$ & $2.68^{+0.12}_{-0.05}$ & $0.62^{+0.11}_{-2.51}$ & $2.99^{+0.04}_{-0.06}$   \tabularnewline
 NGC 1184 & $4.29 \pm 0.64$& $1.84 \pm 0.06 $ & $3.83 \pm 0.21$ & $3.80^{+0.06}_{-0.05}$ & $1.07^{+0.07}_{-2.43}$ & $3.58^{+0.07}_{-0.05}$   \tabularnewline
 NGC 4417 & $1.57 \pm 0.35$& $0.86 \pm 0.03 $ & $3.25 \pm 0.23$ & $1.82^{+0.04}_{-0.05}$ & $0.62^{+0.03}_{-0.02}$ & $2.79^{+0.07}_{-0.06}$   \tabularnewline
 NGC 1380A & $1.70 \pm 0.23$& $0.53 \pm 0.02 $ & $3.36 \pm 0.29$ & $1.80^{+0.04}_{-0.06}$ & $0.98^{+0.32}_{-0.25}$ & $3.71^{+0.07}_{-0.10}$   \tabularnewline
 NGC 1381 & $2.36 \pm 0.56$& $0.88 \pm 0.03 $ & $2.89 \pm 0.20$ & $1.88^{+0.07}_{-0.05}$ & $0.67^{+0.02}_{-0.01}$ & $3.07^{+0.14}_{-0.10}$   \tabularnewline
 \hline
\end{tabular}
\end{minipage}
\end{table*}

\section[]{Main results}

Statistical correlations between structural parameters give us a clue for 
understanding the formation and evolution of spiral galaxies. Here we present 
only some of the correlations of the main parameters of the galaxies. The detailed 
analysis of the properties of bulges and discs will be done in a forthcoming 
article.

\subsection[]{Distributions of structural parameters}

Here we briefly describe the distributions of structural parameters of stellar 
components. In general, we do not take into account any corrections for the 
incompleteness and selection effects of the sample but in special cases we 
give the best fitting parameters obtained for the complete subsample.

\subsubsection[]{Discs}

The distributions of the apparent central surface brightness of discs
for edge-on view
$\mu_\mathrm{0,d}$ corrected for Galactic extinction in the $JHK_s$ filters are 
shown in Fig.~\ref{Distrib}a. The values of this parameter span more than 
3$^m$ arcsec$^{-2}$ in each filter. The distributions shift from the filter 
$J$ to the filter $K_s$ with increasing of mean central surface brightness. 
It is well seen that LSB galaxies were not included in the sample. Thus, one 
has to keep in mind that in fact the number of such galaxies could be high at 
the faint end of the distribution of $\mu_\mathrm{0,d}$ in all three passbands.

Fig.~\ref{Distrib}b shows a large range in scalelengths $h$. Most of the 
sample galaxies have $h = 1-6$ kpc but the number of extended galaxies is 
small. The distributions of the disc scaleheight $z_0$ are shown in 
Fig.~\ref{Distrib}c. There is a fairly narrow distribution, that lies in the 
range $z_0 = 0.4-1.2$ kpc, and a tail of ``thick'' galaxies with 
$z_0 \ga 1.2$ kpc. The shapes of distributions of $h$ and $z_0$ in each 
filter are quite similar and the median values of $h$ and $z_0$ have no 
systematic differences in the infrared bands under consideration 
(Table~\ref{MedDisc}).

The statistical study of the ratio of the radial to the vertical scalelengths 
for galactic discs, $h/z_0$, gives constraints on kinematical and dynamical 
models of galaxies. The distribution of this ratio $h/z_0$ is shown in 
Fig.~\ref{Distrib}d. It is rather flat. The mean ratio 
$\langle h/z_0 \rangle$ is about 3.9 for the $H$ and $K_s$-bands 
(Table~\ref{MedGal}). 
Kregel et al. (2002) found the mean ratio of $\langle h/z_0 \rangle=3.7\pm1.1$ 
(in the $I$-band) which is close to our result. For the sample of 153 galaxies 
composed by BM02 the mean ratio of $h/z_0$ is about 4.8. Although they studied 
galaxies in the $K_s$-band, their sample consists of late-type galaxies and 
the mean value of the ratio $h/z_0$ for these galaxies is larger than the mean 
ratio found in this paper. Apparently the mean $h/z_0$ ratio is higher in 
the blue filters than in the near-infrared bands 
(see for comparison e.g. van der Kruit \& Searle 1982a 
and de Grijs \& van der Kruit 1996).

\subsubsection[]{Bulges}

\begin{center}
 \begin{figure*}
 \begin{center}
  \includegraphics[width=5.4cm, angle=-90]{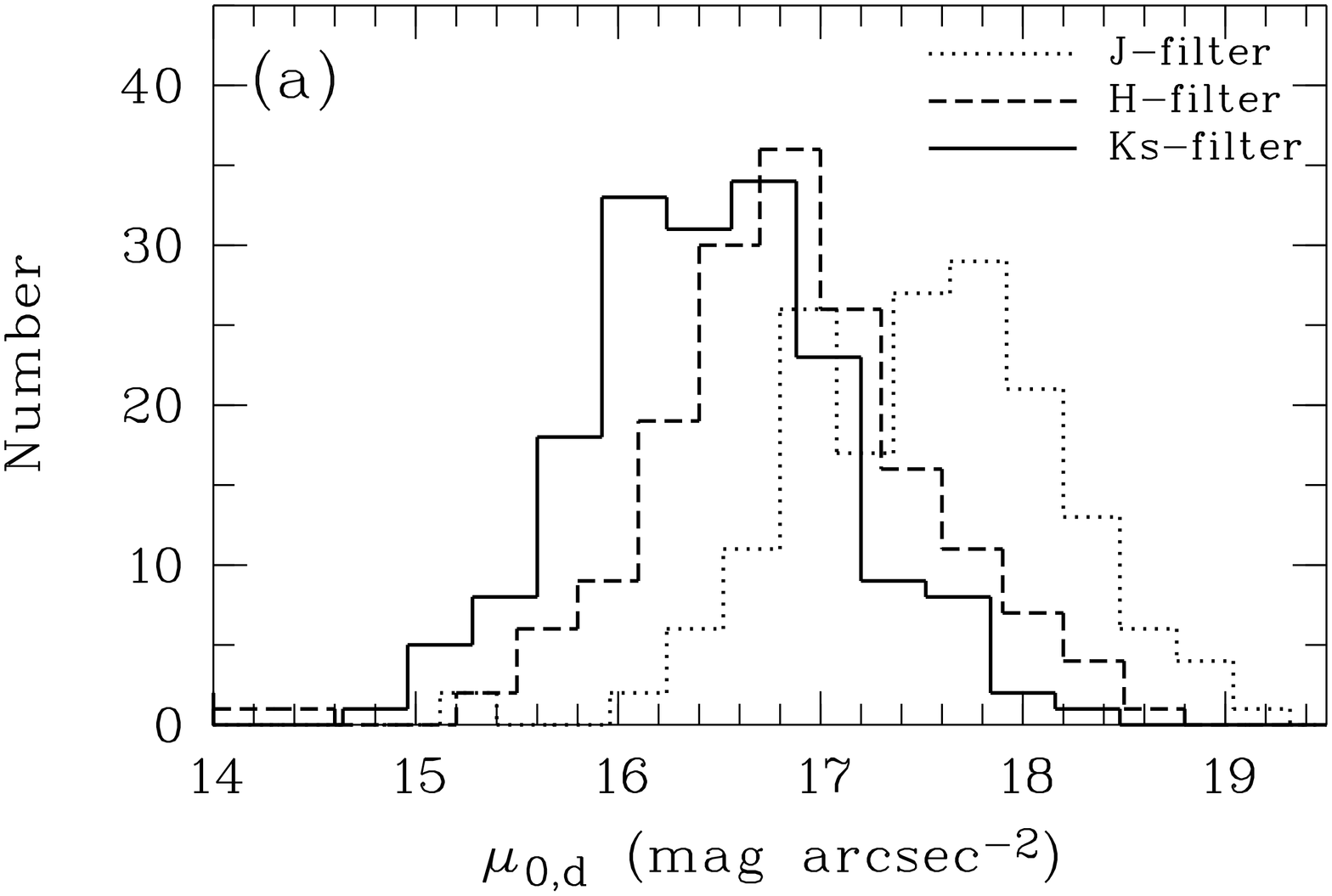}
  \includegraphics[width=5.4cm, angle=-90]{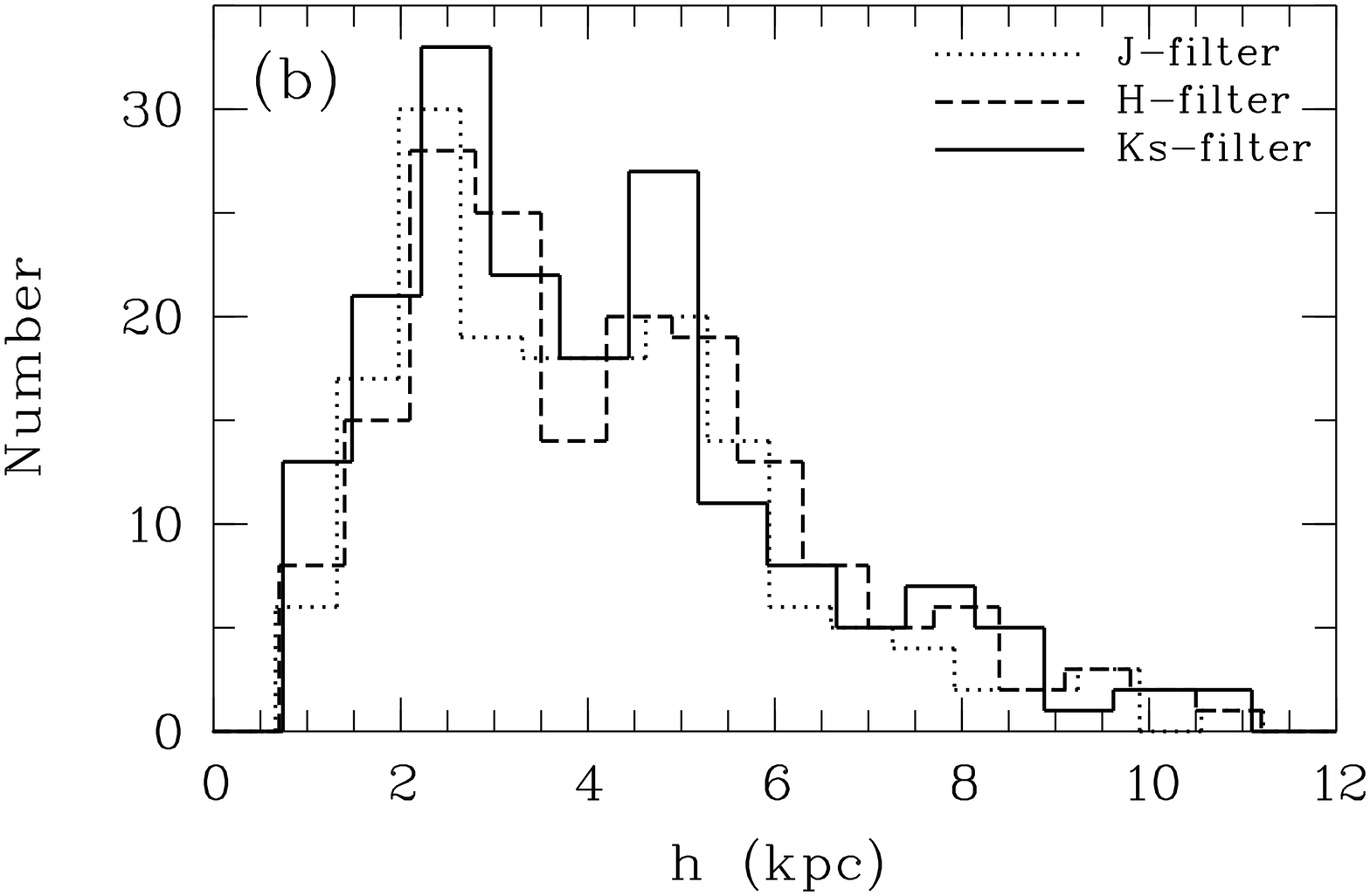}
  \includegraphics[width=5.4cm, angle=-90]{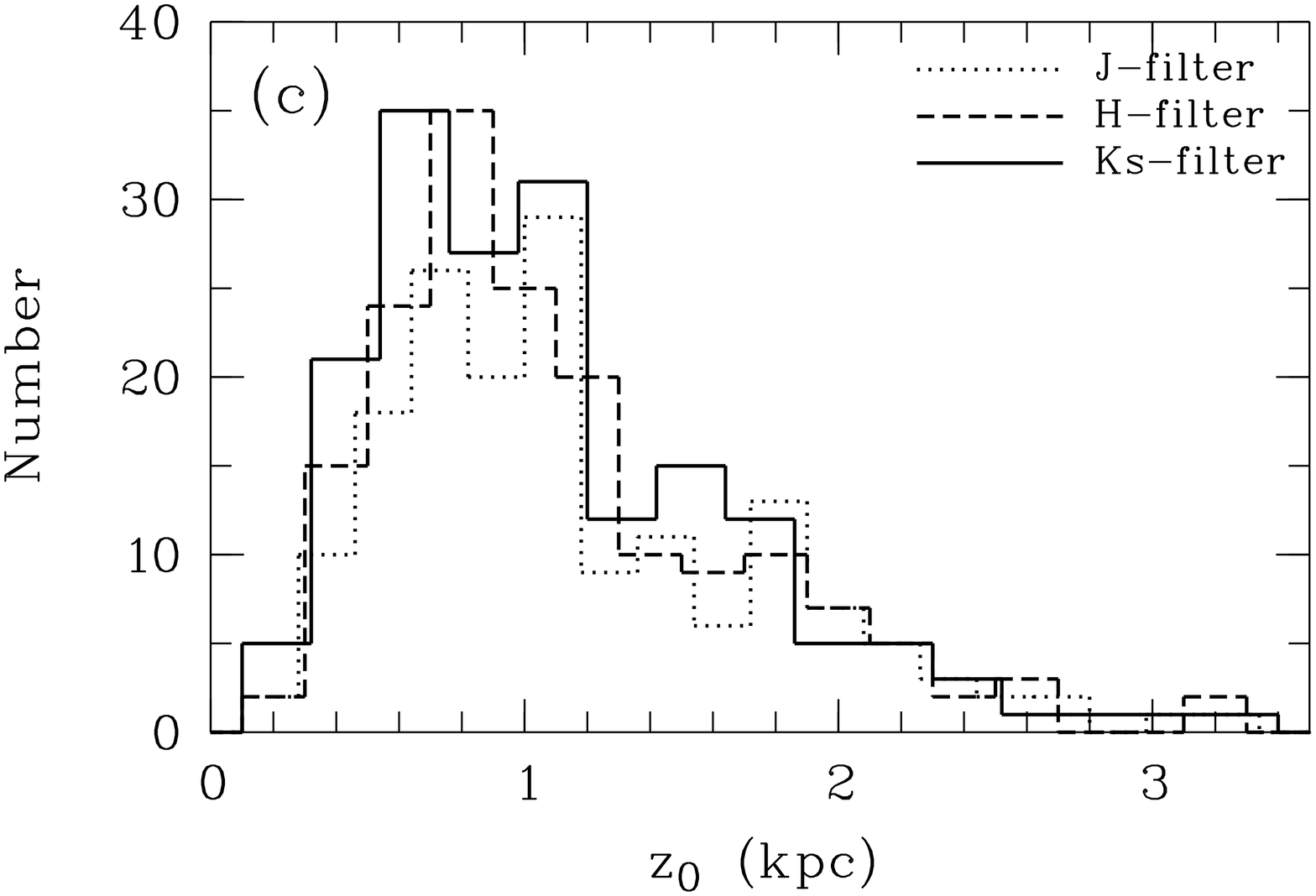}
  \includegraphics[width=5.4cm, angle=-90]{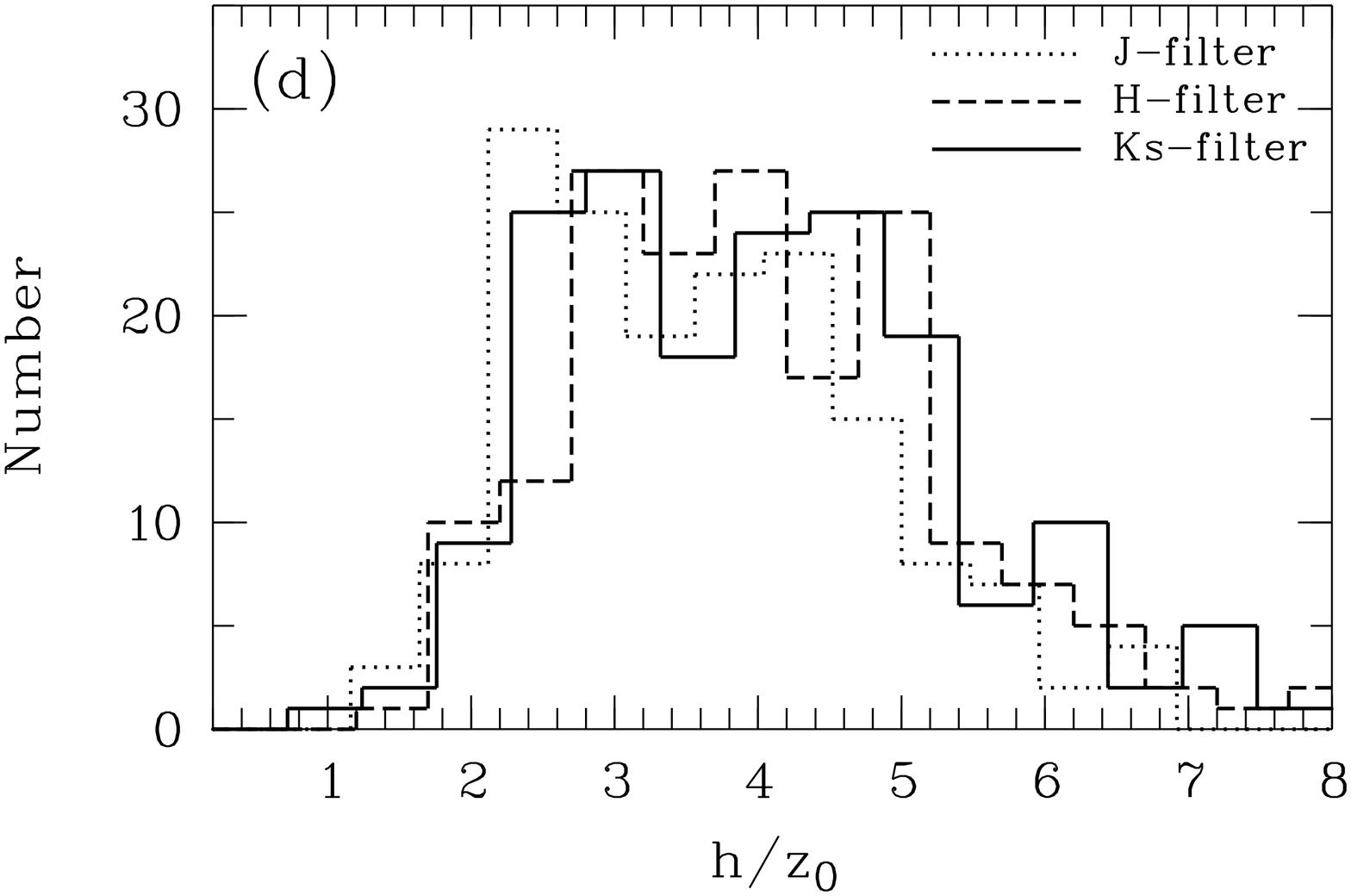}
  \includegraphics[width=5.4cm, angle=-90]{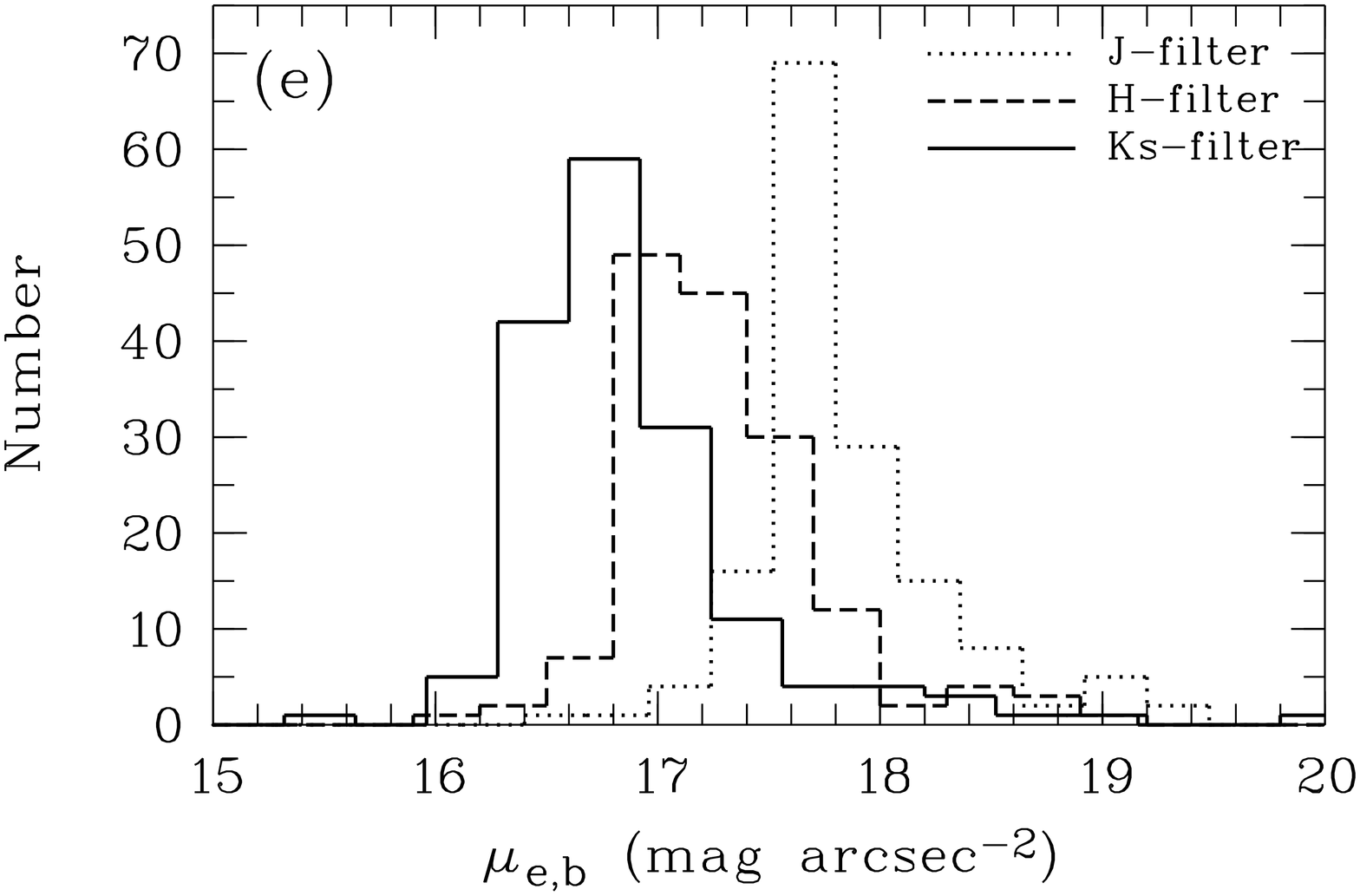}
  \includegraphics[width=5.4cm, angle=-90]{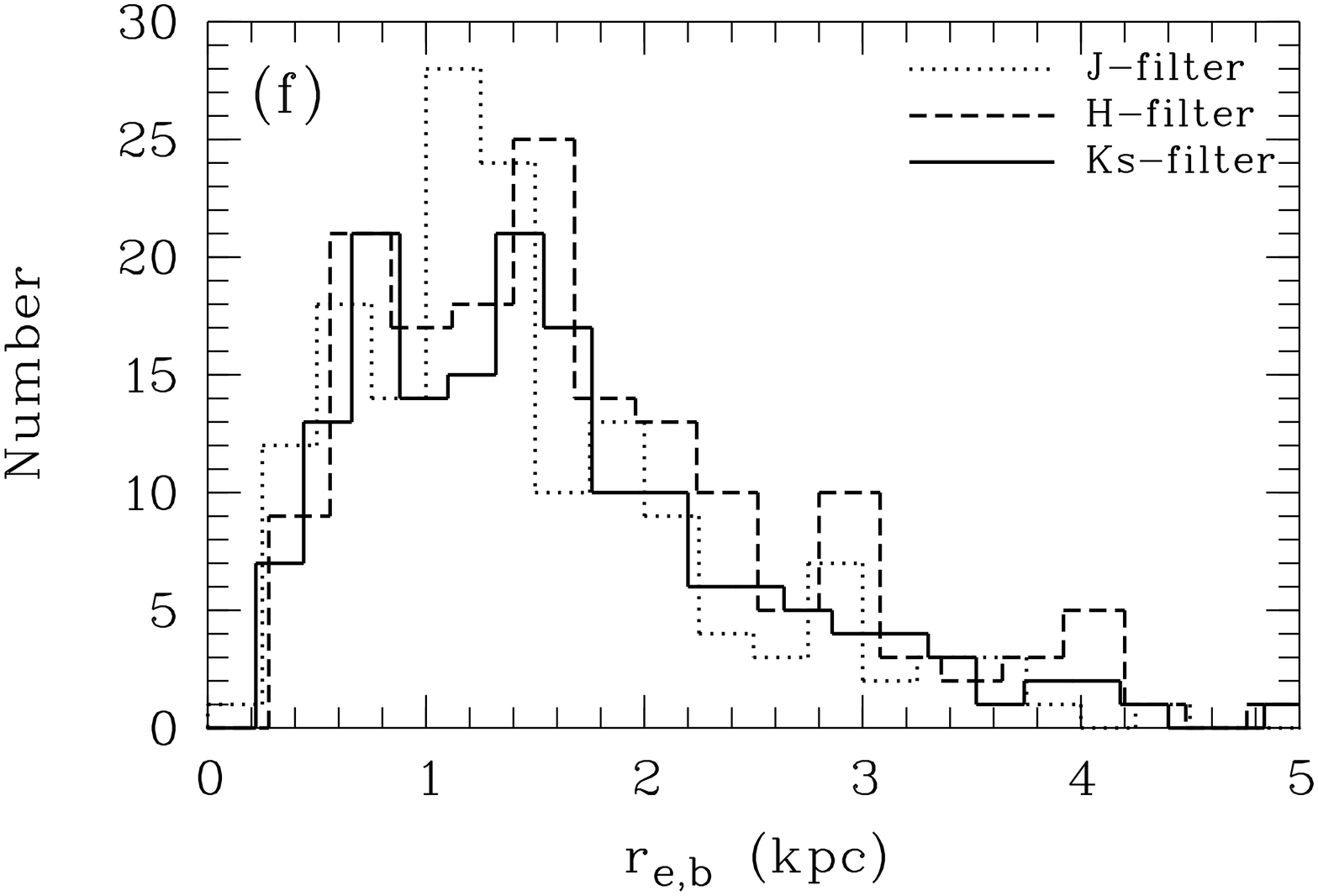}
  \includegraphics[width=5.4cm, angle=-90]{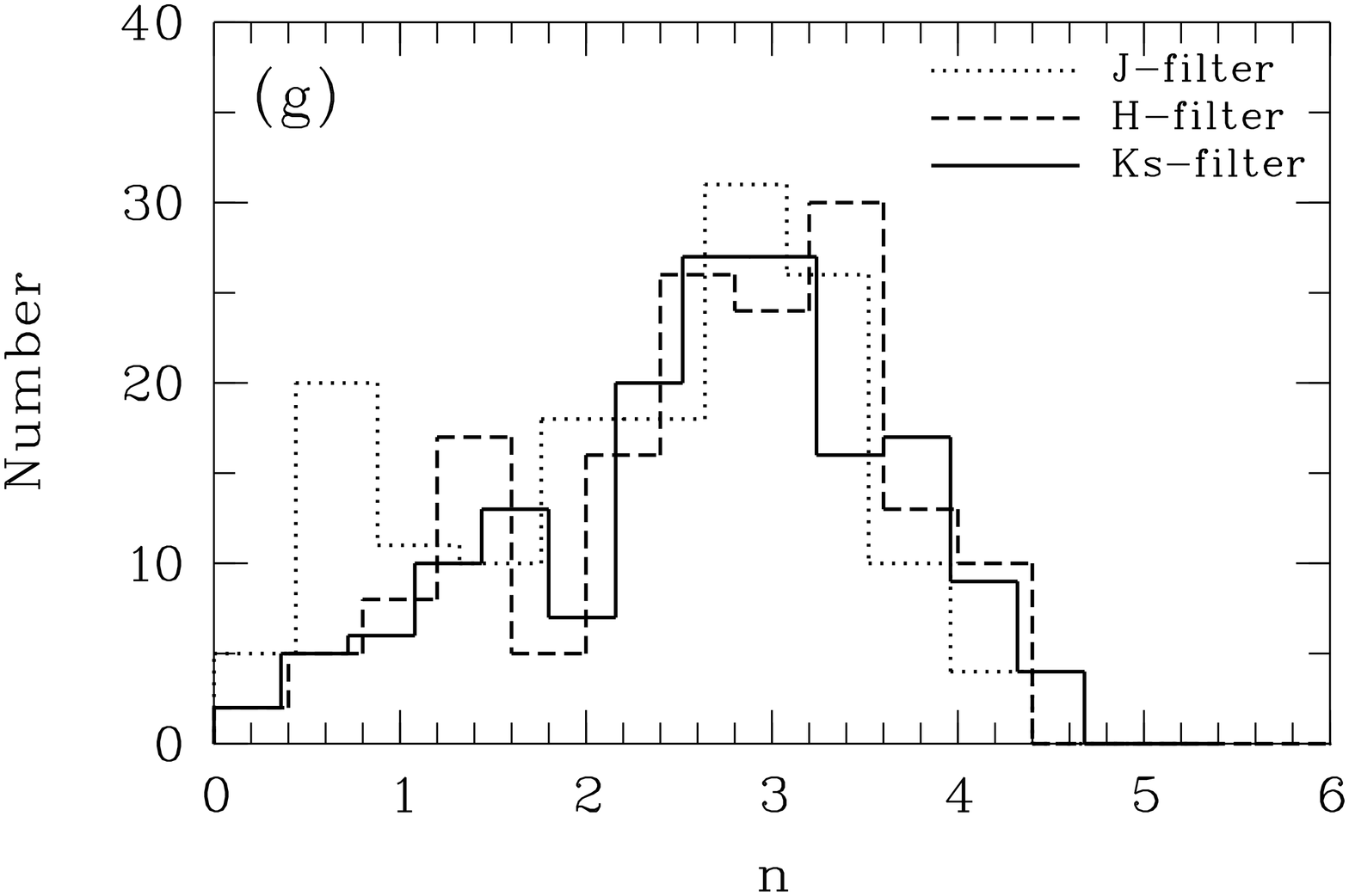}
  \includegraphics[width=5.4cm, angle=-90]{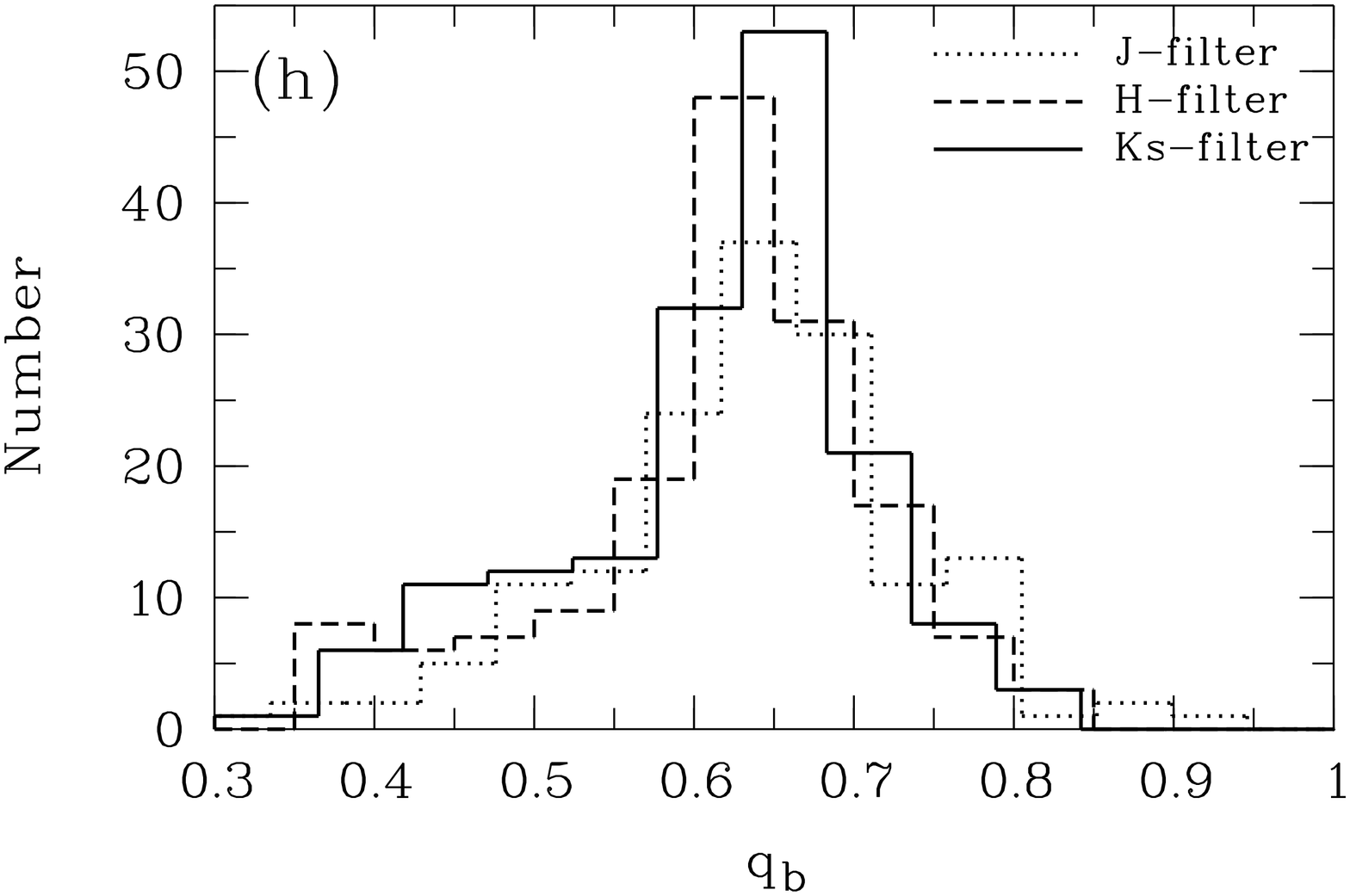}
 \end{center}
 \caption{\small{(a) 
 Distributions of the sample galaxies over 
 (a) the central surface brightness of the disc, 
 (b) the  disc scalelength, 
 (c) the disc scaleheight, 
 (d) the $h/z_0$ ratio, 
 (e) the effective surface brightness of the bulge, 
 (f) the bulge effective radius, 
 (g) the \ser\ index of bulge, 
 (h) the model bulge axis ratio $q_\mathrm{b} = b/a$. 
 In each plot the solid line corresponds to the $K_s$-band, 
 the dashed line shows the distributions corresponds to the $H$-band 
 and the dotted line corresponds to the $J$-band.}}
\label{Distrib}
\end{figure*}
\end{center}

The distributions of the apparent effective surface brightness of bulges are 
shown in Fig.~\ref{Distrib}e. The distributions in each filter have a narrow 
peak in contrast to the distributions of the central surface brightness of 
discs. 

Fig.~\ref{Distrib}f demonstrates the fairly broad distributions of the 
effective radii of the bulges of galaxies. The distributions of the 
\ser\ indices of galaxies are shown in Fig.~\ref{Distrib}g. The 
distributions demonstrate a weak bimodality which may reflect the existence 
of two families of bulges: bulges with $n \ga 2$ and bulges with 
$n \la 2$. This bimodality is known for some galaxy samples 
(e.g. Fisher \& Drory 2008) and probably is real for our sample. As was 
reported earlier (see e.g. Fisher \& Drory 2007, 2008) such a bimodality 
correlates with morphological type of the bulge. Classical bulges have 
$n \ga 2$ and so-called pseudobulges have $n \la 2$, but this \ser\ 
index threshold can be considered only as an approximation to identify 
pseudobulges. Further we are showing the difference in the photometric planes 
for these two types of bulges.

In Fig.~\ref{Distrib}h we demonstrate the distributions of model bulge axis 
ratio $q_\mathrm{b} = b/a$ for edge-on view. In the case of edge-on 
galaxies this parameter describes directly the bulge flatness and the 
intrinsic 3D structure of bulges if they are assumed to be oblate spheroids. 
The Table~\ref{MedGal} gives the median value of this parameter $\sim 0.63$, 
independently of the band. This is in good agreement with 
Moriondo, Giovanardi \& Hunt (1998) and Noordermeer \& van der Hulst (2007). 
Moriondo et al. (1998) derived for their samples of 14 moderately inclined 
galaxies the median intrinsic axis ratio $q_\mathrm{b} = 0.64$ while 
Noordermeer \& van der Hulst (2007) gave the average value of 
$\langle q_\mathrm{b} \rangle = 0.55 \pm 0.12$ for their sample of 21 
early-type disc galaxies decomposed into contributions from a spheroidal 
bulge with a \ser\ profile and a flat disk. Thus, bulges are definitely 
nonspherical and flattened.

In recent years a great deal of new data has made researchers 
revise the assumption that bulges are oblate spheroids. 
However, conclusions about 
intrinsic 3D shapes of bulges and their possible triaxiality are still somewhat 
controversial, despite the fact that such conclusions are thought to be crucial 
for testing different scenarios of galaxy formation. For galaxies at 
intermediate inclination angles misalignment between the bulge and the disk 
major axes indicates that bulges are probably triaxial 
(Bertola, Vietri \& Zeiliger 1991). 
From the observed distribution function of apparent ellipticities and 
misalignment angles Bertola et al. (1991) determined the probability 
distribution function (PDF) of intrinsic 
axial ratios for 32 bulges of their sample. The peak of the PDF falls on the 
intrinsic equatorial axis 
ratio\footnote{We use capital letters $A$, $B$, $C$ to define 
the intrinsic semi-axes of a triaxial system and small letters $a$, $b$ 
for the apparent semi-axes of an image. In axisymmetric case and 
an edge-on view the ratio $b/a$ coincides with the ratio $C/A$.} 
$B/A = 0.95$ and bulge flattening 
$C/A = 0.65$. About one half of all bulges seems to be close to oblate, 
with the remainder being triaxial. Fathi and Peletier (2003) measured the 
bulge deprojected ellipticity in the equatorial plane that was obtained from the 
ellipse fitting the galaxy isophotes within a bulge region. One can estimate 
the median of $B/A$ from their distribution function as $0.7-0.8$ with smaller 
values attributed to more elongated bulges in late-type galaxies. 
M\'endez-Abreu et al. (2008) also focused their attention on the intrinsic 
equatorial ellipticity of bulges. They reconstructed the PDF for the sample 
of 148 unbarred S0-Sb galaxies and obtained the value of 
$\langle B/A \rangle = 0.85$. This is consistent with findings by 
Bertola et al. (1991) and Fathi and Peletier (2003). However, in contrast to 
Fathi and Peletier (2003), M\'endez-Abreu et al. (2008) did not find any 
significant differences in the shape of bulges between samples of early- and 
late-type galaxies. 

\begin{figure}
 \centering
 \includegraphics[width=6cm, angle=-90]{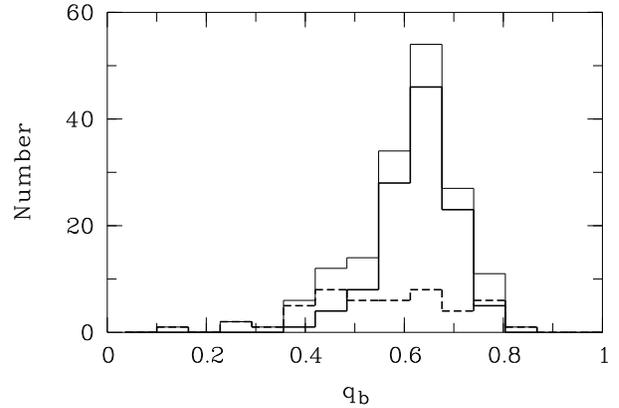}
 \caption{\small{Distribution of the sample galaxies over 
 the model bulge axis ratio $q_\mathrm{b}$ in the $K_s$-band. 
 The thin line indicates the distribution for the whole sample, 
 the solid line  corresponds to the subsample of bulges with $n \ga 2$ and 
 the dashed line shows the distribution for the subsample of bulges with 
 $n \la 2$.}}
\label{Distrib_qb}
\end{figure}

Our sample of edge-on galaxies allows us to distinguish clearly the difference 
in the bulge equatorial ellipticities for early- and late-type galaxies. As a 
\ser\ index is correlated with the Hubble type 
(Andredakis, Peletier \& Balcells 1995; Graham 2001; M\"ollenhof 2004) 
(as well as with the bulge morphology, dividing bulges into classical ones 
and pseudobulges; Fisher \& Drory 2008) we divided our 
sample bulges between those with \ser\ index $n \ga 2$ (early-type 
galaxies or classical bulges) and those with $n \la 2$ (late-type galaxies or 
pseudobulges) using $K_s$ filter. 
As one can see in Fig.~\ref{Distrib_qb} for bulges with 
$n \ga 2$ the distribution of $q_\mathrm{b}$ has a rather narrow peak at 
$q_\mathrm{b} \approx 0.65$. This may reflect the fact that bulges in 
early-type spirals are nearly oblate spheroids with moderate flattening. The 
distribution of $q_\mathrm{b}$ for bulges with $n \la 2$ is very wide, 
spreading from flat bulges up to nearly spherical ones. Such a distribution 
may be attributed to definitely triaxial, near prolate bulges that are seen 
from different projections --- along the major axis and perpendicular to it. 
But we can not exclude that triaxial shape of bulges in late-type 
galaxies may hide the presence of bars that thickened in the vertical 
direction during secular evolution. 
We are planning to investigate this question in more detail in a forthcoming 
paper.

\begin{figure}
 \centering
 \includegraphics[width=6cm, angle=-90]{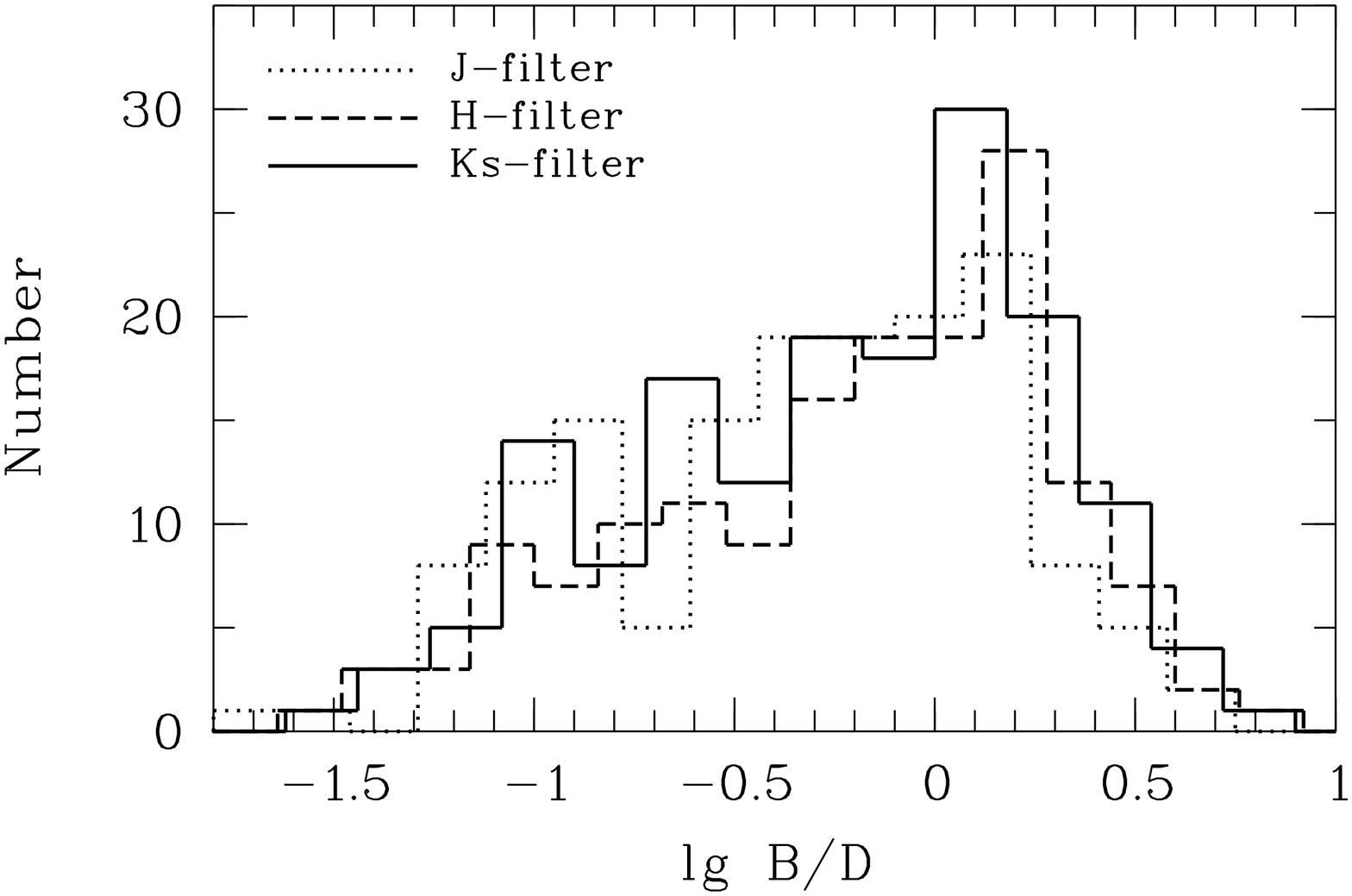}
 \caption{\small{Distribution of the sample galaxies over the ratio 
 bulge-to-disc luminosities $B/D=L_\mathrm{b}/L_\mathrm{d}$.}}
\label{BD}
\end{figure}

\begin{figure*}
 \centering
 \includegraphics[width=5.6cm, angle=-90, clip=]{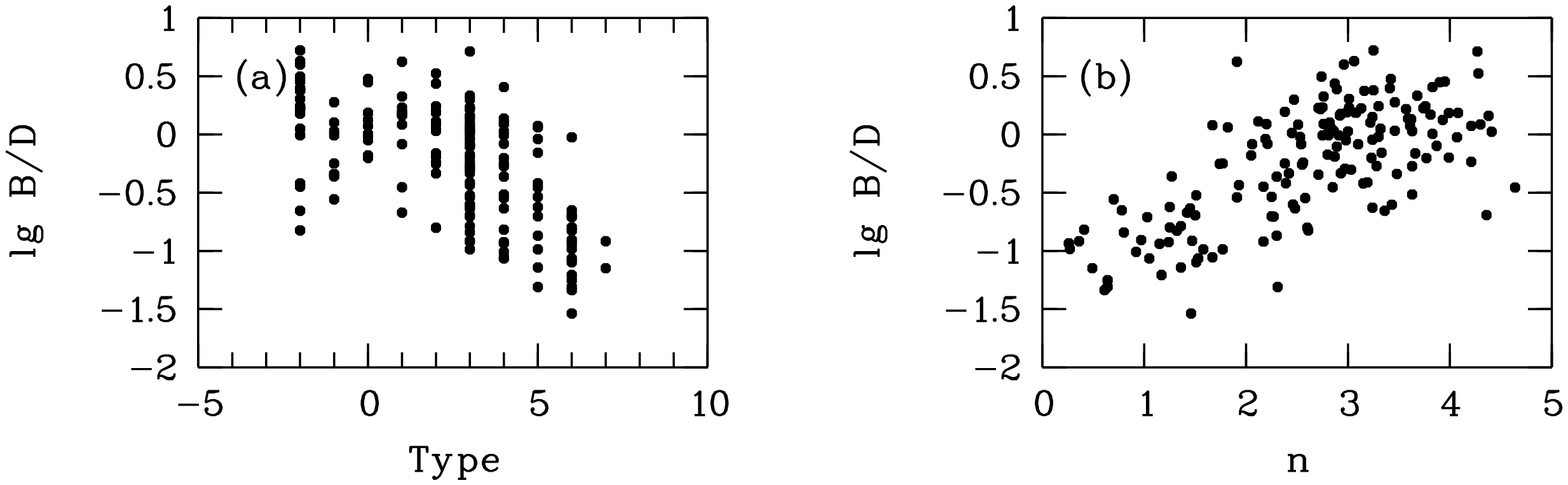}
 \caption{\small{(a) The bulge to disc ratio as a function of 
 the morphological type 
 according to the RFGC (for late type galaxies RFGC 
 provides more detailed dividing into types than LEDA), 
 (b) the bulge to disc ratio versus the \ser\  
 index $n$ ($K_s$ passband).}}
\label{HT}
\end{figure*}

Fig.~\ref{BD} demonstrates the distributions of the ratio of bulge and disc 
luminosities $B/D \equiv L_\mathrm{b}/L_\mathrm{d}$. Our sample contains 
many early-type galaxies with $B/D \ga 0.5$ and late-type galaxies with values 
of $B/D \la 0.2$. For edge-on galaxies the ratio of $B/D$ is a key 
parameter in addition to $n$ for classifying spiral galaxies on the Hubble 
sequence because the spiral arms are not seen in this case (de Jong 1996; 
Graham 2001). Our sample also demonstrates a correlation between the 
ratios of $B/D$ and the Hubble morphological types $T$ taken from LEDA 
(Fig.~\ref{HT}a) that is in agreement with previous work 
(e.g., de Jong 1996; Graham 2001; M\"ollenhoff \& Heid 2001; 
Hunt, Pierini \& Giovanardi 2004; M\"ollenhoff 2004). 
However for edge-on galaxies morphological types are very 
subjective. Comparisons of the Hubble types given by different classifiers 
show an rms uncertainty in the type index of order 2 $T$-units 
(Lahav et al. 1995). Unfortunately there are errors in the bulge/disc 
decomposition. Therefore we use both approaches (type $T$ from LEDA and 
obtained $B/D$) for describing morphological types of edge-on galaxies. In 
some cases we will distinguish early- and late-type galaxies using 
the \ser\ index $n$ because there is a good correlation between 
$B/D$ and $n$ (Fig.~\ref{HT}b). This correlation is not new and was presented 
by many authors (e.g., Andredakis et al. 1995; Graham 2001; 
M\"ollenhoff \& Heid 2001; Hunt et al. 2004).

Details of the statistical properties of the sample discs and bulges are
presented in the Table~\ref{MedDisc}, the Table~\ref{MedBulge} and the 
Table~\ref{MedGal}. The values of $n$ in Table~\ref{MedBulge} 
systematically  increase going from $J$ to  $K_s$ -- this may imply 
that $n$ is systematically underestimated due to extinction affecting 
the central peak of emission at the shorter near-IR wavelengths. 
Another possible explanation is a real change of the surface brightness shape
with wavelength. But, as one can see in Table~\ref{MedBulge},
the quoted changes are within statistical errors. 

\begin{table}
 \centering
 \begin{minipage}{80mm}
 \parbox[t]{80mm}{\caption{The median values of 
 $\mu_{\mathrm{0,d}}$, $h$, $z_0$ and $M_{\mathrm{d}}$ and their quartiles 
 (75 and 25 per cent, respectively)}\label{MedDisc}}
  \begin{tabular}{ccccc}
  \hline
  \hline
  Sp & $\mu_{\mathrm{0,d}}$ & $h$ & $z_0$ & $M_\mathrm{d}$\\ 
     & ($^m/\mathrm{arcsec}^2$)  & (kpc) & (kpc) & $(^m)$ \\ 
  \hline  
   J &  $17.57^{-0.55}_{+0.43}$ & $3.68^{-1.30}_{+1.44}$ & $1.04^{-0.34}_{+0.48}$ & $-22.53^{-0.77}_{+0.83}$ \\ 
   H &  $16.87^{-0.43}_{+0.35}$ & $3.87^{-1.36}_{+1.70}$ & $0.99^{-0.29}_{+0.45}$ & $-23.23^{-0.78}_{+0.82}$ \\ 
K$_\mathrm{s}$ &  $16.44^{-0.41}_{+0.43}$ & $3.66^{-1.27}_{+1.49}$ & $0.97^{-0.30}_{+0.45}$ & $-23.52^{-0.79}_{+0.75}$ \\ \hline          
\end{tabular}
\end{minipage}
\end{table}    

\begin{table}
 \centering
 \begin{center}
 \begin{minipage}{80mm}
 \parbox[t]{80mm}{\caption{The median values of 
 $\mu_{\mathrm{e,b}}$, $r_{\mathrm{e,b}}$, $n$ and $M_{\mathrm{b}}$ and their 
quartiles (75 and 25 per cent, respectively)}\label{MedBulge}}
  \begin{tabular}{ccccc}
  \hline 
  \hline 
  Sp & $\mu_{\mathrm{e,b}}$ & $r_{\mathrm{e,b}}$ & $n$ & $M_\mathrm{b}$ \\ 
     & ($^m/\mathrm{arcsec}^2$)  & (kpc) &    & $(^m)$ \\ 
  \hline  
   J &  $17.68^{-0.13}_{+0.31}$ & $1.23^{-0.54}_{+0.55}$ & $2.35^{-1.36}_{+0.70}$ & $-21.97^{-0.97}_{+1.85}$  \\ 
   H &  $17.13^{-0.17}_{+0.36}$ & $1.48^{-0.63}_{+0.68}$ & $2.65^{-1.06}_{+0.68}$ & $-22.96^{-0.95}_{+1.93}$ \\ 
 K$_\mathrm{s}$ &  $16.69^{-0.17}_{+0.34}$ & $1.35^{-0.59}_{+0.61}$ & $2.74^{-1.44}_{+0.53}$ & $-23.19^{-0.94}_{+2.16}$ \\ 
\hline
\end{tabular}
\end{minipage}
\end{center}
\end{table}  

\begin{table}
 \centering
 \begin{minipage}{55mm}
 \parbox[t]{57mm}{\caption{The median values of 
$h/z_0$, $q_\mathrm{b}$ and $B/D$ and their quartiles 
(75 and 25 per cent, respectively)}\label{MedGal}}
  \begin{tabular}{cccc}
  \hline 
  \hline   
  Sp & $h/z_0$ & $q_\mathrm{b}$ & $B/D$ \\ 
  \hline  
   J & $3.54^{-0.93}_{+0.86}$  & $0.63^{-0.07}_{+0.06}$ & $0.50^{-0.38}_{+0.62}$ \\ 
   H & $3.94^{-0.85}_{+0.92}$& $0.63^{-0.08}_{+0.05}$ & $0.69^{-0.49}_{+0.79}$\\ 
 K$_\mathrm{s}$ & $3.92^{-0.92}_{+0.95}$  & $0.62^{-0.11}_{+0.04}$ & $0.64^{-0.44}_{+0.70}$\\ 
\hline
\end{tabular}
\end{minipage}
\end{table}  

\subsection[]{Scaling relations}

The relationships between bulge and disc parameters might give an insight in 
the ages of the bulge and disc formation and help to reconstruct the 
chronology of these events. Andredakis \& Sanders (1994) were the first to 
report the existence of a correlation between disc and bulge scalelengths for 
their sample of 34 late-type spirals. Just after that, such a correlation was 
confirmed and used for speculating about the scale-free Hubble sequence since 
the relative size of bulge and disc seemed not to depend on the morphological 
type (de Jong 1996; Courteau et al. 1996). De Jong (1996) found the equation 
for the least squares fitted line in the $K$ passband: 
$\lg(r_\mathrm{e,b}) = 0.95\lg(h) - 0.86$ with a standard deviation of 0.17 
and correlation coefficient $r \sim 0.8$. This suggests a linear relation 
between $h$ and $r_\mathrm{e,b}$, i.e. the disc scalelength increases when 
the bulge effective radius increases.

Furthermore it became clear that the ratio $r_\mathrm{e,b} / h$ appeared to be 
strongly influenced by the bulge parametrization 
(Graham \& Prieto 1999; Graham 2001). 
The independence of morphological type and a strong correlation were seen only 
when galaxies were best modeled with a fixed $n = 1$ bulges for all type 
galaxies 
(de Jong 1996; Courteau et al. 1996; Moriondo et. al. 1998; Hunt et al. 2004). 
According to whether bulges were fitted with the $n = 1$ \ser\ model or 
with the de Vaucoulers law, the correlation between scalelengths was clear 
or disappeared completely 
(Andredakis \& Sanders 1994; de Jong 1996; Hunt et al. 2004). 

Using \ser\ bulges with a free shape parameter $n$ as the best fitting 
models 
(M\"ollenhoff \& Heidt 2001; Khosroshahi et al. 2000b; Graham 2001; 
MacArthur, Courteau \& Holtzman 2003) 
or different but fixed $n$ for early- and late-type galaxies 
(Graham \& Prieto 1999) 
confirms this correlation with a fairly large scatter and opens the discussion 
about the variation of the ratio $r_\mathrm{e,b} / h$ with Hubble types 
(Moriondo et al. 1998; Graham \& Prieto 1999; Khosroshahi et al. 2000b; 
Graham 2001; Scodeggio et al. 2002; MacArthur et al. 2003; M\"ollenhoff 2004) 
and about the prematurity of the claims made by 
de Jong (1996) and Courteau et al. (1996) for a scale-free Hubble sequence.

\begin{figure*}
 \centering
 \includegraphics[width=5.4cm, angle=-90]{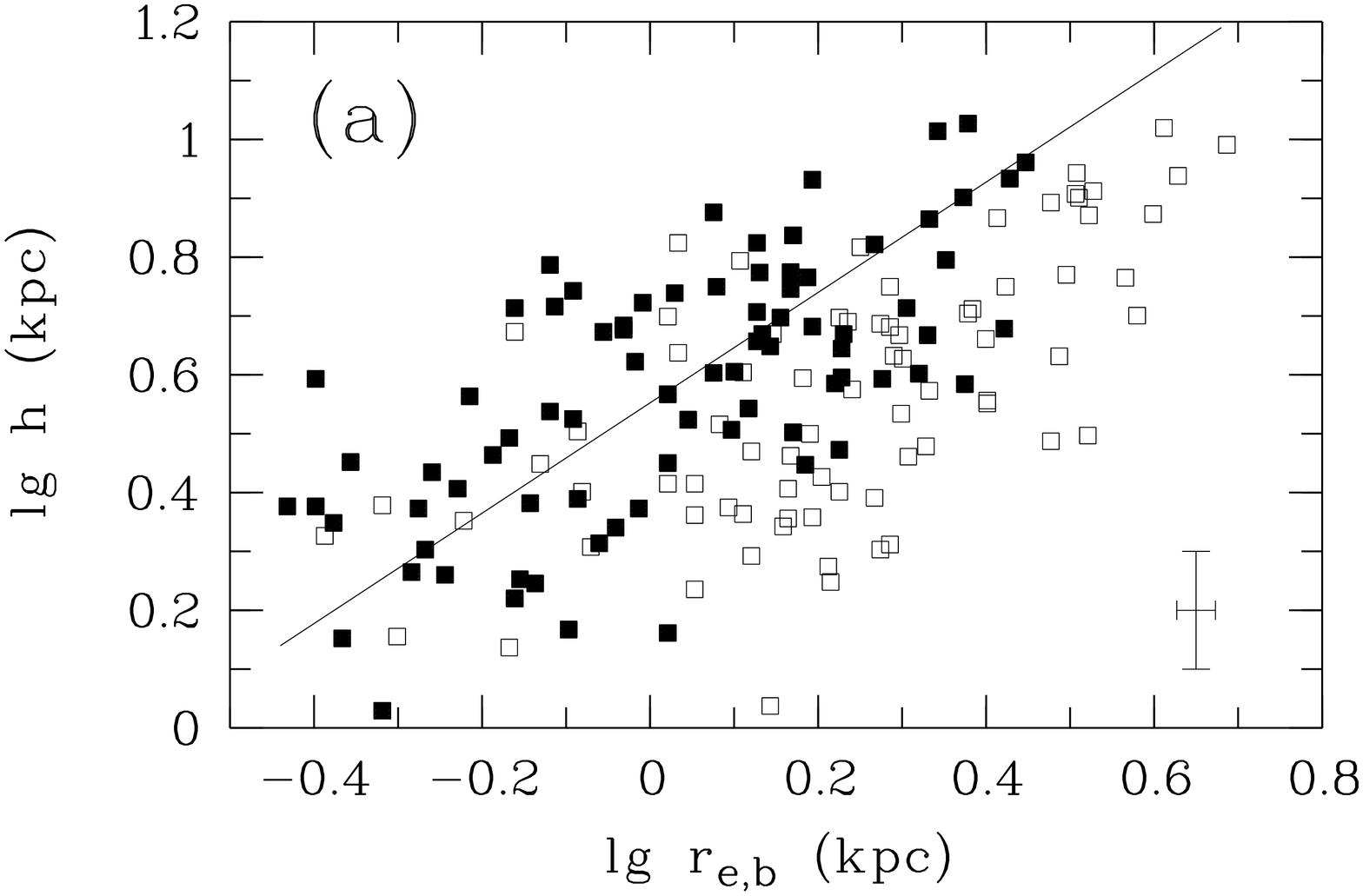}
 \includegraphics[width=5.4cm, angle=-90]{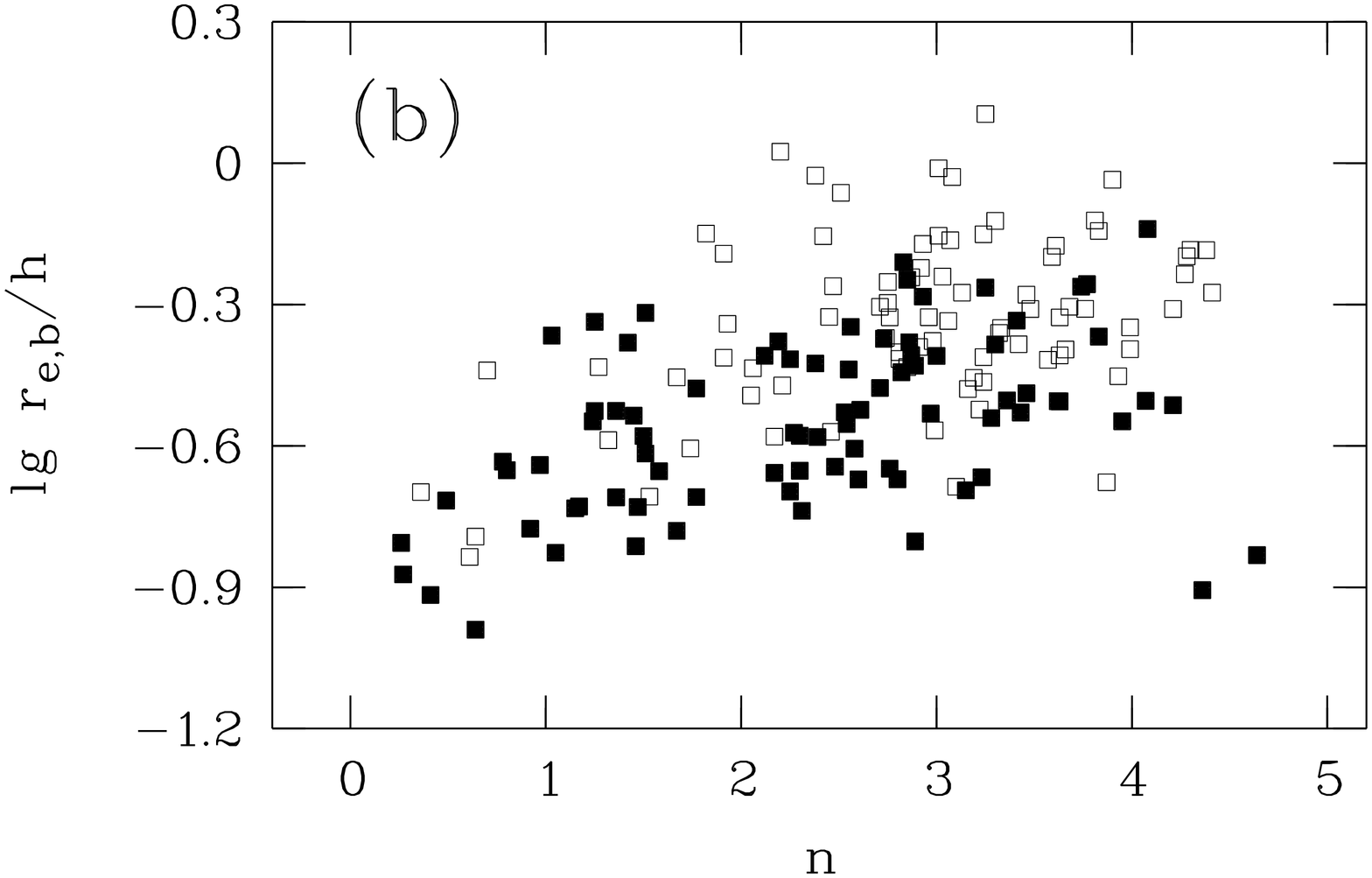} 
 \caption{\small{
 (a) The disc scalelength $h$ vs. the bulge effective radius 
 $r_\mathrm{e,b}$, (b) $r_\mathrm{e,b}/h$ ratio vs. \ser\ index $n$
in the $K_s$-band. Open squares are related to the whole sample and 
the filled squares are  related to the complete sample. The solid 
line corresponds to the regression line for the complete sample. 
The errorbars indicate the median error for all galaxies. }}
\label{Corr1}
\end{figure*}

We also found that there exist fairly strong correlations between structural 
parameters of discs and bulges. The correlation between $h$ and 
$r_\mathrm{e,b}$ is shown in Fig.~\ref{Corr1}a. In this figure and hereafter 
we mark the complete subsample galaxies with filled squares. The regression 
line is plotted only for points of the complete subsample. The equations of 
the linear regression and the correlation coefficients are presented in 
the Table~\ref{Correlations1}. These equations are consistent with a linear 
relation between $h$ and $r_\mathrm{e,b}$.

We found a mean $r_\mathrm{e,b}/ h$ in the $J$-band of $0.39 \pm 0.18$, 
in the $H$-band of $0.42 \pm 0.20$ and in the $K_s$-band of $0.40 \pm 0.20$, 
which are in agreement within errors with the values obtained by 
M\'endez-Abreu et al. (2008) for their sample of 148 unbarred S0--Sb 
galaxies $\langle r_\mathrm{e,b} / h \rangle = 0.36 \pm 0.17$ (in the $J$-band). 

There is some controversial evidence for systematic changes with morphological 
type in the ratio $r_\mathrm{e,b} / h$ 
(Scodeggio et al. 2002; MacArthur et al. 2003; Hunt et al. 2004; 
M\"ollenhoff 2004). 
Despite being difficult to detect, the dependence with Hubble types might be 
real in the sense that the sample of predominantly late-type disk 
galaxies gives $\langle r_\mathrm{e,b} / h \rangle = 0.22 \pm 0.09$ 
(MacArthur et al. 2003; the $H$-band) and the sample of predominantly 
early-type galaxies leads to 
$\langle r_\mathrm{e,b}/ h \rangle = 0.33 \pm 0.17$ 
(Khosroshahi et al. 2000b; the $K$-band).

For edge-on galaxies the reliable indicator of morphological type is 
$B/D$ or $n$. Fig.~\ref{Corr1}b demonstrates the dependence of the ratio 
$r_\mathrm{e,b}/ h$ on the \ser\ shape parameter $n$. There is a 
clear trend for $r_\mathrm{e,b}/ h$ to increase with $n$. The size ratio 
increases from $n \sim 0.5$ to $n \sim 5$ by a factor of $\sim 3-4$. 
Almost all bulges of the sample of MacArthur et al. (2003) have 
$n < 1.2$ and most of the bulges of the sample of Khosroshahi et al. (2000b) 
have $n > 2.2$. Thus, the difference in the mean value of 
$r_\mathrm{e,b}/ h$ for these two samples reflects the same tendency that 
one can see in Fig.~\ref{Corr1}b. The same increase was declared by 
Hunt et al. (2004). They found 
for $n=1$ bulges $\langle r_\mathrm{e,b}/ h \rangle = 0.14 \pm 0.06$, 
for $n=2$ bulges $\langle r_\mathrm{e,b}/ h \rangle = 0.25 \pm 0.17$, 
for $n=3$ bulges $\langle r_\mathrm{e,b}/ h \rangle = 0.35 \pm 0.27$, 
and
for $n=4$ bulges $\langle r_\mathrm{e,b}/ h \rangle = 0.54 \pm 0.54$. 

Since $n$ is an indicator (in statistical sense) of morphological type, 
the changes of $r_\mathrm{e,b}/ h$ with the shape of the bulge implies 
that Hubble sequence is not scale-free. The same conclusion has been 
recently made by Gadotti (2009) for his sample of nearly 1000 galaxies. 
Moreover, he was able to distinguish in his sample two subsamples of 
classical bulges and pseudobulges. Gadotti (2009) have used the Kormendy
relation (Kormendy 1977) for identifying pseudobulges and bulges by their 
loci in the 
$\langle \mu_\mathrm{e,b} \rangle$\footnote{The mean effective surface 
brightness within the circle of effective radius $r_\mathrm{e,b}$.} 
-- $r_\mathrm{e,b}$ plane. The corresponding loci are quite different for 
these two types of bulges. Gadotti (2009) found that $r_\mathrm{e,b}/ h$ for 
galaxies of both subsamples was clearly correlated with the bulge-to-total 
luminosity ratio, although the corresponding relations for pseudobulges and 
classical bulges were offset. As the bulge-to-total luminosity ratio is 
related to the morphological (Hubble) type of a galaxy, the existence of 
such relations rule out the hypothesis of a scale-free Hubble sequence. 
It means that the widely discussed secular evolution scenario of the bulge 
formation should be improved to include the variation of $r_\mathrm{e,b}/ h$ 
with $n$ (or other indicators of the type) even in the case of pseudobulges.

\begin{figure*}
 \centering
 \includegraphics[width=5.4cm, angle=-90]{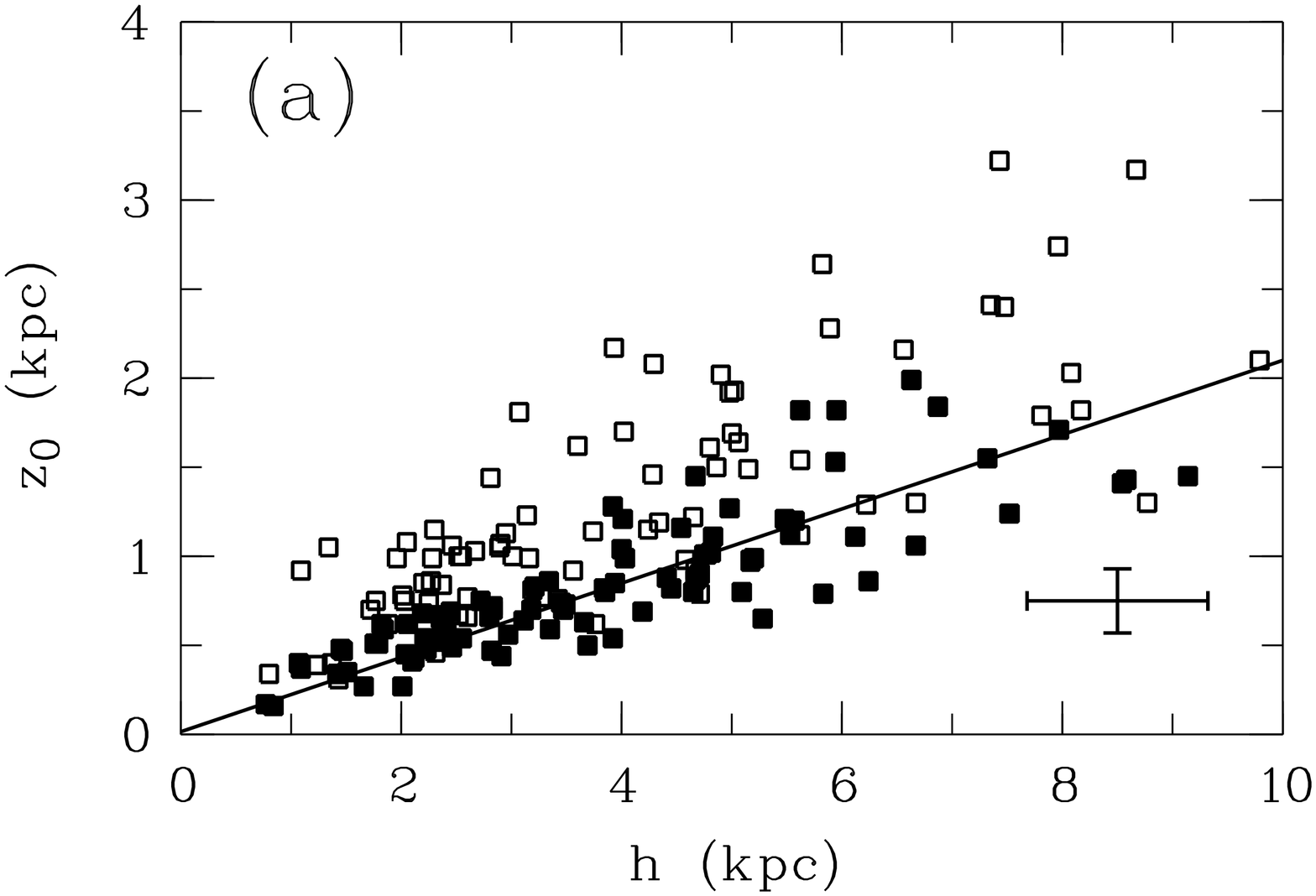}
 \includegraphics[width=5.4cm, angle=-90]{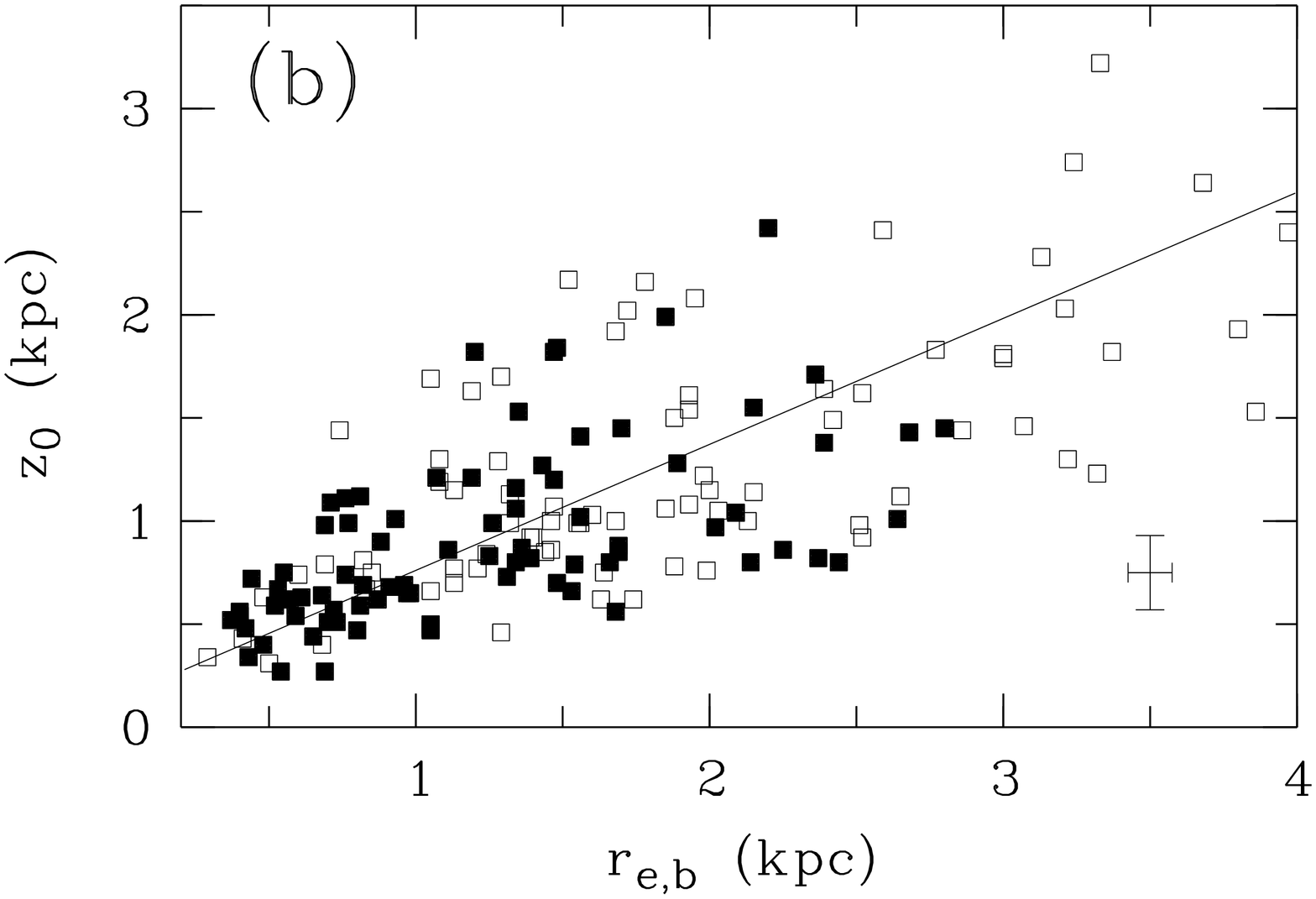} 
 \caption{\small{(a) The disc scaleheight $z_0$ vs. the disc scale length $h$,
 (b) the disc scale height $z_0$ vs. the bulge effective radius 
 $r_\mathrm{e,b}$ ($K_s$). Symbols as in Fig.~\ref{Corr1}.}}
\label{Corr2}
\end{figure*}

The disc flattening can be measured directly only in edge-on galaxies. It 
is commonly expressed as the ratio of disc scaleheight $z_0$ to disc 
scalelength $h$. This ratio shows a weak trend with Hubble type 
(de Grijs 1998), but there is a fairly clear correlation between both scale 
parameters (e.g. Kregel et al. 2002). The result of Kregel et al. (2002) 
can be transformed into the mean value of 
$\langle h/ z_0 \rangle = 3.7 \pm 1.1$ for the volume corrected distribution. 
Our data also demonstrates such a correlation for galaxies of all types
(Fig.~\ref{Corr2}a). It is remarkable that the line of the linear regression 
is drawn through the point close to $(0,0)$.

We used the data of photometric parameters $h$ and $z_0$ in BM02 ($K_s$ filter)
of their 
reliable subsample and found the following equation of linear regression:
$z_0 = (0.210\pm0.026) \, h - (0.017\pm0.258)$ with 
correlation coefficient $r = 0.727$, 
which is very close to our result (Table~\ref{Correlations1}).

Thus, the galaxies with large disc scalelengths have on average larger disc 
scaleheights and larger effective radii of their bulges. However, one 
should keep in mind the fact that the spread in the $h - z_0$ plane is large. 
This scatter is reflected in a wide and remarkably flat distribution of the 
ratio $h/z_0$ presented in Fig.~\ref{Distrib}d. We return to this question 
in Section~\ref{DHalo}.

As is seen in Fig.~\ref{Corr2}b the disc scaleheight tends to increase
with the bulge effective radius. The correlation between $z_0$ and 
$r_\mathrm{e,b}$ is new and was not described earlier. We found the
following ratios of mean $r_\mathrm{e,b}/z_0$: 
$\langle r_\mathrm{e,b}/z_0 \rangle = 1.23 \pm 0.48$ for the $J$-band, 
$\langle r_\mathrm{e,b}/z_0 \rangle = 1.57 \pm 0.62$ for the $H$-band, 
$\langle r_\mathrm{e,b}/z_0 \rangle = 1.49 \pm 0.60$ for the $K_s$-band.

\begin{table}
 \centering
 \begin{minipage}{90mm}
 \parbox[t]{80mm}{\caption{The best fitting results for correlations between
 bulge and disc scale parameters }\label{Correlations1}}
  \begin{tabular}{l}
  \hline
  \hline 
  $J$:\,\,\,\,\,\,$\lg h = (1.011 \pm 0.143) \lg r_\mathrm{e,b} + (0.596 \pm 0.010)$, $r=0.59$  \\
  $H$:\,\, $\lg h = (0.936 \pm 0.113) \lg r_\mathrm{e,b} + (0.533 \pm 0.030)$, $r=0.66$  \\ 
  $K_s$: $\lg h = (0.933 \pm 0.104) \lg r_\mathrm{e,b} + (0.555 \pm 0.021)$, $r=0.68$  \\ 
  \hline
  $J$:\,\,\,\,\,\,$z_0 = (0.237 \pm 0.015) h + (0.010 \pm 0.140)$, $r=0.87$  \\
  $H$:\,\,  $z_0 = (0.208 \pm 0.014) h + (0.030 \pm 0.144)$, $r=0.84$  \\
  $K_s$: $z_0 = (0.207 \pm 0.013) h + (0.027 \pm 0.129)$, $r=0.85$  \\ 
  \hline
  $J$:\,\,\,\,\,\,$z_0 = (0.891 \pm 0.120) r_\mathrm{e,b} + (0.058 \pm 0.356)$, $r=0.62$  \\
  $H$:\,\, $z_0 = (0.680 \pm 0.100) r_\mathrm{e,b} + (0.045 \pm 0.366)$, $r=0.59$  \\
  $K_s$: $z_0 = (0.715 \pm 0.094) r_\mathrm{e,b} + (0.041 \pm 0.325)$, $r=0.61$  \\ 
  \hline
\end{tabular}
\end{minipage}
\end{table} 

\subsection[]{Well-known relationships}

Here we investigate the general scaling relations for spiral
galaxies such as the Tully-Fisher relation, the Fundamental Plane (FP) 
of discs and the Photometric Plane of bulges (PhP). In this subsection we 
study the Tully-Fisher relation and the Fundamental Plane of discs and in 
the next subsection we specially investigate the Photometric Plane of bulges.

\subsubsection[]{The Tully-Fisher relation}

The Tully-Fisher (TF) relation links the luminosity and the maximum 
rotation velocity $v_\mathrm{rot}$ of galaxies. The small scatter 
in the Tully-Fisher relation allows us to use it as a method for 
measurements of distances to spiral galaxies. To determine the slope and the 
zeropoint of the TF relation we used the total luminosities of the sample 
galaxies taken from 2MASS ($J_\mathrm{ext}$, $H_\mathrm{ext}$ or 
$K_\mathrm{ext}$). They were derived from the isophotal magnitudes by 
using the $K_s$-band 20 mag/arcsec$^2$ elliptical radius ($J_{20}$, $H_{20}$, 
$K_{20}$) and then were extrapolated to roughly four times the disc 
scalelength $h$ (for the detailed description see Jarrett et al. 2000 and 
Cutri et al. 2006).

Unfortunately, measured rotation velocities are not known for all the 
sample galaxies. In the $J$-band the number of galaxies with known rotational 
velocities is 107, in the $H$-band there are 110 galaxies and in the 
$K_s$-band there are 116 objects with known velocities. The maximum rotational 
velocity corrected for inclination $v_\mathrm{rot}$ was taken 
from LEDA.

\begin{center}
 \begin{figure*}
 \begin{center}
  \includegraphics[width=5.4cm, angle=-90]{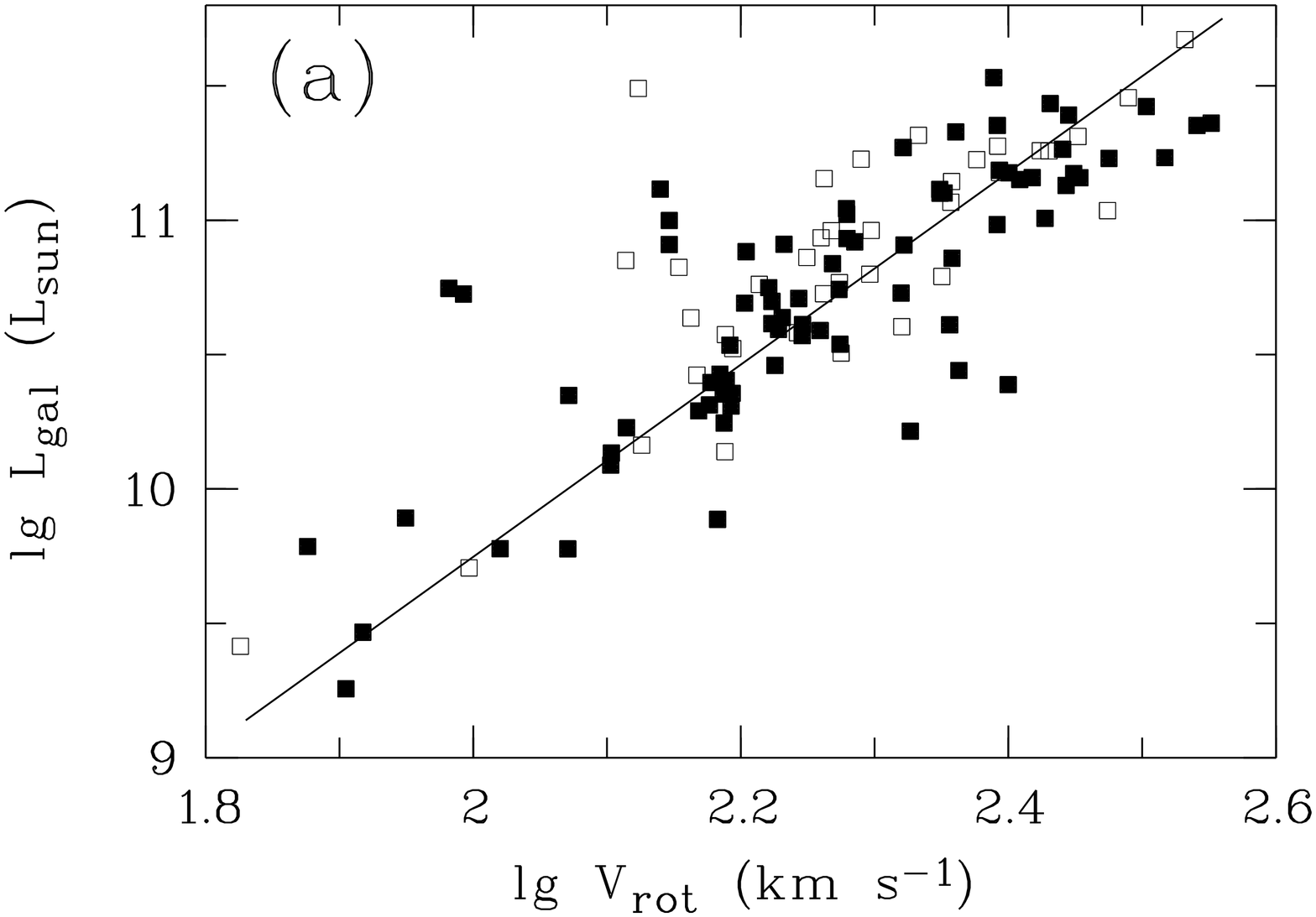}
  \includegraphics[width=5.4cm, angle=-90]{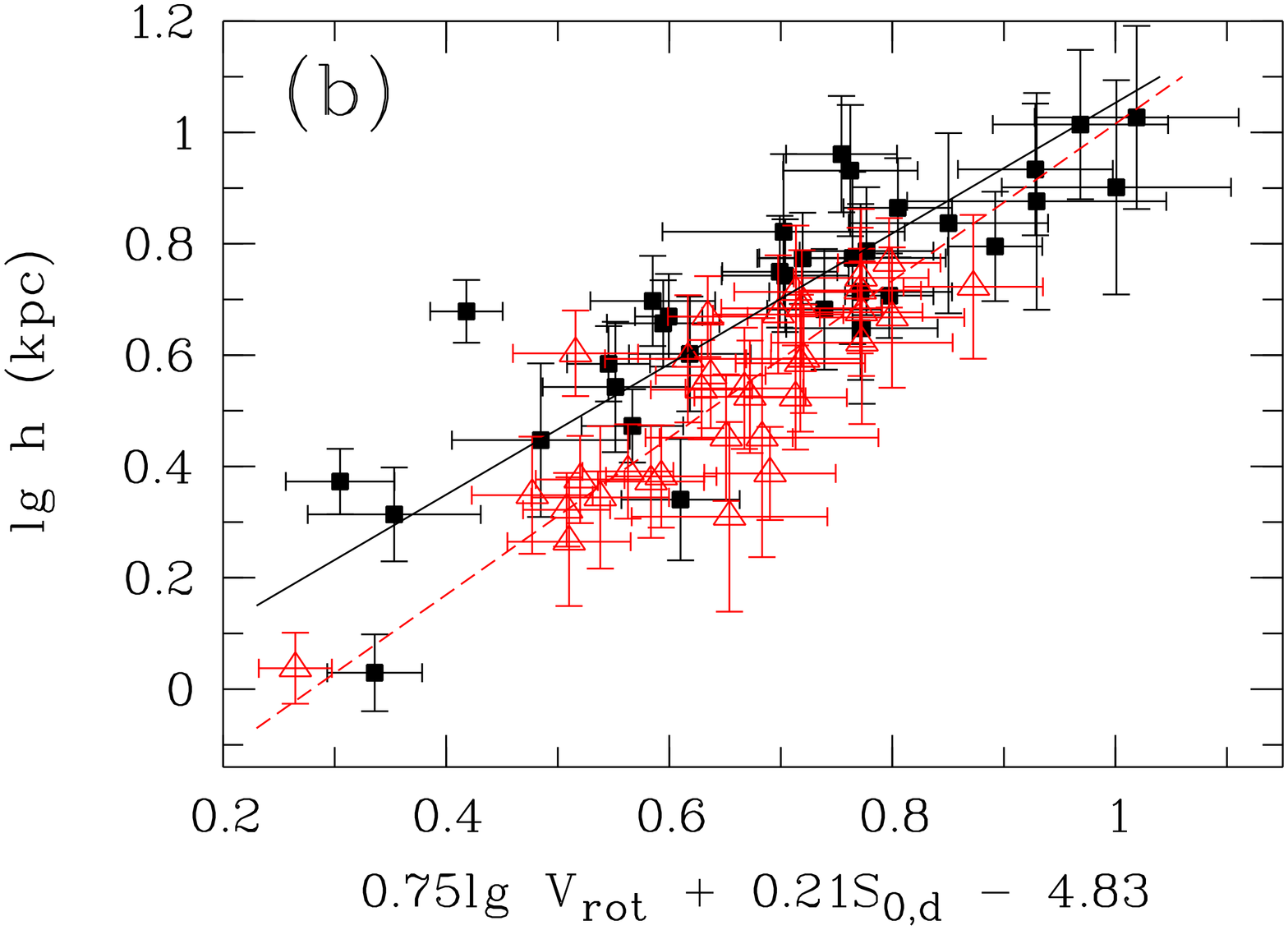}
 \end{center}
 \caption{\small{
 (a) The Tully-Fisher relation in the $K_s$-band. Symbols as in 
 Fig.~\ref{Corr1}.
 (b) The disc fundamental plane for galaxies with $n\ga2.2$ in 
 the $K_s$-band (filled squares). It is shown that galaxies with $n\la2.2$ 
 compose a different fundamental plane. These galaxies are plotted 
 with open  triangles. Only galaxies of the complete sample are shown.}}
 \label{discCorr}
 \end{figure*}
\end{center}

The absolute magnitude in different bands was calculated as
\begin{equation}
M_{\lambda} = m_{\lambda} - A_{\lambda}-25-5\lg(D_\mathrm{L}) \,,
\end{equation}
where $m_{\lambda}$ is the total apparent magnitude in each passband 
($J_\mathrm{ext}$, $H_\mathrm{ext}$ or $K_\mathrm{ext}$), $D_\mathrm{L}$ is 
the luminosity distance and $A_{\lambda}$ is the Galactic extinction taken from 
Schlegel et al. (1998). The luminosity is expressed as
\begin{equation}
\lg L_{\lambda} = 0.4(M^{\lambda}_{\sun}-M_{\lambda}) \,,  
\end{equation}
where $M^J_{\sun}=3.64$ mag, $M^H_{\sun}=3.32$ mag and $M^{K}_{\sun}=3.28$
mag (Binney \& Merrifield 1998).
We fitted the TF relation coefficients by taking the bisector between the
lines of linear regression obtained through the standard LSQ and inverse LSQ
methods. The best fitting results are presented 
in the Table~\ref{Correltions2}. The $K_s$-band TF relation is plotted in 
Fig.~\ref{discCorr}a.

There are many papers with the TF relation in the NIR bands for early- and 
late-type galaxies. Here we compare our results with those of some works 
based on 2MASS observations. The results by  
Karachentsev et al. (2002), Courteau et al. (2007),  
de Rijcke et al. (2007) and Masters, Springob \& Huchra (2008) are presented 
in Table~\ref{TF}.
 
As was shown by Masters et al. (2008) the TF relation coefficients depend 
on galaxy morphology. The relationship is steeper for later type spirals than 
earlier type spirals. The TF zeropoint is dimmer for late-type galaxies. We 
did not divide our sample on morphological types to get the TF relation for 
various galactic types. Our sample contains an almost equal number of 
early- and late-type galaxies. That is why the coefficients found 
are somewhat different from those given in other papers.

\begin{table*}
 \centering
 \begin{minipage}{140mm}
  \parbox[t]{140mm} {\caption{Comparison between some published 2MASS TF 
  relations $\lg L = a \lg v_\mathrm{rot} + b$}\label{TF}}
  \begin{tabular}{lccccccl}
  \hline 
  \hline
 Reference    & \multicolumn{2}{c}{$J$-band} & \multicolumn{2}{c}{$H$-band} & \multicolumn{2}{c}{$K$-band} & Comments \\    
              & a & b & a & b & a & b & \\  \hline 

This work &  3.43 & 2.67 & 3.44 & 2.83 & 3.55 & 2.66 & all types   \tabularnewline
Karachentsev et al. (2002) & 3.28 & & 3.48 & & 3.61 & & Based on $J_\mathrm{ext}$, $H_\mathrm{ext}$ and $K_{s,\mathrm{ext}}$, RFGC galaxies \tabularnewline
Courteau et al. (2007) & 2.5  & 4.0 & 2.6 & 4.0 & 2.6 & 3.9 &  64\% Sa galaxies \tabularnewline
de Rijcke et al. (2007) &   &   &  &  & 3.46 & 2.44 & early-type galaxies \tabularnewline
Masters et al. (2008) & 3.63 & 1.05 & 3.61 & 1.30 & 4.01 & 0.36 & Sc galaxies \tabularnewline
Masters et al. (2008) & 3.12 & 2.32 & 3.12 & 2.53 & 3.45 & 1.76 & Sb galaxies \tabularnewline
Masters et al. (2008) & 2.44 & 4.11 & 2.43 & 4.31 & 2.78 & 3.48 & Sa galaxies \tabularnewline
 \hline
\end{tabular}
\end{minipage}
\end{table*}

\subsubsection[]{The Fundamental Plane of discs}

Following Moriondo et al. (1999) we fit a relation for disc parameters as:
\begin{equation} \lg h = a \lg v_\mathrm{rot} + b \, S_\mathrm{0,d} + c \, , 
\label{ScaleRel}
\end{equation}
involving the disc scalelength, the maximum rotational velocity, and the 
deprojected central surface brightness $S_0$, analogous to the FP of 
elliptical galaxies. To define coefficients of the linear approximation we used 
the standard least squares method. From the complete subsample we excluded 7 
galaxies: NGC~7232 and NGC~5775 are interacting, NGC~7814 is a pure S0 galaxy 
with very faint disc, NGC~4222 and NGC~100 are low surface brightness 
galaxies, NGC~5023 and NGC~4183 have very large errors of disc parameters. 
The results of the fitting are given in Table~\ref{Correltions2}. The 
scatter of points is rather large which may be caused by observational 
uncertainties and decomposition errors. The coefficients derived in this paper 
are comparable with those reported by Karachentsev (1989) for the $I$-band: 
$a=1.4$, $b=0.28$; and by Moriondo et al. (1999) for the $H$-band: 
$a=1.31 \pm 0.19$, $b=0.25 \pm 0.04$. Our coefficients as well as the 
coefficients derived by Karachentsev (1989) and Moriondo  et al. (1999) are 
not consistent with what would be expected on the basis of the virial theorem 
and a universal mass-to-light ratio ($a=2$ and $b=0.4$). 

We divided galaxies 
into galaxies with $n\ga2.2$ and galaxies with $n\la2.2$. In spite of the 
fact that the shape parameter $n$ is related to the bulge model, the 
fundamental planes of discs, which tie only disc parameters and the 
maximum rotational velocity of gas, are different for galaxies with different 
bulges. Perhaps it is due to observational errors and errors of decomposition 
for galaxies of different morphological types (for early-type galaxies 
rotational velocities and parameters of faint discs are less well defined  than 
for late-type galaxies). Another reason may be the real difference of discs 
in galaxies with low and high density bulges. This effect requires a more 
careful examination.
\begin{table*}
 \centering
 \begin{minipage}{95mm}
 \parbox[t]{95mm}{\caption{The Tully-Fisher relations and disc fundamental
planes for complete sample galaxies in the $J$, $H$, and $K_s$-bands. The
fundamental planes are presented for all complete sample galaxies, for
galaxies with $n\ga2.2$ and for galaxies with $n\la2.2$.}
 \label{Correltions2}}
  \begin{tabular}{l}
  \hline
  \hline 
  $J$:\,\,\,\,\,\,$\lg L = (3.43 \pm 0.29) \lg v_\mathrm{rot} + (2.67 \pm 1.39)$, $r=0.82$  \\
  $H$:\,\,  $\lg L = (3.44 \pm 0.28) \lg v_\mathrm{rot} + (2.83 \pm 1.32)$, $r=0.82$  \\
  $K_s$: $\lg L = (3.55 \pm 0.30) \lg v_\mathrm{rot} + (2.66 \pm 1.41)$, $r=0.81$  \\ \hline
  
  For all galaxies of the complete sample with known velocities: \\ \hline
  $J$:\,\,\,\,\,\,$\lg h = (1.017 \pm 0.101) \lg v_\mathrm{rot} + (0.173 \pm 0.021)S_\mathrm{0,d}-(5.04 \pm 0.11)$ \\
  $H$:\,\, $\lg h = (1.044 \pm 0.092) \lg v_\mathrm{rot} + (0.190 \pm 0.020)S_\mathrm{0,d}-(5.28 \pm 0.11)$\\
  $K_s$: $\lg h = (1.077 \pm 0.104) \lg v_\mathrm{rot} + (0.202 \pm 0.023)S_\mathrm{0,d}-(5.53 \pm 0.12)$\\ \hline
  
  For galaxies of the complete sample with $n\ga2.2$: \\ \hline
  $J$:\,\,\,\,\,\,$\lg h = (0.729 \pm 0.265) \lg v_\mathrm{rot} + (0.204 \pm 0.054)S_\mathrm{0,d}-(4.92 \pm 0.14)$ \\
  $H$:\,\, $\lg h = (0.790 \pm 0.149) \lg v_\mathrm{rot} + (0.193 \pm 0.029)S_\mathrm{0,d}-(4.70 \pm 0.11)$\\
  $K_s$: $\lg h = (0.750 \pm 0.149) \lg v_\mathrm{rot} + (0.209 \pm 0.031)S_\mathrm{0,d}-(4.83 \pm 0.12)$\\ \hline

  For galaxies of the complete sample with $n\la2.2$: \\ \hline
  $J$:\,\,\,\,\,\,$\lg h = (1.218 \pm 0.122) \lg v_\mathrm{rot} + (0.213 \pm 0.027)S_\mathrm{0,d}-(6.29 \pm 0.09)$ \\
  $H$:\,\, $\lg h = (1.224 \pm 0.149) \lg v_\mathrm{rot} + (0.230 \pm 0.025)S_\mathrm{0,d}-(6.45 \pm 0.08)$\\
  $K_s$: $\lg h = (1.281 \pm 0.146) \lg v_\mathrm{rot} + (0.231 \pm 0.026)S_\mathrm{0,d}-(6.53 \pm 0.08)$\\ \hline

\end{tabular}
\end{minipage}
\end{table*}

\subsection[]{Photometric Plane and \ser\ index}

\begin{center}
 \begin{figure*}
 \begin{center}
  \includegraphics[width=5.4cm,  angle=-90]{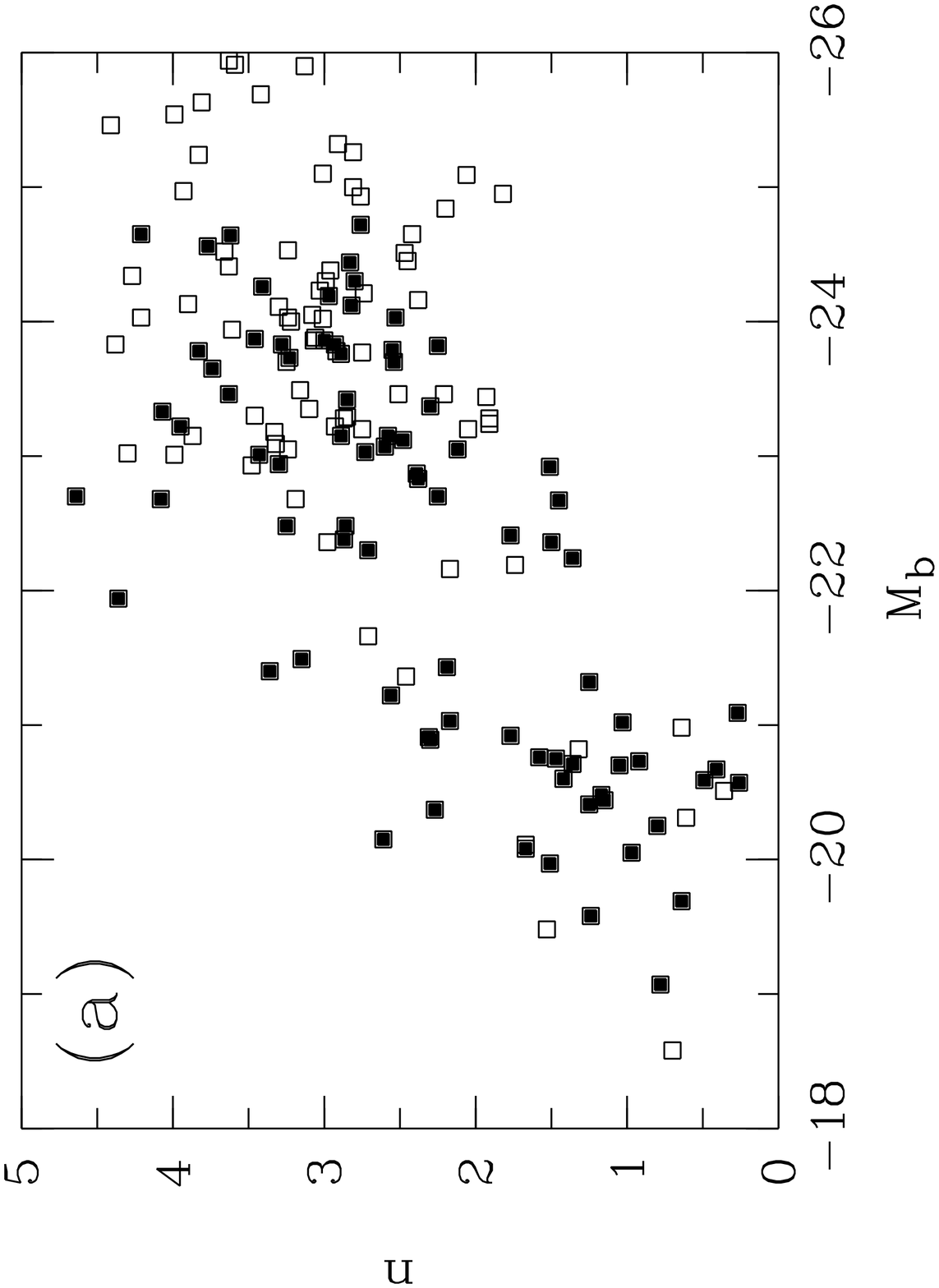}
  \includegraphics[width=5.4cm,  angle=-90]{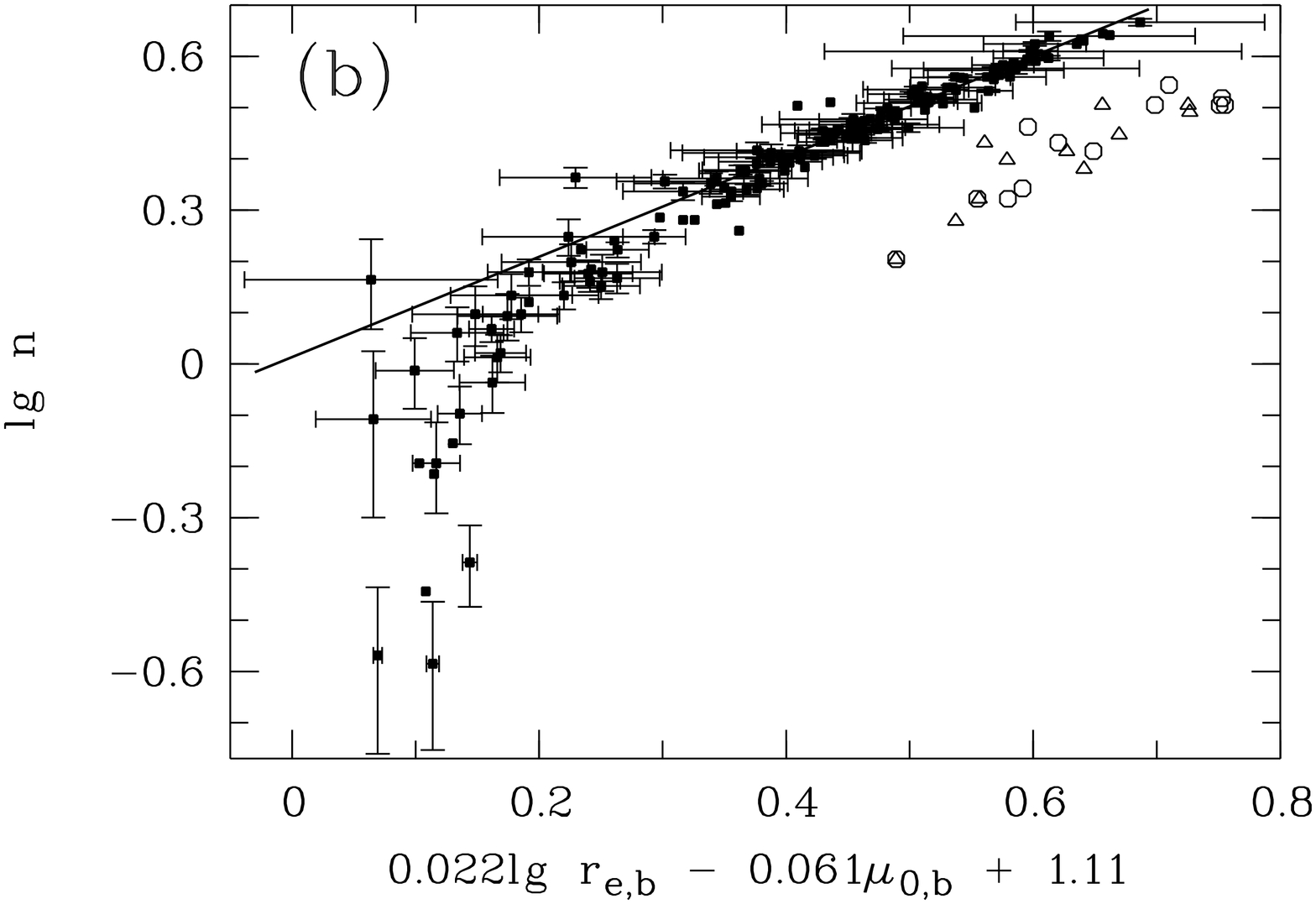}
 \end{center}
 \caption{\small{
 (a) Bulge \ser\ index $n$ vs. bulge absolute magnitude in the $K_s$-band. 
 Symbols as in Fig.~\ref{Corr1}. 
 (b) The Photometric Plane  built for bulges with $\lg n \ge 0.2$ of the 
 complete subsample in the $K_s$-band. The points corresponding to bulges 
 with $\lg n\la0.1$ do  not lie in this plane. The model merger remnants
 of disc galaxies are shown with open circles and triangles 
 (Aceves et al. 2006). The model data are shifted by a constant along the 
 horizontal axis. }}
 \label{BulgeCorr}
 \end{figure*}
\end{center}

The \ser\ index $n$ represents the parameter quantifying the 
radial concentration of the stellar distribution of bulges. 
The detailed structure of 
bulges, their formation and evolutionary status is now widely discussed. 
Here we present a correlation between bulge absolute magnitude 
$M_\mathrm{b}$ and $n$. The plot is shown in Fig.~\ref{BulgeCorr}a. The 
presented correlation is well-known 
(e.g. Andredakis et al. 1995; Graham 2001; Hunt et al. 2004; 
M\"ollenhoff \& Heidt 2001; M\"ollenhoff 2004) 
and seen in all bands --- from optical to near-infrared. It shows that 
exponential bulges are fainter than the de Vaucouleurs' bulges, although the 
scatter in the plot is large.

There are a lot of trends of galactic structural parameters with 
the \ser\ index of a bulge $n$, but the most prominent difference in 
parameters arises when one divides the sample under investigation into 
the subsample of classical bulges and pseudobulges. As was shown above, 
perhaps, the difference between classical bulges and pseudobulges is 
reflected 
in the intrinsic flattening and the 3D shape of bulges, in the ratio 
$r_\mathrm{e,b}/ h$ and in building fundamental planes for discs.

The division of bulges into classical bulges and pseudobulges was described in 
many works (see Kormendy \& Kennicutt 2004 for review). Classical bulges 
appear to be similar to elliptical galaxies and have similar properties. 
It is assumed that these systems were built via minor and major merging. 
The galaxies with pseudobulges are classified by the presence of some features 
within a bulge, e.g. nuclear bars, nuclear spirals and nuclear rings 
(Erwin \& Spark 2002; Carollo et al. 1997; Fisher \& Drory 2008). 
Pseudobulges are thought to be formed via disc instabilities and secular 
evolution and have disc-like apparent flattening. One of the main differences 
between classical bulges and pseudobulges is in the shape parameter $n$. 
Pseudobulges as usual have $n\la2$ and classical bulges have $n\ge2$ with 
little-to-no overlap.

The images of galaxies under investigation have low resolution and inner 
features inside a bulge are smoothed. That is why we were searching for
correlations between obtained structural parameters of the bulges with 
different \ser\ index $n$, not with morphological features. Here we describe 
the difference between bulges and pseudobulges while building the photometric 
plane for them. 

Khosroshahi et al. (2000a,b) found a tight correlation of the \ser\ indices 
$n$ with the central surface brightness $\mu_\mathrm{0,b}$ and the effective 
radius of a bulge $r_\mathrm{e,b}$. They called this relationship the 
Photometric Plane. 

Following Khosroshahi et al. (2000a,b) we performed the least-squared fit 
which expresses $\lg n$ as a linear combination of two other parameters: 
$\lg n = a \lg r_\mathrm{e} + b \, \mu_\mathrm{0,b} + c$. From the complete 
subsample we exclude galaxies with very small bulges and 2 bright flat 
galaxies of late types: NGC~4244 and NGC~4517. 

For the complete sample the best fitting results are presented in 
Table~\ref{FP}. Khosroshahi et al. (2000a,b) found the Photometric Plane for 
26 bulges of predominantly early-type galaxies in the $K$-band: 
$\lg n = (0.130 \pm 0.040)\lg r_\mathrm{e,b} - 
(0.073 \pm 0.011)\mu_\mathrm{0,b} + (1.21 \pm 0.11)$. 
M\"ollenhoff \& Heidt (2001) obtained for their sample of 41 predominantly 
late-type spirals in the $JHK$ bands: 
$\lg n = 0.187 \lg r_\mathrm{e,b} - 0.081\mu_\mathrm{0,b} + 1.34 \pm 0.10$ 
with the correlation coefficient $r = 0.91$. Finally, 
M\'endez-Abreu et al. (2008) gave 
$\lg n = 0.17(\pm 0.02)\lg r_\mathrm{e,b} - 
0.088(\pm 0.004)\mu_\mathrm{0,b} + 1.48(\pm 0.05)$ for their sample of 148 
unbarred S0--Sb galaxies in the $J$-band. Within error bars, the differences 
between our coefficients for all our complete sample and those of 
Khosroshahi et al. (2000a,b), M\"ollenhoff \& Heidt (2001) 
and M\'endez-Abreu et al. (2008) are not significant, at least in the $K$-band.

The surprising result was the non-linear correlation of the PhP for galaxies 
with various values of $n$ (Fig.~\ref{BulgeCorr}b). From our main sample we 
extracted the subsample of galaxies with $\lg n \ga 0.2$ ($n \ga 1.58$) and 
obtained the best-fit Photometric Plane (see Table~\ref{FP}). It occurred that 
the galaxies with $\lg n \ga 0.2$ lying on the PhP, that is built for these 
galaxies, has a small scatter, but the galaxies with $\lg n \la 0.1$ 
($n \la 1.26$) do not correspond to this plane and form their own 
plane (see Table~\ref{FP}). Inside the window $0.1 <\lg n\la 0.2$ there are 
points which do not correspond to these two planes. The regression of the PhP 
is non-linear, but we approximate this dependence as a superposition of two 
linear regressions. The definite difference between the photometric planes 
of the bulges with different shape parameters $n$ is shown in 
Fig.~\ref{BulgeCorr}b. The error bars are plotted as an indicator of the 
really different location of bulges and pseudobulges in the photometric 
planes. At a first glance the errors in the leftmost region of the plot are 
large, but bulges with very small $n$ are clearly recognized in surface 
brightness profiles of galaxies. What more, the relationship for bulges in 
the rightmost region of the plot begins to deviate from a linear trend 
for values of $n$ significantly above 1.
The main difference between our sample and the samples of the above-mentioned 
authors is the range of the shape parameters $n$. Our sample contains a 
substantial amount of bulges with $\lg n \la 0.1$. For other samples the low 
boundary for $n$ lies at $\lg n \ga 0.2$, that is why the curvature of the 
Photometric Plane towards small values of $n$ for bulges was not noticed 
earlier. 

This result is quite new and surprising. We realise that
the points on the lower-left of the figure have bulges with $n < 1$, which 
are somewhat controversial and need further investigation, which will be done 
in a future paper. But a similar planar relation associated with the 
Photometric Plane was found for bright ellipticals,
dwarf ellipticals and lenticular galaxies 
(Khosroshahi et al. 2000a; Khosroshahi et al. 2004; Ravikumar et al. 2006; 
Barway et al. 2009). If the range of the \ser\ parameter $n$ for ellipticals 
stretches to very small values of $n$, the curvature is clearly observed 
(Khosroshahi et al. 2004; Ravikumar et al. 2006).

The origin of the tight relation betveen $n$ and the linear combination 
of $\lg r_\mathrm{e}$ and $\mu_\mathrm{0,b}$ as well as the curvature of this 
relation are not completely clear. Aceves, Vel\'asquez \& Cruz (2006) argued 
that the curvature was related to an intrinsic property of a \ser\ profile. 
They wrote an expression for the total luminous matter associated with a 
\ser\ profile in log-space, involving $n$, $\lg r_\mathrm{e}$ and 
$\mu_\mathrm{0,b}$, and showed that for a set of galaxies with equal 
luminosities there were non-constant terms in this expression. These terms 
introduce a systematic change in a PhP-like expression.

It is not clear whether two distinct parts of a curve in 
Fig.~\ref{BulgeCorr}b reflect the different origin of bulges lying in 
the corresponding regions. However, the collisionless merger remnants of 
disc galaxies obtained in $N$-body simulations and fitted by a \ser\ profile 
fairly well reproduce the slope of the PhP in the region of $\lg n \ga 0.2$ 
(Aceves et al. 2006). We put on the data of Aceves et al. (2006) in our 
Fig.~\ref{BulgeCorr}b and found the numerical data are consistent 
with our observational ones.

The curvature of the PhP towards small values of $n$ may reflect the 
quite different nature of such bulges, formed, for example, via 
secular evolution of discs. One of the signs of secular evolution is the 
presence of a bar or other structures (such as rings, double exponential discs). 
We have made the assumptions about the existence of a bar or other 
structures in our sample galaxies on the basis of the 1D surface brightness 
profiles along the major axis and on the residual images where additional 
components, that were not included in the decomposition, become apparent. 
The galaxies with or without bars are represented among the galaxies with 
$\lg n \ga 0.2$ (classical bulges) as well as among the galaxies with 
$\lg n \la 0.2$ (pseudobulges). So the shape of the PhP may describe the 
deeper nature of these objects. M\'arquez et al. (2001) have suggested a 
theoretical explanation of the PhP of ellipticals. They calculated the 
specific entropy of elliptical galaxies (Entropic Surface), took into account 
a scaling relation between the potential energy and mass (Energy-Mass Surface) 
and showed that the PhP arose as an intersection line of these two planes 
demonstrating a curvature consistent with the observational data in the 
limited range of values that has been considered for $n$. All these arguments may 
be valid for bulges of all types too. 

\begin{table*}
 \centering
 \begin{minipage}{90mm}
 \parbox[t]{90mm}{\caption{The Photometric Planes for bulges of the complete 
 sample} 
 \label{FP}}
  \begin{tabular}{l}
  \hline
  \hline 
  For all galaxies of complete sample\\
  $J$:\,\,\,\,\,\,$\lg n = (0.053 \pm 0.059) \lg r_\mathrm{e,b} - (0.101 \pm 0.004)\mu_\mathrm{0,b} + (1.63 \pm 0.06)$ \\
  $H$:\,\,  $\lg n = (0.106 \pm 0.047) \lg r_\mathrm{e,b} - (0.084 \pm 0.004)\mu_\mathrm{0,b} + (1.37 \pm 0.05)$ \\
  $K_s$: $\lg n = (0.160 \pm 0.058) \lg r_\mathrm{e,b} - (0.084 \pm 0.005)\mu_\mathrm{0,b} + (1.34 \pm 0.07)$ \\  \\ \hline

  $\lg n\geq0.2$\\
  $J$:\,\,\,\,\,\,$\lg n = (0.030 \pm 0.018) \lg r_\mathrm{e,b} - (0.067 \pm 0.002)\mu_\mathrm{0,b} + (1.23 \pm 0.03)$ \\
  $H$:\,\,  $\lg n = (0.018 \pm 0.014) \lg r_\mathrm{e,b} - (0.062 \pm 0.002)\mu_\mathrm{0,b} + (1.14 \pm 0.02)$ \\
  $K_s$: $\lg n = (0.022 \pm 0.013) \lg r_\mathrm{e,b} - (0.061 \pm 0.002)\mu_\mathrm{0,b} + (1.11 \pm 0.02)$ \\  \\ \hline

  $\lg n\la0.1$\\
  $J$:\,\,\,\,\,\,$\lg n = (0.137 \pm 0.134) \lg r_\mathrm{e,b} - (0.142 \pm 0.039)\mu_\mathrm{0,b} + (2.31 \pm 0.67)$  \\
  $H$:\,\,  $\lg n = (0.201 \pm 0.162) \lg r_\mathrm{e,b} - (0.172 \pm 0.051)\mu_\mathrm{0,b} + (2.76 \pm 0.83)$  \\
  $K_s$: $\lg n = (0.467 \pm 0.214) \lg r_\mathrm{e,b} - (0.204 \pm 0.075)\mu_\mathrm{0,b} + (3.21 \pm 1.19)$ \\  \\ \hline
   \hline
\end{tabular}
\end{minipage}
\end{table*}

\subsection[]{The dark halo and disc flattening}
\label{DHalo}

In spite of a good correlation between the disc scalelength $h$ and 
the disc scaleheight $z_0$, the scatter of the ratio $h/z_0$ is fairly 
large. The ratio $h/z_0$ characterizes a relative thickness of a galaxy, 
or its flattening. Fig.~\ref{Distrib}d shows that $h/z_0$ varies from 
$\sim 8$ for thin galaxies to $1-2$ for plump discs. Disc flattening appears 
to correlate with the morphological type (e.g. de Grijs 1998) and the 
global HI content (Reshetnikov \& Combes 1997). Late-type and gas-rich 
galaxies are, on average, thinner than early-type systems, but this 
correlation is rather weak. Although the discs seem to be larger and thicker 
with a higher maximum rotation (van der Kruit \& de Grijs 1999; 
Kregel et al. 2002, 2005), the ratio $h/z_0$ is not correlated with the 
rotation velocity (Zasov et al. 2002; Kregel et al. 2002, 2005).

Zasov et al. (2002) and Kregel et al. (2005) found that the disc flattening 
shows a definite trend with the ratio of dynamical mass to disc luminosity 
provided the dynamical mass is defined as the total mass enclosed within the 
sphere of radius that is equal to four disc scalelengths: 
$\mathrm{M}_\mathrm{tot} = 4 h v_\mathrm{rot}^2 / G$. 
The total mass includes the 
mass of the disc $\mathrm{M}_\mathrm{d}$, the dark halo 
$\mathrm{M}_\mathrm{h}$ 
and the bulge $\mathrm{M}_\mathrm{b}$. So, the ratio of dynamical mass to disc 
luminosity determines the relative mass of a spherical component or the 
relative mass of the dark halo for bulgeless galaxies. The found correlation 
implies that relatively thinner discs in bulgeless galaxies 
tend to be embedded in more massive dark haloes.

Kregel et al. (2005) argued that the analytical collapse model of 
disc galaxy formation 
(White \& Rees 1978; Fall \& Efstathiou 1980; Blumenthal 1986; 
Mo, Mao \& White 1998), 
the local Toomre's (1964) stability criterion for a disc and the suggestion 
about vertical dynamical equilibrium, provided the 
constant value of the ratio of the vertical to the radial velocity dispersion 
$\sigma_\mathrm{z} / \sigma_\mathrm{R}$, lead to 
the correlation between the disc flattening and the ratio of dynamical 
mass to disc luminosity. 

Zasov et al. (2002) suggested that discs are marginally stable against 
the growth of perturbations in their planes and bending perturbations. 
This stability against bending perturbations 
means that the ratio $\sigma_\mathrm{z} / \sigma_\mathrm{R}$ must be larger 
than some analytical threshold: 
$\sigma_\mathrm{z} / \sigma_\mathrm{R} \ga 0.3-0.4$ 
(Polyachenko \& Shukhman 1977; Araki 1985). $N$-body simulations give 
a somewhat larger value for this threshold --- 
$\sigma_\mathrm{z} / \sigma_\mathrm{R} \simeq 0.6-0.8$ 
(e.g. Sotnikova \& Rodionov 2003, 2006). For a disc in vertical dynamical 
equilibrium Zasov, Makarov \& Mikhailova (1991) and Zasov et al. (2002) 
deduced a simple expression for $h/z_0$. We rewrite it in the following form:
\begin{equation}
 \frac{h}{z_0} 
 \sim 
 \frac{1}{\left[Q (\sigma_\mathrm{z} / \sigma_\mathrm{R})\right]^2} 
 \frac{\mathrm{M}_\mathrm{tot}}{\mathrm{M}_\mathrm{d}} 
 \sim
 \frac{f_{\lambda}}{\left[Q (\sigma_\mathrm{z} / \sigma_\mathrm{R})\right]^2} 
 \frac{\mathrm{M}_\mathrm{tot}}{L_\mathrm{d}}\, ,
\label{dark}
\end{equation}
where $Q$ is Toomre's (1964) stability parameter, and $f_{\lambda}$ is 
the mass-to-luminosity ratio in a fixed band. The linear relation 
between the ratio $h/z_0$ and the ratio of dynamical mass (or dark halo mass) 
to disc luminosity arises for marginally stable discs with $Q$ and 
$\sigma_\mathrm{z} / \sigma_\mathrm{R}$ near their thresholds.

To verify a linear trend (\ref{dark}), obtained analytically, 
Zasov et al. (2002) studied two different samples of edge-on bulgeless 
galaxies with known structural parameters in the $R$ and $K_s$ bands. They used 
the HI line width $W_{50}$ as a measure of dynamical mass and concluded that 
discs become thinner with increasing mass fraction of their dark halos. For 
our sample, which contains a great many galaxies with massive bulges, we 
performed the same analysis. 

\begin{center}
 \begin{figure*}
  \includegraphics[width=6cm, angle=-90]{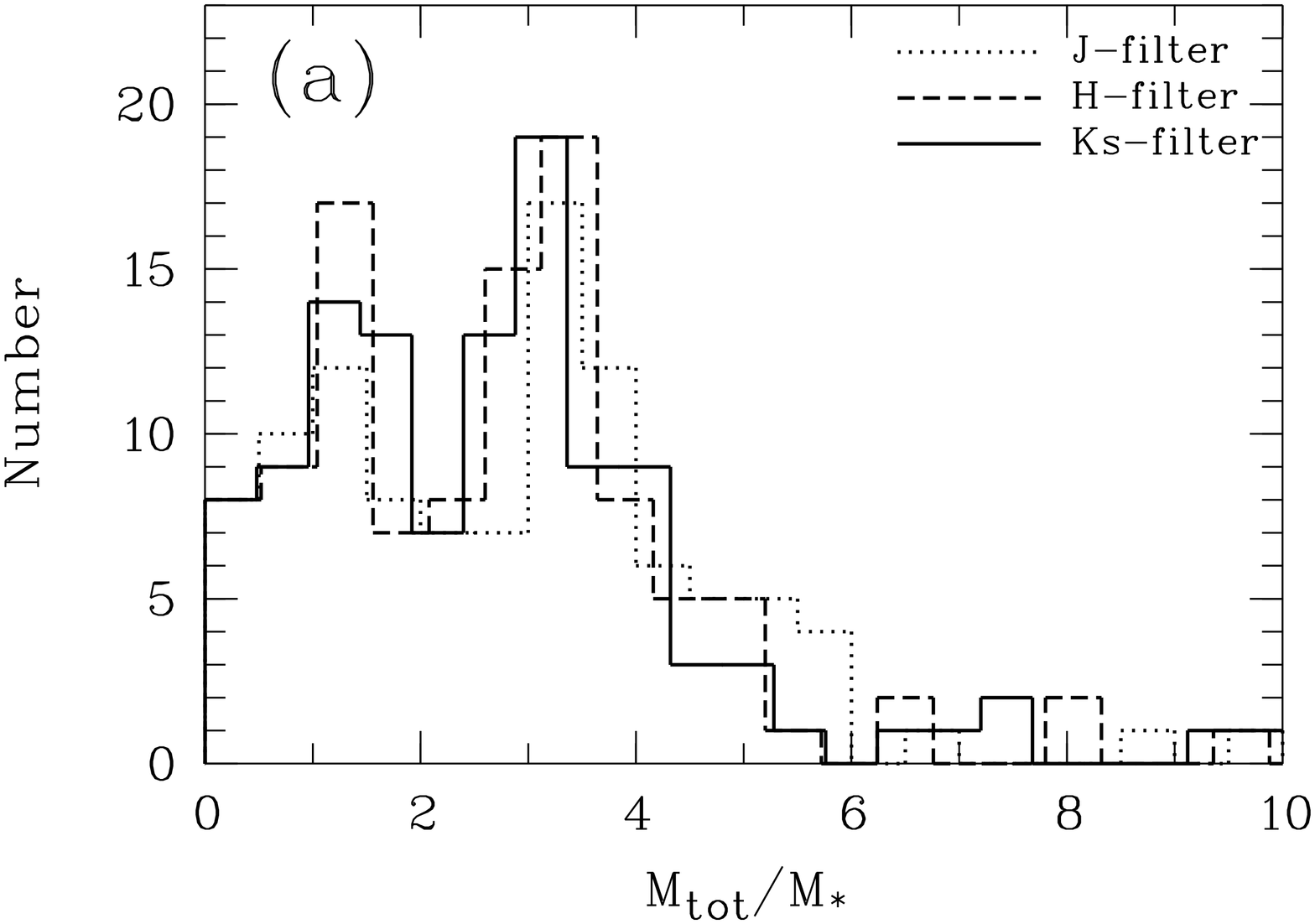}
  \includegraphics[width=6cm, angle=-90]{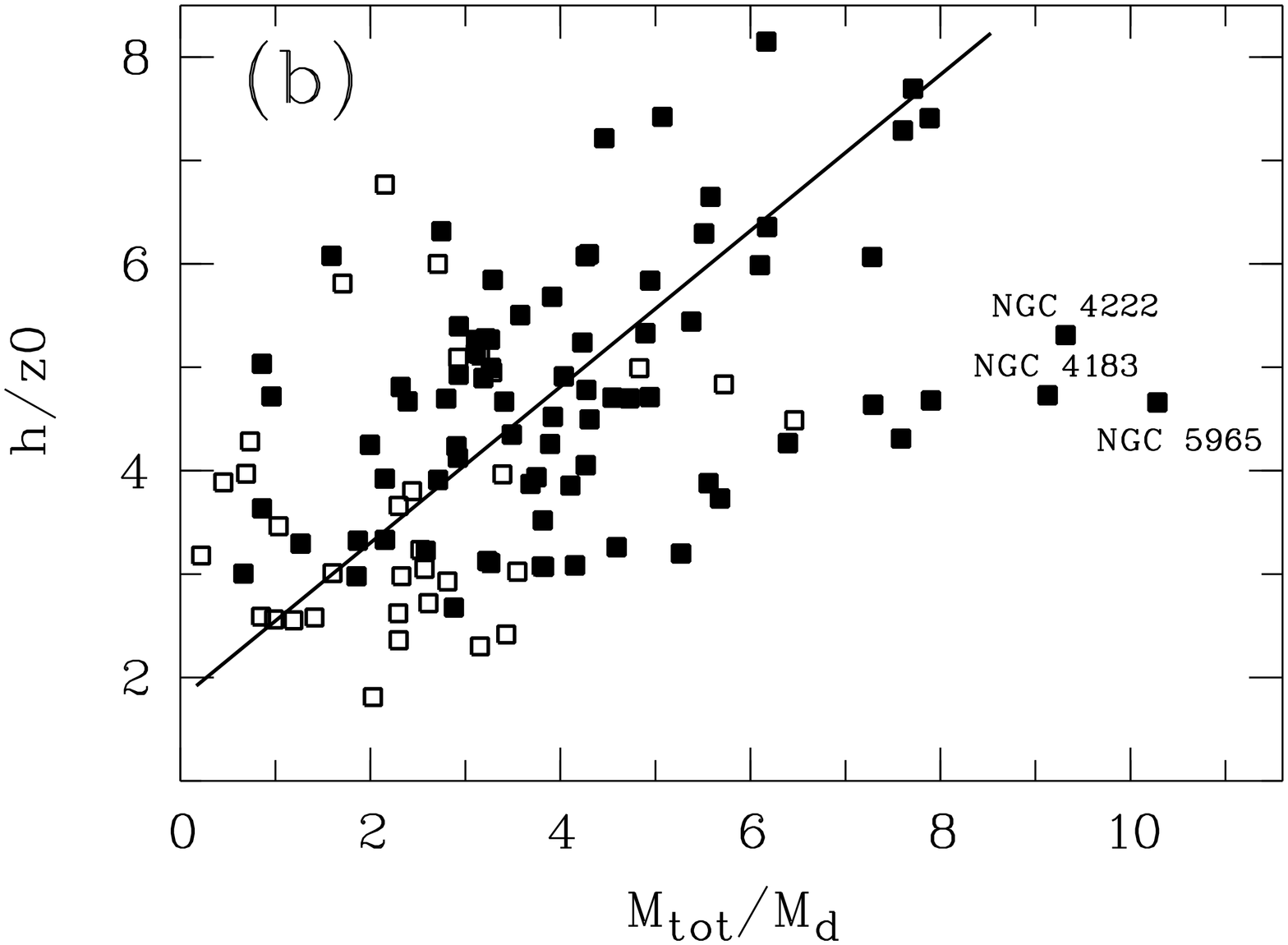}
 \caption{\small{
 (a) Distribution of the sample galaxies over the ratio of dynamical mass 
 to stellar mass of the galaxy. 
 (b) The ratio of $h/z_0$ as a function of the ratio of dynamical mass to 
 disc mass. Symbols as in Fig~\ref{Corr1}.
 }}
 \label{DHCorr}
 \end{figure*}
\end{center}

Fig.~\ref{DHCorr}a shows the distribution of the sample galaxies over the ratio 
of the dynamical mass to the stellar mass 
$\mathrm{M}_{*} = \mathrm{M}_\mathrm{d} + \mathrm{M}_\mathrm{b}$. 
We adopted mass-to-light 
ratios $f_J = 1.5$, $f_H = 1.0$, and $f_{K_s} = 0.8$ $\mathrm{M}_\odot / L_\odot$ 
(McGaugh et al. 2000). These mass-to-light ratios are consistent with the 
maximum disk best fitting the rotation curves for bright galaxies 
(e.g. Bottema 1999). There is a hint of bimodality in Fig~\ref{DHCorr}a. 
This may reflect the presence of two different 
families of galaxies with different bulges in our sample.

The distribution of the ratio $\mathrm{M}_\mathrm{tot}/\mathrm{M}_{*}$ 
ranges over the 
values from $\sim 1$ to $\sim 8-10$. The distribution of the ratio $h/z_0$
is varying in the same range (Fig.~\ref{Distrib}d). As the 
relation~(\ref{dark}) deals not only with the contribution of the dark halo to the 
total mass but with the relative mass of spherical components, including 
a bulge, we choose the ratio $\mathrm{M}_\mathrm{tot}/\mathrm{M}_\mathrm{d}$ 
to demonstrate 
a trend for the disc flattening with the relative mass of a spherical 
component (Fig.~\ref{DHCorr}b and Table~\ref{CorrelationsDH}). We excluded 
6 points from fitting of 
$\mathrm{M}_\mathrm{tot}/\mathrm{M}_\mathrm{d}$ vs. $h/z0$: 
NGC~7814, NGC~4222, NGC~5965, NGC~4183 and NGC~100, that have faint
disc, and NGC~5981 with a prominent bar.

The scatter of points in Fig.~\ref{DHCorr}b is rather large. It may be 
caused by uncertainties in velocities and mass-to-light ratio determinations, 
decomposition errors and disc relaxation processes, that can thicken a 
disc just above the threshold for marginal stability. But some of this 
scatter must be real. Sotnikova \& Rodionov (2005) concluded that the 
presence of a compact bulge is enough to suppress the bending instability 
that leads to the disc thickness increasing. A series of $N$-body simulations 
with the same total mass of a spherical component (dark halo $+$ bulge) 
were performed. The final disc thickness was found to be much smaller 
in the simulations, where a dense bulge is present, than in the simulations 
with bulgeless systems. The results of $N$-body simulations of discs starting 
from an unstable state were summarized by Sotnikova \& Rodionov (2006). They 
plotted the ratio $z_0 / h$ versus 
$(\mathrm{M}_\mathrm{h} + \mathrm{M}_\mathrm{b})/\mathrm{M}_\mathrm{d}$ 
and showed that there is a 
clear scatter in this relation, in spite of the same model mass for a 
spherical component ($\mathrm{M}_\mathrm{h} + \mathrm{M}_\mathrm{b}$). 
We are planning to 
compare in details our observational data with the results of simulations in a 
forthcoming paper.

\begin{table}
 \centering
 \begin{minipage}{90mm}
 \parbox[t]{90mm}{\caption{Correlations between 
 $z_0/h$ and $\mathrm{M}_\mathrm{tot}/\mathrm{M}_\mathrm{d}$}
 \label{CorrelationsDH}}
  \begin{tabular}{l}
  \hline
  \hline 
  $J$:\,\,\,\,\,\,$h/z_0 = (0.634 \pm 0.100)\mathrm{M}_\mathrm{tot}/\mathrm{M}_\mathrm{d} + (1.63 \pm 1.16)$, $r=0.58$  \\
  $H$:\,\, $h/z_0 = (0.696 \pm 0.120)\mathrm{M}_\mathrm{tot}/\mathrm{M}_\mathrm{d} + (1.76 \pm 1.51)$, $r=0.51$  \\
  $K_s$: $h/z_0 = (0.756 \pm 0.135)\mathrm{M}_\mathrm{tot}/\mathrm{M}_\mathrm{d} + (1.79 \pm 1.92)$, $r=0.46$  \\ \hline
 \hline
\end{tabular}
\end{minipage}
\end{table}

\section[]{Summary and conclusions}

We constructed the largest sample of edge-on spiral galaxies with  performed
2D bulge/disc decomposition of 2MASS galaxies images in $J$-, $H$- and
$K_s$-passband using the BUDDA v2.1 package. We extracted global bulge and
disc parameters for all objects. This first paper describes our edge-on 
galaxy sample. The main point of this work is in the first carefully fitting 
of edge-on spiral galaxies of early and late morphological types in 
3 near-infrared bands.

Edge-on galaxies are of great 
interest because they provide a unique possibility to obtain information 
about the vertical structure of a disk and a bulge. The disc scaleheight 
$z_0$ together with the disc scalelength $h$ determine the relative thickness 
of a stellar disc for each galaxy. This ratio as well as the bulge flattening 
$q_\mathrm{b}$ may put constraints on the disc and bulge formation processes 
and their secular evolution.

The results of the sample parameter analysis, that are specific for 
edge-on galaxies, can be summarized as follows:
\begin{enumerate}
\item Our data demonstrate a clear correlation between the disc scaleheight 
$z_0$ and the disc scalelength $h$ for all type galaxies. The distributions 
of the sample galaxies over the ratio of $h/z_0$ have a median value of 
3.5 for $J$-band and 3.9 for $H$ and $K_s$-bands. But there is a substantial 
scatter of this ratio (the distribution is fairly flat). The ratio $h/z_0$ 
stretches over a large range --- from very thin galaxies to plump ones. 
There is a correlation between the relative thickness of stellar discs 
$h/z_0$ and the relative mass of a spherical component, including a dark halo. 
This correlation was known previously for bulgeless galaxies (Zasov et al. 2002)
or samples with predominantly bulgeless galaxies (Kregel et al. 2005) 
and was argued to arise from marginally stable discs. 
Our sample is much larger than the samples of Zasov et al. (2002) and 
Kregel et al. (2005) and more reliable for statistical analysis. Here, 
we confirm the correlation under discussion. What is more, we do this 
not for bulgeless galaxies but for galaxies with massive bulges. We conclude
that the {\it total} mass contained in spherical components 
may be one of the factors that determines the final steady state disc thickness.

\item The bulge flattening in the vertical direction $q_\mathrm{b}$ divides 
the sample into two different families --- triaxial, nearly prolate bulges 
and close to oblate bulges with moderate flattening. The \ser\ index 
threshold $n \simeq 2$ can be used to identify these two bulge types. 

\item We found a correlation between $z_0$ and $r_\mathrm{e,b}$. It is new 
and has not been described previously. It means that the disc flattening is linked 
with the bulge structure.
\end{enumerate}

Many of our results are in good agreement with the results of other authors, 
but we found several new relations.

\begin{enumerate}
\item In accordance with previous studies we found that the scale parameters 
of bulges and discs are tightly correlated. The disc scalelength 
linearly increases when the bulge effective radius increases. However, there is 
a clear trend for the ratio $r_\mathrm{e,b}/ h$ to increase with $n$. As $n$ 
is an indicator of the Hubble type, such a trend unambiguously rules out the 
widely discussed hypothesis of a scale-free Hubble sequence. 

\item There is a hint that the fundamental planes of discs, which link only 
disc parameters and the maximum rotational velocity of gas, are different 
for galaxies with different bulges. This may indicate a real difference of 
discs in galaxies with low and high density bulges.

\item The most surprising result arises from the investigation of the 
Photometric Plane of sample bulges. The bulges with $n\ga2$ populate a
narrow strip in their Photometric Plane. However, there is a difference 
in behavior of this plane for bulges with $n\ga2$ and $n\la2$. 
The plane is not flat and has a prominent curvature towards small values of 
$n$. For bulges this fact was not noticed earlier. This result may 
be due to the physical distinction between classical bulges and 
pseudobulges.
As the Photometric Plane is proposed to be used as a distance indicator, 
the correct shape of the plane is of no little interest.
\end{enumerate}

Some of results were described only briefly, but it is clear that our sample 
is very useful for further detailed studying and modelling of edge-on 
spiral galaxies. In the forthcoming papers we are planning to discuss 
properties of edge-on galaxies in more detail in the context of their 
bulge and disc formation and joint evolution.

\section*{Acknowledgments}

We thank R.E. de Douza, D.A. Gadotti and S. dos Anjos for getting the last 
version of code BUDDA. We especially thank D.A. Gadotti for the helpful
comments to this program. We are indebted to the referee, Phil James,
for his constructive comments and insightful suggestions which helped to 
improve the quality and the presentation of the paper.

This work was supported by the Russian Foundation for Basic Research 
(grant 09-02-00968) and by a grant from President of the RF for support of 
Leading Scientific Schools (grant NSh-1318.2008.02).

This research has made use of the NASA/IPAC Extragalactic Database (NED)
which is operated by the Jet Propulsion Laboratory, California Institute
of Technology, under contract with the National Aeronautics and Space
Administration. We made use of the LEDA database 
(http://leda.univ-lyon1.fr).


\appendix

\begin{table}
 \begin{minipage}{140mm}
  \caption{The Sample.}
  \label{Table1}
  \begin{tabular}{clllllll}
  \hline 
  \hline
  \# & Galaxy & 2MFGC & Type & $D$ \\ 
      &     &    &      &   (Mpc)   \\ 
      &(1)&(2)&(3)&(4) \\  \hline
1 & UGC 529 & 633 & S0-a & 70.4  \tabularnewline
2 & NGC 504 & 1059 & S0 & 53.2  \tabularnewline
3 & NGC 527 & 1070 & S0-a & 74.7  \tabularnewline
4 & NGC 565 & 1133 & Sa & 57.0  \tabularnewline
5 & UGC 1120 & 1193 & Sab & 58.8  \tabularnewline
6 & ESO 353-G020 & 1202 & S0-a & 61.7  \tabularnewline
7 & UGC 1166 & 1243 & S0-a & 60.8  \tabularnewline
8 & ESO 013-G024 & 1422 & Sab & ~ ---  \tabularnewline
9 & UGC 1531 & 1557 & Sb & 105.0  \tabularnewline
10 & IC 207 & 1660 & S0-a & 62.1  \tabularnewline
11 & NGC 861 & 1758 & Sb & 106.0  \tabularnewline
12 & UGC 1938 & 1930 & Sbc & 82.0  \tabularnewline
13 & NGC 955 & 1965 & Sab & 17.0  \tabularnewline
14 & NGC 960 & 1984 & Sb & 63.1  \tabularnewline
15 & MCG-02-07-038 & 2044 & S0 & 62.2  \tabularnewline
16 & NGC 1029 & 2102 & S0-a & 46.0  \tabularnewline
17 & UGC 2304 & 2260 & Sbc & 93.7  \tabularnewline
18 & MCG-02-09-013 & 2605 & S0-a & 61.2  \tabularnewline
19 & ESO 005-G004 & 4913 & Sb & 26.1  \tabularnewline
20 & NGC 1381 & 2983 & S0 & 22.2  \tabularnewline
21 & NGC 1394 & 3025 & S0 & 55.5  \tabularnewline
22 & NGC 1401 & 3030 & S0 & 18.8  \tabularnewline
23 & NGC 4256 & 9696 & Sb & 35.7  \tabularnewline
24 & MCG-02-08-012 & 2274 & Sc & 68.8  \tabularnewline
25 & NGC 1529 & 3366 & S0 & 68.7  \tabularnewline
26 & UGC 7388 & ~ --- & SBb & 89.8  \tabularnewline
27 & ESO 251-G028 & 3869 & Sb & 158.0  \tabularnewline
28 & NGC 4866 & 10288 & S0-a & 31.2  \tabularnewline
29 & ESO 555-G023 & 4860 & S0-a & ~ ---  \tabularnewline
30 & ESO 555-G032 & 4905 & Sb & ~ ---  \tabularnewline
31 & ESO 087-G004 & 5097 & Sbc & 118.0  \tabularnewline
32 & UGC 4012 & 6163 & Sb & 130.0  \tabularnewline
33 & UGC 4190 & 6390 & S0 & 69.9  \tabularnewline
34 & UGC 4507 & 6803 & Sb & 141.0  \tabularnewline
35 & UGC 4526 & 6822 & Sab & 62.4  \tabularnewline
36 & ESO 433-G012 & 7223 & Sbc & 115.0  \tabularnewline
37 & ESO 373-G013 & 7473 & S0-a & 40.7  \tabularnewline
38 & NGC 2946 & 7484 & SBb & 122.0  \tabularnewline
39 & UGC 5345 & 7714 & Sb & 53.5  \tabularnewline
40 & IC 615 & 8108 & Sb & 133.0  \tabularnewline
41 & ESO 501-G056 & 8260 & S0 & 53.0  \tabularnewline
42 & NGC 3431 & 8449 & SABb & 75.3  \tabularnewline
43 & MCG-01-28-005 & 8496 & Sab & 91.0  \tabularnewline
44 & CGCG 267-039 & 8658 & Sb & 88.2  \tabularnewline
45 & NGC 3539 & 8691 & S0-a & 132.0  \tabularnewline
46 & NGC 1184 & 2685 & S0-a & 30.8  \tabularnewline
47 & NGC 4149 & 9578 & Sb & 43.4  \tabularnewline
48 & MCG-02-32-006 & 9725 & SBc & 60.0  \tabularnewline
49 & NGC 4686 & 10067 & Sa & 69.5  \tabularnewline
50 & MCG-02-33-076 & 10273 & Sab & 47.0  \tabularnewline
51 & UGC 8119 & 10315 & Sbc & 117.0  \tabularnewline
52 & IC 844 & 10368 & S0 & 43.6  \tabularnewline
53 & MCG-01-33-076 & 10374 & Sc & 58.4  \tabularnewline
54 & IC 5096 & 16137 & Sbc & 41.2  \tabularnewline
55 & IC 881 & 10662 & Sa & 96.9  \tabularnewline
56 & NGC 5119 & 10737 & S0-a & 43.6  \tabularnewline
57 & UGC 8462 & 10813 & Sab & 132.0  \tabularnewline
58 & UGC 8704 & 11102 & Sa & 140.0  \tabularnewline
59 & NGC 5290 & 11096 & Sbc & 37.4  \tabularnewline
60 & NGC 5553 & 11621 & Sa & 63.8  \tabularnewline
   \hline
\end{tabular}
\end{minipage}
\end{table} 

\appendix

\begin{table}
 \begin{minipage}{140mm}
  \caption{continued.}
  \label{Table2}
  \begin{tabular}{clllllll}
  \hline 
  \hline
  \# & Galaxy & 2MFGC & Type & $D$ \\ 
      &     &    &      &   (Mpc)   \\ 
      &(1)&(2)&(3)&(4) \\  \hline
61 & ESO 384-G014 & 11249 & S0 & 57.2  \tabularnewline
62 & IC 1110 & 12300 & Sa & 46.0  \tabularnewline
63 & ESO 387-G029 & 12429 & Sab & 60.6  \tabularnewline
64 & UGC 11060 & 14135 & Sa & 61.1  \tabularnewline
65 & NGC 4235 & 9679 & Sa & 37.4  \tabularnewline
66 & UGC 11614 & 15682 & Sb & 105.0  \tabularnewline
67 & IC 5181 & 16757 & S0 & 24.1  \tabularnewline
68 & NGC 7232 & 16791 & S0-a & 23.1  \tabularnewline
69 & ESO 471-G024 & 17877 & S0-a & 113.0  \tabularnewline
70 & NGC 482 & 1009 & Sab & 85.0  \tabularnewline
71 & NGC 522 & 1102 & Sbc & 32.8  \tabularnewline
72 & NGC 585 & 1169 & Sa & 68.9  \tabularnewline
73 & NGC 653 & 1286 & Sab & 69.9  \tabularnewline
74 & NGC 684 & 1382 & Sb & 44.1  \tabularnewline
75 & NGC 1596 & 3625 & S0 & 20.5  \tabularnewline
76 & IC 2085 & 3680 & S0-a & 13.4  \tabularnewline
77 & NGC 2295 & 5413 & Sab & 26.9  \tabularnewline
78 & ESO 311-G012 & 6172 & S0-a & 18.2  \tabularnewline
79 & NGC 3203 & 7996 & S0-a & 37.5  \tabularnewline
80 & NGC 3404 & 8433 & SBab & 66.8  \tabularnewline
81 & NGC 5047 & 10588 & S0 & 88.7  \tabularnewline
82 & ESO 321-G010 & 9597 & Sa & 46.5  \tabularnewline
83 & ESO 416-G025 & 2246 & Sb & 65.0  \tabularnewline
84 & NGC 3390 & 8406 & Sb & 45.7  \tabularnewline
85 & IC 127 & 1149 & Sb & 23.1  \tabularnewline
86 & NGC 3628 & ~ --- & Sb & 16.2  \tabularnewline
87 & NGC 3717 & ~ --- & Sb & 28.2  \tabularnewline
88 & NGC 5965 & 12550 & Sb & 46.8  \tabularnewline
89 & NGC 4013 & 9412 & Sb & 14.5  \tabularnewline
90 & ESO 446-G018 & 11466 & Sb & 67.5  \tabularnewline
91 & ESO 512-G012 & 11898 & Sb & 49.5  \tabularnewline
92 & NGC 4710 & 10109 & S0-a & 19.6  \tabularnewline
93 & NGC 7814 & ~ --- & Sab & 9.5  \tabularnewline
94 & ESO 240-G011 & ~ --- & Sc & 35.7  \tabularnewline
95 & NGC 4565 & ~ --- & Sb & 20.7  \tabularnewline
96 & NGC 5746 & ~ --- & SABb & 26.4  \tabularnewline
97 & NGC 891 & ~ --- & Sb & 4.3  \tabularnewline
98 & NGC 5775 & 12067 & SBc & 25.7  \tabularnewline
99 & NGC 4217 & 9661 & Sb & 16.8  \tabularnewline
100 & NGC 5908 & ~ --- & Sb & 45.6  \tabularnewline
101 & NGC 4244 & ~ --- & Sc & 6.8  \tabularnewline
102 & NGC 4570 & 9935 & S0 & 28.1  \tabularnewline
103 & NGC 5907 & 12346 & Sc & 10.0  \tabularnewline
104 & NGC 3564 & ~ --- & S0 & 42.5  \tabularnewline
105 & NGC 4266 & 9707 & SBa & 25.6  \tabularnewline
106 & IC 4202 & 10456 & Sbc & 98.5  \tabularnewline
107 & NGC 4703 & 10103 & Sb & 64.5  \tabularnewline
108 & ESO 443-G042 & 10371 & Sb & 43.5  \tabularnewline
109 & NGC 5308 & 11118 & S0 & 29.1  \tabularnewline
110 & NGC 5365A & 11257 & SBb & 40.5  \tabularnewline
111 & IC 1048 & 11928 & Sb & 25.2  \tabularnewline
112 & UGC 243 & 293 & Sb & 66.6  \tabularnewline
113 & ESO 315-G020 & 7524 & Sb & 68.9  \tabularnewline
114 & NGC 669 & 1340 & Sab & 59.5  \tabularnewline
115 & NGC 973 & 2024 & Sb & 62.4  \tabularnewline
116 & IC 335 & 2973 & S0 & 20.7  \tabularnewline
117 & IC 1970 & 2982 & Sb & 15.9  \tabularnewline
118 & NGC 1380A & 2985 & S0 & 20.0  \tabularnewline
119 & ESO 362-G011 & 4306 & Sbc & 18.8  \tabularnewline
120 & NGC 1886 & 4383 & Sbc & 24.3  \tabularnewline
   \hline
\end{tabular}
\end{minipage}
\end{table} 

\appendix

\begin{table}
 \begin{minipage}{140mm}
  \caption{continued.}
  \label{Table3}
  \begin{tabular}{clllllll}
  \hline 
  \hline
  \# & Galaxy & 2MFGC & Type & $D$ \\ 
      &     &    &      &   (Mpc)   \\ 
      &(1)&(2)&(3)&(4) \\  \hline
121 & NGC 2310 & 5487 & S0 & 17.9  \tabularnewline
122 & NGC 2654 & 6926 & SBab & 19.8  \tabularnewline
123 & NGC 2862 & 7315 & SBbc & 58.9  \tabularnewline
124 & NGC 3126 & 7847 & Sb & 73.2  \tabularnewline
125 & NGC 3692 & 8974 & Sb & 28.3  \tabularnewline
126 & NGC 3957 & 9342 & S0-a & 27.1  \tabularnewline
127 & NGC 4417 & 9805 & S0 & 15.8  \tabularnewline
128 & NGC 1247 & 2619 & Sbc & 50.9  \tabularnewline
129 & NGC 5981 & 12588 & Sbc & 24.5  \tabularnewline
130 & NGC 4026 & 9425 & S0 & 15.3  \tabularnewline
131 & NGC 3501 & ~ --- & Sc & 20.0  \tabularnewline
132 & NGC 5023 & 10525 & Sc & 8.3  \tabularnewline
133 & NGC 360 & 761 & Sbc & 30.1  \tabularnewline
134 & UGC 1817 & 1825 & Scd & 47.2  \tabularnewline
135 & MCG-02-10-009 & 3079 & Sc & 27.7  \tabularnewline
136 & UGC 3326 & 4605 & Sc & 55.1  \tabularnewline
137 & UGC 3474 & 5237 & Sc & 49.1  \tabularnewline
138 & NGC 3044 & 7660 & SBc & 22.3  \tabularnewline
139 & NGC 3279 & 8207 & Sc & 23.7  \tabularnewline
140 & NGC 4222 & 9670 & Sc & 7.7  \tabularnewline
141 & ESO 533-G004 & 16767 & Sc & 31.2  \tabularnewline
142 & NGC 4330 & 9747 & Sc & 25.8  \tabularnewline
143 & ESO 564-G027 & 7159 & Sc & 33.8  \tabularnewline
144 & ESO 531-G022 & 16371 & Sbc & 42.1  \tabularnewline
145 & IC 4393 & 11612 & Sc & 40.5  \tabularnewline
146 & ESO 263-G015 & 7900 & Sc & 38.0  \tabularnewline
147 & NGC 4835A & 10246 & Sc & 49.3  \tabularnewline
148 & ESO 288-G025 & 16587 & Sbc & 30.7  \tabularnewline
149 & IC 2058 & 3483 & Scd & 18.6  \tabularnewline
150 & ESO 201-G022 & 3389 & Sc & 55.4  \tabularnewline
151 & IC 4871 & 15015 & SABc & 25.1  \tabularnewline
152 & IC 4484 & 11988 & Sc & 61.3  \tabularnewline
153 & ESO 340-G008 & 15435 & Sc & 36.1  \tabularnewline
154 & UGC 12533 & 17534 & Sb & 75.8  \tabularnewline
155 & ESO 509-G019 & 10818 & Sbc & 144.0  \tabularnewline
156 & ESO 506-G002 & 9712 & Sbc & 57.9  \tabularnewline
157 & NGC 4183 & 9620 & Sc & 15.8  \tabularnewline
158 & NGC 4517 & 9881 & Sc & 20.1  \tabularnewline
159 & NGC 2357 & 5811 & Sc & 33.0  \tabularnewline
160 & UGC 5173 & 7514 & Sb & 87.6  \tabularnewline
161 & NGC 100 & 282 & Sc & 6.7  \tabularnewline
162 & ESO 121-G006 & 4933 & Sc & 17.6  \tabularnewline
163 & NGC 2424 & 6083 & SBb & 47.3  \tabularnewline
164 & IC 2207 & 6207 & Sc & 66.7  \tabularnewline
165 & ESO 563-G014 & 6848 & SBcd & 27.1  \tabularnewline
166 & UGC 6012 & 8487 & Sbc & 88.8  \tabularnewline
167 & ESO 507-G007 & 10056 & Sbc & 75.2  \tabularnewline
168 & IC 3799 & 10098 & Scd & 54.4  \tabularnewline
169 & MCG-03-34-041 & 10613 & Sc & 40.2  \tabularnewline
170 & NGC 5073 & 10645 & SBc & 41.4  \tabularnewline
171 & IC 4351 & ~ --- & Sb & 39.8  \tabularnewline
172 & NGC 5529 & 11577 & Sc & 41.4  \tabularnewline
173 & NGC 5714 & 11872 & Sc & 32.1  \tabularnewline
174 & ESO 340-G026 & 15509 & Sc & 71.2  \tabularnewline
175 & UGCA 150 & 7144 & SABb & 29.3  \tabularnewline
   \hline
\end{tabular}

\end{minipage}
 \parbox[t]{73mm}{ Columns: 
(1) Name (the first in NED). 
(2) Number in 2MFGC (2MASS Flat Galaxy Catalog. 
(3) Morphological type taken from LEDA (Lyon/MeudonExtragalactic Database). 
(4) Angular-size distance ($H_0=73$~(km~sec$^{-1}$)/Mpc, $\Omega_\mathrm{M}=0.27$, 
$\Omega_{\Lambda}=0.73$).} 
\end{table} 

\label{lastpage}

\end{document}